%% file: main.tex
\documentclass[twocolumn]{aastex701}

\accepted{for publication in \apjs, December 17, 2025}

\input{preamble}

\begin{document}
\title{The Second CHIME/FRB Catalog of Fast Radio Bursts}
\shorttitle{CHIME/FRB \cattwo}
\shortauthors{CHIME/FRB Collaboration}

\collaboration{1000}{The CHIME/FRB Collaboration}

\input{authors.tex}

\correspondingauthor{Emmanuel Fonseca}
\email{emmanuel.fonseca@mail.wvu.edu}
\correspondingauthor{Seth R.~Siegel}
\email{sethrsiegel@gmail.com}

\begin{abstract}
We present a catalog of \nfrbtot fast radio bursts (FRBs) observed with the Canadian Hydrogen Intensity Mapping Experiment (CHIME) telescope between \start and \finish. These bursts originate from \nsource unique sources, including \nfrbrep bursts from \nrepeater known repeating sources. For each FRB, the catalog provides a $O(10')$ estimate of sky location along with corresponding measurements of cumulative exposure time and survey sensitivity over the observing period. It includes a total-intensity dynamic spectrum between 400 and 800 MHz at 0.983~ms resolution.  From this spectrum, we constrain a model of the burst morphology and measure key parameters such as arrival time, intrinsic temporal width, dispersion measure, scattering time, and flux density.  This second catalog includes all FRBs from the first catalog, with every event reprocessed using a uniform and improved analysis framework.  We show that previously published inferences remain valid under the updated measurements.  We assess consistency of the detection rate across observational parameters, present initial distributions of burst properties, and outline ongoing and future studies that will use this catalog to investigate the nature of FRBs and their utility as astrophysical and cosmological probes.
\end{abstract}


\section{Introduction}
\label{sec:intro}

The Canadian Hydrogen Intensity Mapping Experiment (CHIME) has served as an unparalleled detector of fast radio bursts (FRBs) since the start of its operations in 2018 \citep{abb+18}. Within a few years, the CHIME/FRB backend has dramatically increased the number of known FRB sources from tens to hundreds \citep{aab+21}; this includes a significant expansion of sources that have been observed to emit multiple FRBs (henceforth, ``repeaters''), growing from a single known source to many dozens \citep[e.g.,][]{abb+19b,abb+19c,fab+20,abb+23}.  This growth in the FRB population allows for both per-source and population inferences on viable progenitor classes \citep[e.g.,][]{jam23}, the cosmological distribution of FRB sources \citep[e.g.,][]{smb+23,jpm+22a,jpm+22b}, and the impact of inter- and intra-galactic plasma on FRB dispersion and morphology \citep[e.g.,][]{ckr+22,cr22,kdl+24,sls+25}. The CHIME/FRB experiment has been pivotal in advancing FRB science by enabling the large-scale, untargeted detection of bursts across most of the radio sky. Despite imperfect localization, its $\sim$1000 detections per year -- observed with the same instrument and well characterized selection effects -- provide an unprecedented dataset for studying FRBs.

One way to solve the mystery of FRBs \citep[see][for reviews]{cc19,phl21} is by identifying rare sources that reveal key phenomenology and place strong constraints on progenitor models.  Because of its high detection rate, CHIME/FRB is particularly well suited for discovering such rare sources.  Examples include: an FRB from a Galactic magnetar \citep{abb+20,brb+20}; periodic activity from an FRB source \citep{aab+20}; and an FRB exhibiting both polarization position angle variations resembling those of Galactic radio pulsars \citep{mbe+25} and scintillation that constrains the emission region to be similar in scale to a neutron-star magnetosphere \citep{npb2024}. These measurements suggest or fully implicate neutron stars as the origins of at least the aforementioned FRBs. However, the gap in radio luminosity between observed Galactic radio bursts and those of the bulk of the FRB population, particularly those at cosmological distances, is still a puzzling factor of $10^6 - 10^9$, leaving open the possibility of additional progenitor classes.  Host localizations provide further support for this possibility:  FRB 20200120E was found in a globular cluster in M81 \citep{bgk+21,kmn+22}, an environments not typically associated with young neutron stars. Nonetheless, neutron stars may still form in such clusters through alternative channels such as accretion-induced collapse or white dwarf mergers \citep{blt+11,kfpr23}, suggesting that the progenitor may still be a neutron star, albeit with a non-standard formation history. In contrast, FRB 20240209A was localized to the outskirts of a massive, quiescent elliptical galaxy \citep{ssl+25,edf+25}, where even these alternative pathways are less well established. Together, these findings suggest that while neutron stars are likely responsible for many FRBs, a diversity of formation channels — and possibly progenitor types — may be needed to explain the full population.

Another way to constrain FRB models is by using large samples to perform full population synthesis studies \citep[e.g.,][]{jpm+22b,smb+23,wv2024,md2025}. The First CHIME/FRB Catalog \citep{aab+21}, hereafter \catone, presented 536 FRBs detected between \start and \finishone.  Population studies resulting from \catone verified the consistency of the sky distribution with isotropy \citep{jcc+21}, demonstrated that repeater and apparent non-repeater bursts have systematically different properties \citep{pgk+21}, showed that FRB scattering times are larger than expected if their host galaxies had interstellar media similar to the Milky Way's \citep{ckr+22}, and found a statistically significant cross-correlation between FRB sky locations and galaxies in the redshift range $0.3-0.5$, additionally providing evidence for a population of FRBs with large host dispersion measures \citep{rsl+21}.  Studies using \catone also compared FRB rates with other transient populations \citep{aab+21},  {measured FRB} energy distributions \citep[e.g.,][]{smb+23,tll+23}, assessed how well FRBs trace cosmic star formation \citep[e.g.,][]{jpm+22a, zz2022, smb+23, llz2024, wv2024}, analyzed the intrinsic FRB spectrum \citep{cjl+25}, investigated the use of machine learning techniques to classify sub-populations of FRBs \citep[e.g.,][]{lzz2023, zlz2023, qzy+25}, and estimated the repeater fraction \citep{jam23,yglh24}.  A large sample of FRBs also enables their use as novel cosmic probes of the baryonic content of the intergalactic and circumgalactic media \citep[e.g.,][]{cr22,cbg+23,wm23,crs+25}, and facilitates searches for associations with other transient phenomena \citep[e.g.][]{ctj+23,csk+24,pbf24} and multi-messenger counterparts \citep[e.g.,][]{mwj+23, lvk+23_chimefrb_O3a, lz24}.

We report here on continued sky monitoring by CHIME/FRB and present the Second CHIME/FRB Catalog, hereafter \cattwo, comprising \nfrbtot bursts detected between \start and \finish.  This represents an 8.5-fold increase in the number of bursts relative to \catone, which is fully included in this release. All bursts have been processed using a uniform and improved analysis framework, which we describe in detail.  We confirm that several key population-level inferences from the first catalog remain valid when revisited with the updated measurements and expanded sample. We assess the reliability of our results through internal consistency checks, including an examination of detection rate across various observational parameters. Finally, we present the observed distributions of burst properties and explore their correlations.  This catalog is intended to serve as a foundation for a wide range of studies into FRB origins, propagation effects, and potential use as cosmological probes.

The paper is organized as follows.  \S\ref{sec:obs} describes the CHIME/FRB observations, providing a brief overview of the system and discussing sky exposure, system sensitivity, and FRB verification. \S\ref{sec:catalog} details the catalog contents, including event names, sky localizations, and measurements of burst morphology and signal strength, along with the methods used to obtain them.  In \S\ref{sec:validation}, we assess systematic effects and test the internal consistency of the dataset. \S\ref{sec:discussion} presents the observed distributions of FRB properties. We conclude in \S\ref{sec:conclusions} with a summary of ongoing and future work enabled by this dataset.

\section{Observations}
\label{sec:obs}

The CHIME telescope and its FRB detection system have been described in detail elsewhere \citep{abb+18,abb+22}; here, we provide a brief summary.  The telescope is located on the grounds of the Dominion Radio Astrophysical Observatory (DRAO) near Penticton, British Columbia, Canada. It consists of four immovable 20 m × 100 m parabolic cylinders oriented N-S, with each cylinder axis populated with 256 dual-linear-polarization antennas sensitive in the range 400–800 MHz. Every day CHIME maps the entire sky north of $\dec > -9.5^{\circ}$.  Sources with  $\dec > +70^{\circ}$ are visible twice per day as they transit on opposite sides of the North Celestial Pole (NCP), while those very close to the NCP remain continuously within the field of view. The 2048 antenna signals are amplified, digitized, and split into 1024 frequency channels at 2.56~$\mu\mbox{s}$ time resolution by the ``F-Engine'' portion of CHIME’s correlator \citep{bbc+16}. These signals are then sent to the GPU-based ``X-Engine.'' The X-Engine performs a spatial correlation and polarization sum, forming 1024 independent total-intensity sky beams covering the primary beam \citep[256 N-S × 4 E-W;][]{nvp+17}.  The X-engine also performs an up-channelization to 16k frequency channels with 24.4~kHz resolution, while downsampling in time to 0.983~ms time resolution; these data are then transmitted to the CHIME/FRB backend to be used in the real-time FRB search.

We search each beam’s data in real time for FRBs using a triggering software pipeline consisting of four stages termed L0, L1, L2/L3 and L4 \citep[see][for details]{abb+18}. L0 does beamforming and up-channelization within the X-Engine. L1 is executed on a separate CPU-based cluster and is the primary FRB search pipeline, with radio frequency interference (RFI) mitigation \citep{rs23} and a highly optimized tree-style dedispersion, spectral-weighting, and peak-search algorithm (called ``bonsai'').  L1 nodes buffer intensity data that can be saved upon detection of a candidate FRB. L2/L3 combines results from all beams and groups detections, identifying likely unique events, and further rejecting RFI; this part of the pipeline also identifies known sources, and verifies a source’s extragalactic nature by ensuring its observed DM is greater than those predicted by the NE2001 and YMW16 models of electron number density \citep{ne2001,ymw17}. Metadata headers for FRB candidates are stored in the L4 database. Raw intensity data buffered by L1 are saved to disk for offline analysis. Since December 2018, the CHIME/FRB real-time pipeline has triggered the recording of raw telescope voltage data (hereafter referred to as baseband data) from CHIME’s 1024 dual-polarization antennas upon the detection of a bright FRB candidate. The CHIME/FRB baseband subsystem and its analysis pipeline are presented by \citet{mmm+21}. The analysis and results of the baseband data for \catone were presented by \citet{aaa+24}; a similar analysis of baseband data recorded for \cattwo FRBs will be presented in a forthcoming paper.

Since October 2021, the CHIME/FRB Virtual Observatory Event (VOEvent) Service has been issuing low-latency public alerts, broadcasting approximately two CHIME/FRB detections per day. These alerts are designed to facilitate rapid follow-up observations and are typically transmitted within 10–15 seconds of detection \citep{voe_25}. Because of this rapid turnaround, the metadata in VOEvents are not human-verified, and occasional retractions are therefore unavoidable. Of the VOEvents issued during the period covered by \cattwo, 1109 correspond to bursts validated and published here, while 89 were subsequently retracted, corresponding to a false-positive rate of 9\%. In addition, VOEvents do not include many of the quantities presented in this catalog, such as burst morphology parameters, flux densities, fluences, sensitivity thresholds, and exposure estimates. The official CHIME/FRB catalogs therefore complement the VOEvent stream by providing homogeneous, quality-controlled datasets that enable robust population-level analyses not possible from VOEvents alone.

\begin{figure*}[p]
\centering
\includegraphics[width=\textwidth]{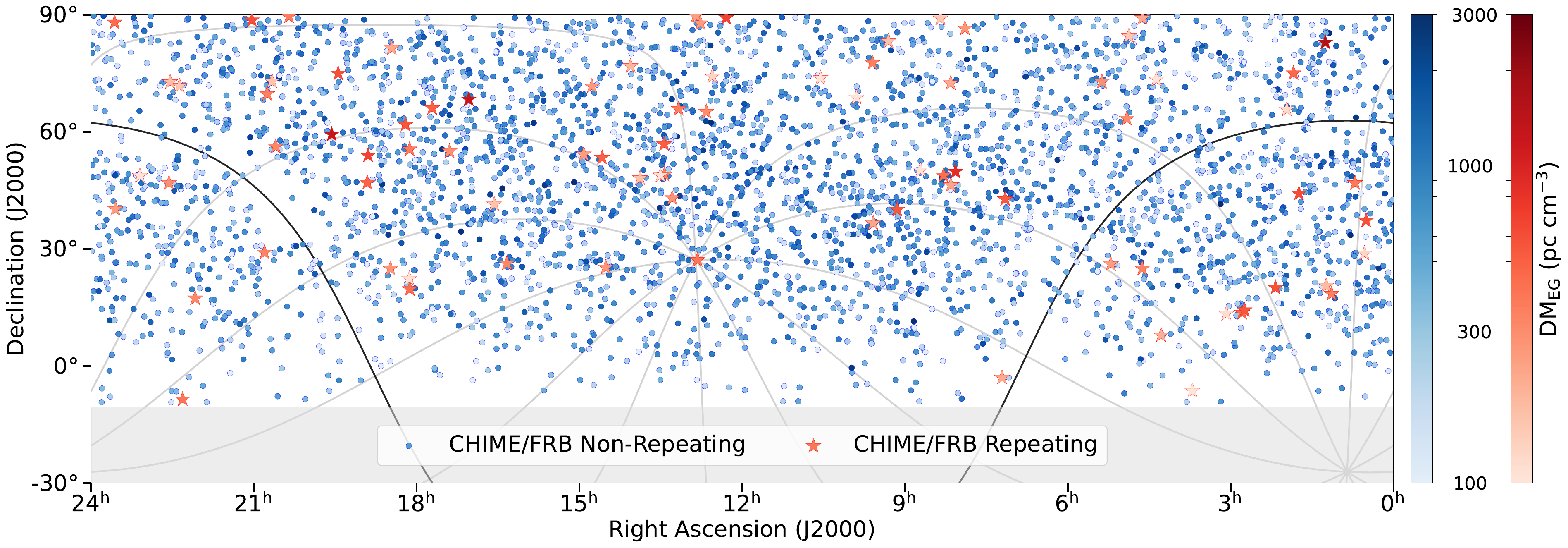}
\caption{Sky distributions of \nfrbnorep non-repeating sources and \nrepeater repeating sources (Cook et al., \inprep) are shown as blue circles and red stars, respectively. Marker color indicates the extragalactic \dm, obtained by subtracting the Galactic DM -- estimated using the NE2001 model \citep{ne2001} and a 30 pc cm$^{-3}$ contribution from the Galactic halo \citep{cbg+23} -- from the measured DM. For repeating sources, we show the inverse-variance-weighted average of the localizations and extragalactic DM values from all associated bursts (see \S\ref{sec:repeater} for details on the methodology used to identify repeating FRBs). The gray shaded region at the bottom indicates declinations outside the CHIME/FRB field of view ($\dec \leq -9.5^{\circ}$). The black line marks the Galactic plane, while thin gray lines denote graticules of constant Galactic longitude, spaced by $30^{\circ}$. As a transit telescope, CHIME provides relatively uniform exposure in this Mercator projection of equatorial coordinates. \label{fig:coords}}
\end{figure*}

The following criteria were applied to candidate events identified by the real-time system to produce the final list of cataloged events:
\begin{enumerate}
    \item Must have occurred between \start and \finish (UTC, inclusive).

    \item Must have a signal-to-noise ratio (\snr) measured by the real-time pipeline above a threshold, which was gradually lowered as enhancements to the real-time pipeline and visualization software reduced false positives and improved the capacity for human inspection (see Criterion~6):
        \begin{itemize}
            \item $\snr \geq 10$ from \start to 13 December 2018.
            \item $\snr \geq 9$ from 13 December 2018 to 1 September 2019.
            \item $\snr \geq 8.5$ from 1 September 2019 to \finish.
        \end{itemize}
    with the following exceptions:
        \begin{itemize}
            \item From 19 May 2019 to 1 January 2022, events that the real-time pipeline identified as potential repeat bursts from a known source were considered if they had $\snr \geq 8$.  Note that in five cases the event was later determined to be unrelated to the known source but is nevertheless included in the catalog: FRB 20200109C, 20200204H, 20200423E, 20200915B, and 20201001A.
            \item From 20 May 2020 to 1 January 2022, events with $\snr \geq 8.0$, $312 \ \mbox{pc~cm}^{-3} \leq \dm \leq 352 \ \mbox{pc~cm}^{-3}$, and $\dec \geq 21^{\circ}$ were considered due to the possibility of being associated with SGR~1935+2154.  In four cases, the event was later determined to be unrelated to SGR~1935+2154 and was included in the catalog: FRB 20200609B, 20200701G, 20201002B, and 20201017B.
            \item From 30 October 2020 to 1 January 2022, events with \(7.5 \leq \snr < 8.5\) were forwarded to citizen scientists for evaluation (see \S\ref{sec:citizen}).
        \end{itemize}

    \item Must have total-intensity callback data.  A small number of candidate events identified by the automated pipeline lack saved intensity data due to errors in the callback system. These events are excluded from the catalog, as their verification and further analysis are not possible.  This is estimated to account for approximately 1\% of all candidate events.  A more detailed assessment will be performed as part of future population studies, particularly those aimed at constraining the all-sky FRB rate.

    \item Must lack an association with any known Galactic sources, as determined by the criteria outlined in \S\ref{sec:repeater}.

    \item Must have a dispersion measure (DM) exceeding the maximum Galactic DM along the line of sight, as predicted by both the \cite{ne2001} and \cite{ymw17} models. This criterion is discussed further in \S\ref{sec:repeater}.

    \item Must be independently verified as an FRB candidate through visual inspection of the dynamic spectrum and evaluation of metadata by two human reviewers.  In cases of disagreement (187 events), one of three experienced members of the catalog team made the final determination.  Of these, roughly half (97 events) were ultimately classified as FRBs.

    \item Must be confirmed as unique and not a duplicate of previously cataloged bursts. In 35 cases, events in the same beam or neighboring beams were not grouped correctly by the L1 or L2/L3 pipeline due to differences in arrival time and/or DM. These were later flagged as duplicates, and only the event with the higher \snr was included.
\end{enumerate}
In total, \nfrbtot events met these seven criteria and are included in \cattwo.  The sky distribution of these bursts is shown in Figure~\ref{fig:coords}.

\subsection{Sky Exposure}
\label{sec:exposure}

\begin{figure*}[p]
    \includegraphics[scale=0.7]{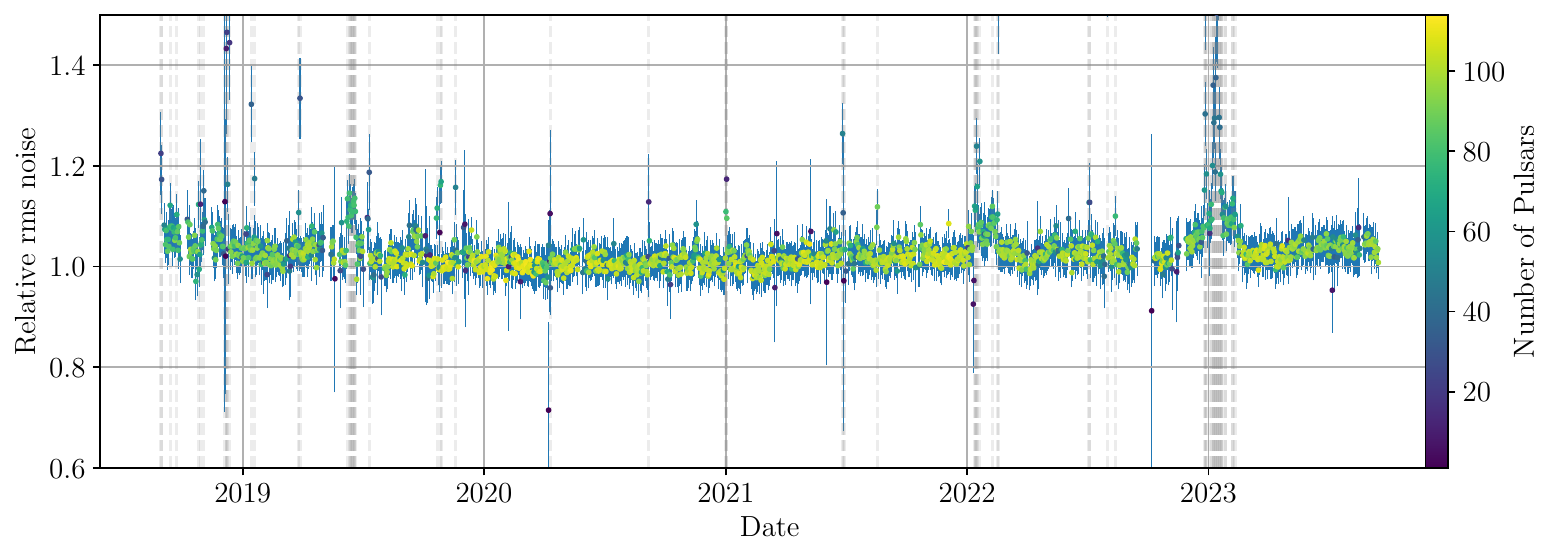}
    \caption{Daily sensitivity variation from \expstart to \finish. The vertical axis represents the relative RMS noise, $\sigma_{\rm pulsar}(t)$, computed from the \snr of individual pulses detected from a selection of known pulsars transiting overhead \citep{aab+21}. The color bar indicates the number of distinct pulsars detected per day. Dashed gray lines mark days where the RMS noise variation exceeded 10\%; these days, comprising approximately 4.5\% of the total observing time, were excluded from the final exposure calculation. Reduced sensitivity observed during early 2023 corresponds to heavy snowfall at the site in January and February of that year. }
    \label{fig:sens}
\end{figure*}

\begin{figure*}[p]
\centering
\gridline{\fig{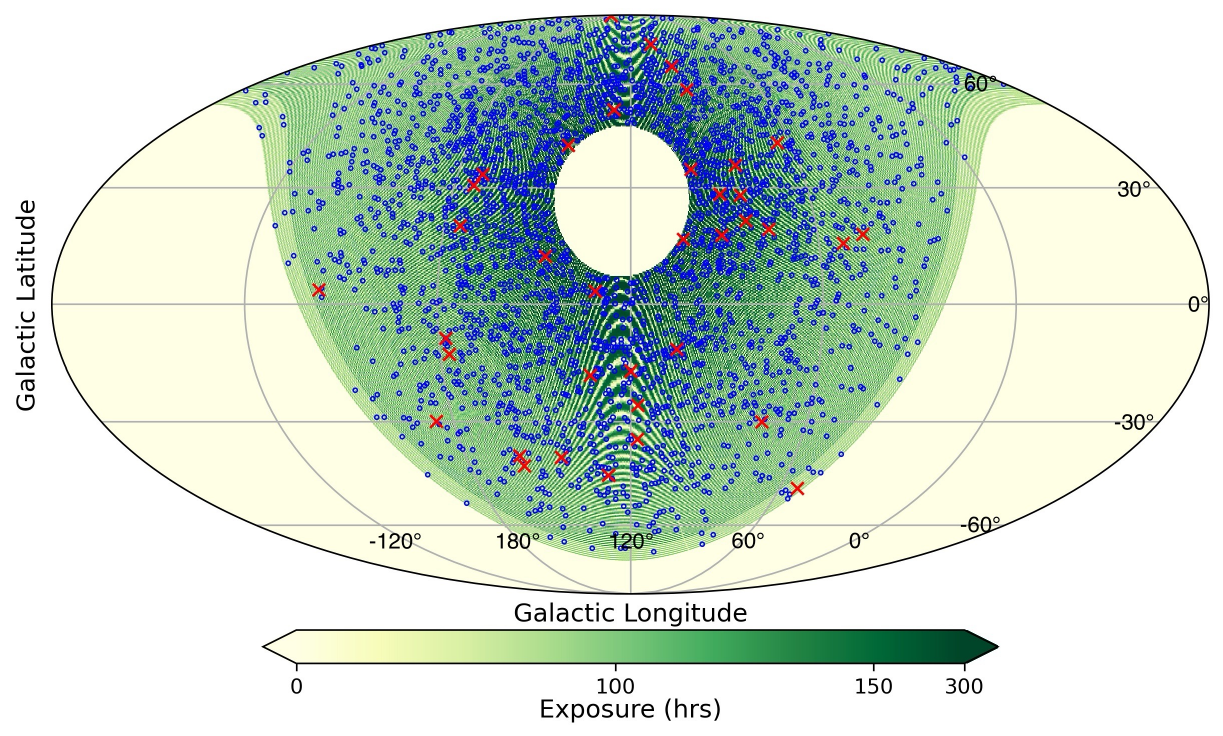}{0.85\textwidth}{}}
\gridline{\fig{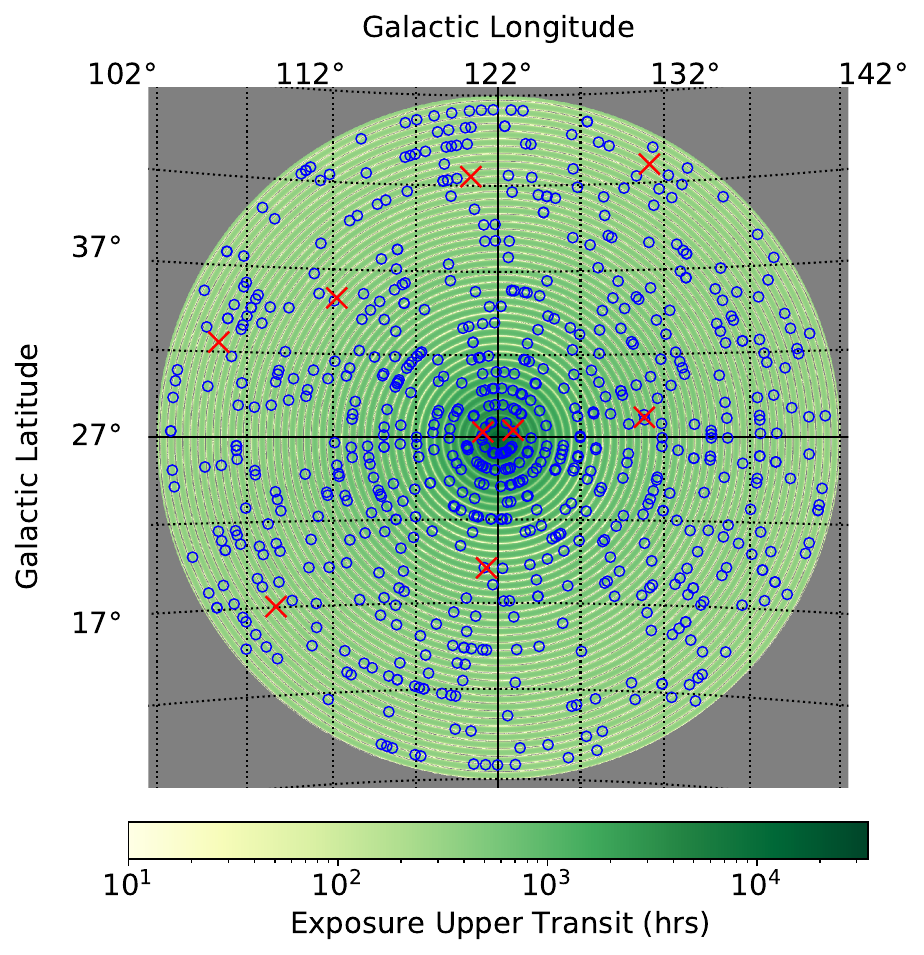}{0.425\textwidth}{}
          \fig{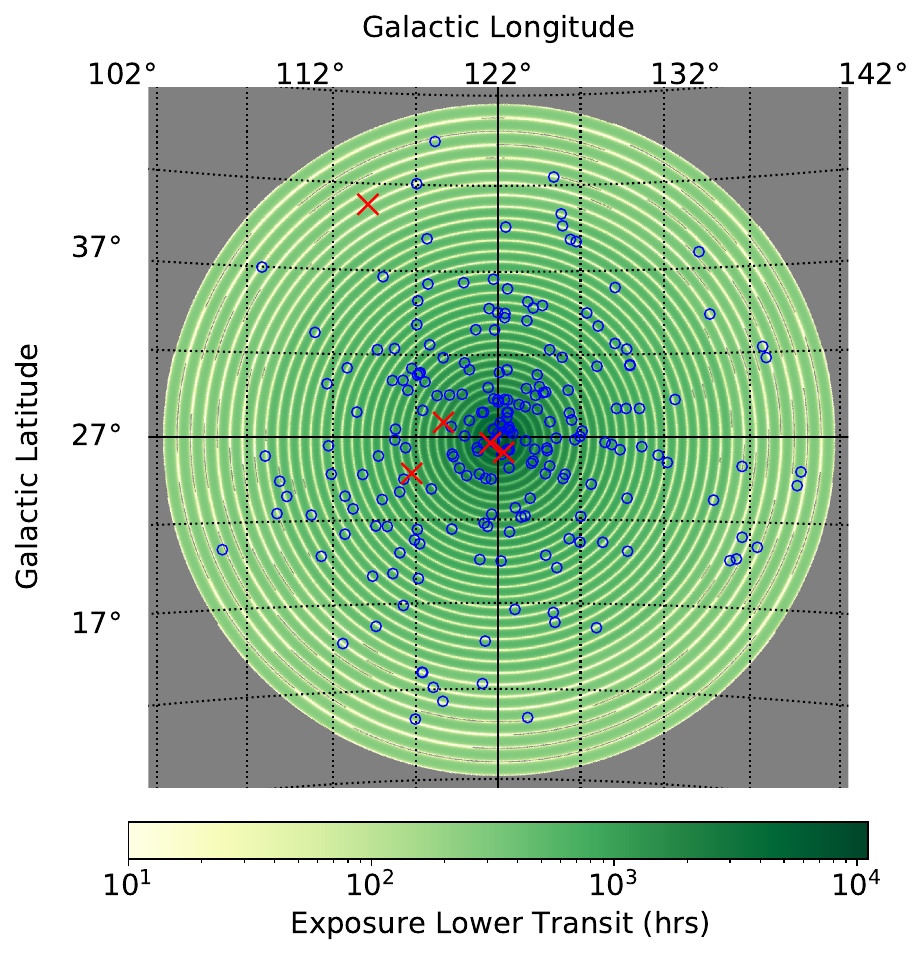}{0.425\textwidth}{}}
    \caption{Galactic coordinate sky maps, depicting the locations of all non-repeating (blue circles) and repeating (red crosses) FRB sources observed during the period from \expstart to \finish, overlaid on the total exposure of the CHIME/FRB system during that time. The top panel displays the sky locations that transit through the primary beam of the telescope once per day (with $\dec < 70^{\circ}$), while the bottom panels illustrate the upper (left) and lower (right) transit exposures for sources that transit through the primary beam twice per day (with $\dec > 70^{\circ}$). The bottom panel maps are centered on the North Celestial Pole and utilize a logarithmic color scale. We detect fewer FRBs in the lower transit due to reduced sensitivity.}
    \label{fig:exposure}
\end{figure*}

We define the exposure at each sky position as the total amount of time that region spent within the full width at half maximum (FWHM) of a CHIME formed beam, evaluated at 600 MHz (the center of the observing band), during periods when both the telescope and the relevant data-processing path were operational and the system was achieving nominal sensitivity. We provide the resulting exposure as an all-sky map. For each event in \cattwo, we also compute a localization-weighted exposure by integrating the exposure map over a uniform grid of positions within the 90\% confidence region of the event's sky localization, using the localization probability map as weights (see \S\ref{sec:localization}). This yields an estimate of the total observing time applicable to each source. Our methodology follows the approach introduced in \cite{aab+21}, though we provide a review of the procedure below.

CHIME is a transit telescope oriented N-S and nominally operates 24 hours per day.  However, the telescope is occasionally offline due to maintenance or power outages.  Even when online, parts of the system may be degraded; for example, a failed GPU node in the X-engine can reduce the number of available frequency channels for the FRB search.  For this reason, we track a variety of metrics related to the downtime of each stage of the observing system \citep[see][for more details]{aab+21}.  A major improvement in our exposure estimation for \cattwo, relative to \catone, is the automation of monitoring for the L2/L3 and L4 stages, now performed at a cadence of 4 seconds, replacing manual checks previously done every few hours.  Based on this monitoring, we find the system was fully operational for 76\% of the 1837 days of observation between \expstart and \finish. Note that exposure estimates are not available for the commissioning period between \start and \expstart. A total of 13 FRBs were detected during commissioning; these are flagged in the catalog tabular data with \texttt{excluded\_flag} = \True and should be omitted from statistical analyses

In addition to tracking uptime, we monitor daily variations in telescope sensitivity -- e.g., due to snow on the reflector or temporarily enhanced RFI -- using a set of known Galactic pulsars.  As these sources transit the CHIME field of view, we record the \snr of individual pulses and use this to estimate the average RMS noise of the instrument (see \S3.3 in \citet{cha22}).  Both the number of pulsars detected each day and the inferred RMS noise are shown in Figure~\ref{fig:sens}.  Days on which the RMS noise exceeds 10\% of the median value (computed over the first 920 days of operation: September 2018 - March 2021) are excluded from the exposure calculation.  This threshold differs from the approach used in \catone, which excluded days with RMS more than 1$\sigma$ above the median value computed over the \catone observation period.  We adopt a fixed 10\% threshold to ensure that the same days are consistently removed in future pipeline runs and avoid the over-exclusion of early data, when the system was less sensitive. In total, $4.5\%$ of days are excluded due to high RMS noise. Including such intervals would bias population-level inferences by introducing an inconsistent selection function. A total of 121 FRBs were detected on days with elevated RMS noise; these are flagged with \texttt{excluded\_flag} = \True and should be omitted from statistical analyses.

We generate exposure maps by combining the operational and sensitivity masks with the CHIME beam model\footnote{\url{https://chime-frb-open-data.github.io/beam-model/}}, considering only sky regions within the FWHM of a formed beam at 600 MHz as detectable.  For \cattwo, the exposure maps were generated using a 12-second integration time (compared to 4 seconds in \catone) to reduce computational cost.  Figure \ref{fig:exposure} shows the exposure map for the full observing period (\expstart to \finish), split between sky regions with $\dec < 70^{\circ}$ (top) and $\dec > 70^{\circ}$ (bottom).  The positions of  non-repeaters (blue) and repeaters (red) from the catalog are marked.  The region with $\dec > 70^\circ$ transits through the field of view twice per sidereal day: once during the ``upper'' transit, when the source is south of the NCP and at higher elevation, and once during the ``lower'' transit, when the source is north of the NCP and at lower elevation.  These transits correspond to the bottom left and bottom right panels of the figure, respectively.  The two exposures cannot be summed directly because the response of both the primary and formed beams differ between the two transits.  Exposure maps for both transits are available for download from \data.

The concentric ring pattern in the exposure map highlights regions where the estimated exposure is zero, corresponding to locations that fall between the FWHM of adjacent formed beams at 600~MHz. Although the instrumental sensitivity is non-zero in these areas and bursts may be detected there, our maps assign zero exposure to sky positions that do not transit within a formed beam’s FWHM. This modeling choice was made to maintain consistency with previous CHIME/FRB publications and to ensure a well-defined and reproducible exposure criterion.  Population synthesis efforts should assess detectability using this definition of exposure and the CHIME/FRB beam model, which describes the instrument’s sensitivity across the sky.

\begin{figure*}[t]
    \includegraphics[scale=0.43]{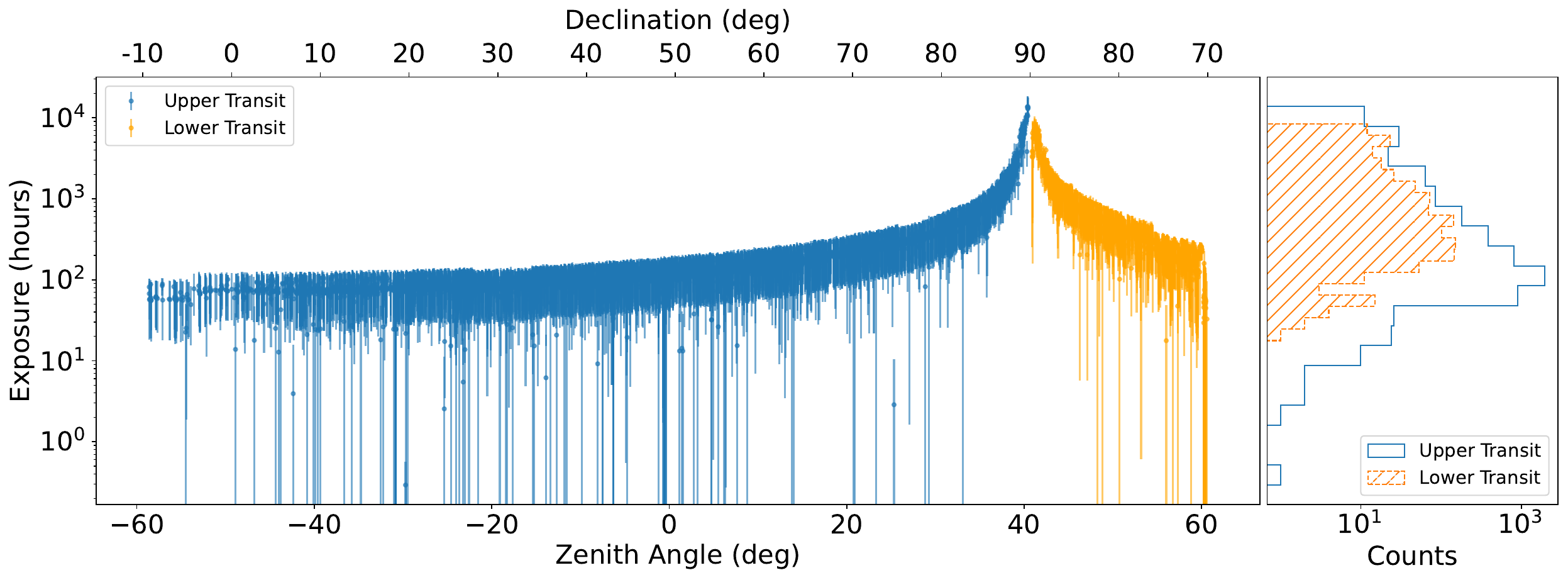}
    \caption{Total exposure per source in \cattwo shown for the observational period from \expstart to \finish. The left panel displays the exposure during upper and lower transits as a function of both declination and zenith angle. FRBs with $\dec > 70^{\circ}$ undergo two transits through CHIME's field of view, with the transits exhibiting different durations and consequently different exposure times and sensitivities. The right panel presents the distribution of exposure for sources observed during both upper and lower transits. The error bars account for uncertainties in source localization, as discussed in \S\ref{sec:localization}.}
    \label{fig:exp_dec}
\end{figure*}

Figure~\ref{fig:exp_dec} shows the integral of the exposure map over the localization probability map for each cataloged event, evaluated separately for the upper and lower transits. The large error bars reflect uncertainties in event localization (see \S\ref{sec:localization}). Some sources exhibit below-average exposure for their declination range, primarily because a significant portion of their localization probability lies in regions between adjacent formed beams, where the estimated exposure is zero.

\subsection{Sensitivity Threshold}
\label{sec:sensitivity}

For each burst in \cattwo, we report sensitivity thresholds estimating the minimum fluence required for a burst to be detected under observing conditions similar to those encountered during the survey. These thresholds represent 68\% and 95\% completeness limits—that is, the fluences above which 68\% and 95\% of bursts from the same sky location would have been detected. We note that these thresholds are lower limits, as they are derived from measured fluences that are themselves lower limits (see \S\ref{sec:flux}). The median 95\% threshold across the \cattwo sample is approximately 3.5~Jy~ms, reflecting the typical sensitivity of the survey during this period.

We use the Monte Carlo simulation procedure employed for \catone to determine these thresholds; the details are discussed by \cite{jcf+19} and \cite{abb+19c}, though we summarize the main components here. In keeping with past CHIME/FRB works, the simulation accounts for three sources of sensitivity variation that affect burst detectability: (i) daily fluctuations in instrument sensitivity due to weather, RFI, and calibration quality; (ii) variations in the primary and formed beam response across the source’s transit; and (iii) differences in FRB spectral energy distributions (SEDs), including central frequency and bandwidth.

For each Monte Carlo realization, we simulate an FRB by jointly drawing a random date, a position along the detected source’s transit, and an SED, consistent with the exposure and best-fit morphological parameters described in \S\ref{sec:exposure} and \S\ref{sec:fitburst}. We then evaluate the three sensitivity factors for that simulated FRB. The first factor is determined from pulsar observations (see \S\ref{sec:exposure}), which monitor day-to-day instrumental sensitivity via \snr measurements of known calibration pulsars. The second is derived from the CHIME/FRB beam model, evaluated at the simulated FRB’s sky position and integrated over frequency. The third is computed by simulating an FRB spectrum—assumed to be Gaussian with configurable bandwidth and central frequency—multiplying it by the beamformer-units-to-Jansky calibration factor, and summing over frequency. The product of these three terms yields a relative sensitivity scale factor for the simulated FRB compared to the detected burst.

To determine the corresponding fluence threshold, we begin by drawing a fluence from a Gaussian distribution centered on the measured fluence of the detected burst, with a width set by its measurement uncertainty. Assuming a linear relationship between fluence and S/N, we compute a fluence-to-S/N conversion factor using the drawn fluence and the detected burst’s S/N. We then apply this factor to the system’s nominal S/N detection threshold to obtain a fluence threshold for that realization. Finally, we divide this value by the simulated sensitivity scale factor to account for variations in observing conditions across the exposure.

The final quoted sensitivity thresholds are defined as the 68th and 95th percentiles of the distribution of fluence thresholds obtained from the Monte Carlo realizations. For repeating sources, we account for intra-source burst variability by selecting a randomly chosen burst from the source as the reference burst in each realization. The completeness thresholds, aggregated across all bursts from the repeater, are reported identically for each burst in the catalog tabular data. For sources at declination greater than $70^{\circ}$, we report sensitivity thresholds for both the lower and upper transit.

\subsection{Filtering Out Galactic Pulsars and RRATs and Identifying Repeating FRBs}
\label{sec:repeater}
\paragraph{Known Galactic Sources} Millisecond bursts are received from known Galactic radio transients (pulsars and RRATs); they are filtered from the sample in real time by our `Known Source Sifter' (KSS).  By comparing the estimated position, DM, and their respective uncertainties of each detected burst, the KSS calculates the probability of association for all sources in our database within 5 times the estimated DM uncertainty and 5 degrees of angular separation from the header localization of each detected signal  \citep[for a full description see][]{ple21}. The known source database was initially populated with all published pulsars and FRBs, and is continuously updated with new CHIME/FRB discoveries both extragalactic (see repeater discussion below) and Galactic (the most up-to-date list can be found at \url{https://www.chime-frb.ca/galactic },  \citealt{dcm+23}). We also manually update the known source database with new pulsar and FRB discoveries by other telescopes. We rerun the KSS with the most up-to-date known source database and remove all bursts from the sample with association probability to a known Galactic pulsar greater than 0.68.
 
As discussed further in \S\ref{sec:localization}, we occasionally detect events from known pulsars in the sidelobes of the telescope's formed beams, which the KSS has difficulty associating with their sources given the realtime pipeline's inaccurate position estimate. During the human-verification stage of each candidate event, we query any known sources from our database that could appear with the event's apparent position to help identify such sidelobe detections. For any event with measured DM below (i) the 99.7\% upper bound on the maximum DM predicted for the Milky Way's WIM disk \citep{occ20} or (ii) 100 pc cm$^{-3}$, we re-examine these events to filter out additional Galactic known source `contaminants' by Galactic sources. This verification allows us to remove $\sim10$ additional detections of bright pulsars in our sidelobes from the extragalactic sample, primarily $O({\rm kJy})$ giant pulses from the Crab pulsar (PSR B0531+21).

\paragraph{DM Selection} We require candidates to have DM exceeding the maximum Galactic DM along the line of sight, as predicted by both the NE2001 \citep{ne2001} and YMW16 \citep{ymw17} models. Both models suffer from known biases and unquantified uncertainties, and in some directions can significantly overestimate the Galactic DM contribution \citep{cbv+09,rcc+25}. Even without accounting for the $\sim 30$ pc cm$^{-3}$ contribution from our Galactic halo \citep{cbg+23}, we are thus likely incomplete to nearby sources in a direction-dependent way. We do call back total intensity data for events that fail this DM criterion, albeit with a higher S/N threshold given the much more frequent contamination of this sample by known pulsars and RFI. Because this contamination renders the astrophysical interpretation of individual pulses highly ambiguous, we currently exclude these events from the catalog.

 \paragraph{Known Repeaters} We include and flag repeat bursts associated with established repeating FRB sources \citep{abb+19b,abb+19c,fab+20,bgk+21,lac+22,R117_atel,abb+23} in this catalog, but only the first known burst from a given repeater is included in population analyses. Our repeat bursts are identified either from the KSS and/or from an offline in-house implementation \citep{dcm+23} of the DBSCAN algorithm \citep{eksx96} which is an unsupervised machine-learning clustering algorithm. This clustering algorithm, when run on the additional two years of data presented in this catalog compared to \cite{abb+23}, identified 92 unknown clusters after removing those contributed by known pulsars or where the events are otherwise deemed non-astrophysical. We find 33 have a probability of chance coincidence lower than our threshold to report as significant repeaters using the method of \cite{cle+24}. We present this sample and the calculation, as well as population analyses, elsewhere (Cook et al., \inprep).  Repeat bursts from these newly classified repeaters are also identified in the \cattwo tabular data by listing the name of the associated repeater under the \texttt{repeater\_name} column.

\subsection{Bursts identified by Zooniverse citizen scientists}
\label{sec:citizen}

From 30 October 2020 to 1 January 2022, we collected intensity data for FRB candidates with $7.5 \leq \snr < 8.5$. Instead of being sent to the CHIME/FRB expert reviewers, this low $\snr$ sample was classified by citizen scientist volunteers on the Zooniverse platform\footnote{\url{https://www.zooniverse.org/projects/mikewalmsley/bursts-from-space}}. Such candidates were classified by at least 15 citizen scientists. Candidates that were classified as astrophysical by 13 or more volunteers were sent to CHIME/FRB experts for final verification. From this sample, 57 bursts were expert-verified as FRB candidates and included in this catalog.  They can be identified by searching for events where the \texttt{citizen\_science\_flag} is set to \True.  These events should be excluded from statistical studies of the population, as their selection differs from that of the other events.

\section{CHIME/FRB \cattwo}
\label{sec:catalog}

\input{table_excerpt}

Here we describe the contents of \cattwo. The following subsections outline the methods used to measure and report each of the properties included in the catalog. The complete catalog is provided in machine-readable format with the online version of this article and through the \portal. Each row of the catalog corresponds to a sub-burst of an FRB, and each column contains a property of that sub-burst. A detailed description of each column is provided in Appendix~\ref{app:datafields}. In addition to these tabulated properties, we release several associated data products, including total-intensity dynamic spectra, localization contours, and full-sky exposure estimates. These supplementary data products are archived at the Canadian Advanced Network for Astronomical Research (CANFAR): \data.

Table~\ref{table:catalog_excerpt} presents a small excerpt from the catalog. Some FRBs exhibit complex temporal structure, with emission composed of multiple sub-bursts. FRBs with multiple detected sub-bursts are represented by multiple rows in the catalog, ordered by time of arrival and indexed by the \texttt{sub\_num} column. Some properties, such as name, $\ra$, $\dec$, DM, and fluence, remain the same across sub-bursts, while others, such as time of arrival, width, and spectral parameters, differ between sub-bursts. As an example, an FRB consisting of two sub-bursts, FRB~20190702B, appears in the last two rows of Table~\ref{table:catalog_excerpt}. The excerpt also includes FRB~20190702A, a repeat burst from the established source FRB~20180908B, demonstrating how repeaters are recorded in the catalog.

For a small number of events, certain properties could not be successfully determined.  In such cases, the corresponding table entries are populated with a \missing value, accompanied by a note explaining the reason for the missing data.  Table~\ref{table:missing} summarizes the number of events missing each property, with details on the causes of these missing values provided in the corresponding subsections of either \S\ref{sec:obs} or \S\ref{sec:catalog}.  It also lists the number of cataloged events flagged for occurring during periods of reduced instrument sensitivity, being detected in the sidelobes of the primary beam, being first identified by citizen scientists, or undergoing non-standard processing.  Details of these flags are described in \S\ref{sec:obs} and \S\ref{sec:catalog}.

\input{table_missing_flagged}

\subsection{Event Names}
\label{sec:tns}

The Transient Name Server (TNS)\footnote{\url{https://www.wis-tns.org/}} supplies and maintains the official FRB naming scheme, as used in the publication of the first catalog of FRBs from CHIME/FRB \citep{aab+21}. The naming scheme consists of the discovery date in YYYYMMDD format, and proceeds alphabetically according to the submission time to the TNS. Repeating FRB sources are referred to by their \textit{source name} -- the TNS name corresponding to their first detection\footnote{\textit{First detection} refers to the first burst from a repeating source to have a subsequently detected burst associated with it. This avoids naming complications introduced by the discovery of precedent bursts from archival data of a repeating source.}, and all subsequent detections of the source are given their own TNS name, but these are typically only used for identifying individual bursts from repeaters.

\subsection{Event Localization}
\label{sec:localization}

\begin{figure*}[p]
	\centering
\includegraphics[width=0.8\textwidth]{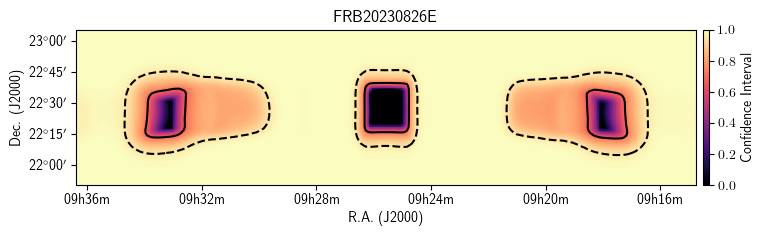} \\ 
\includegraphics[width=0.8\textwidth]{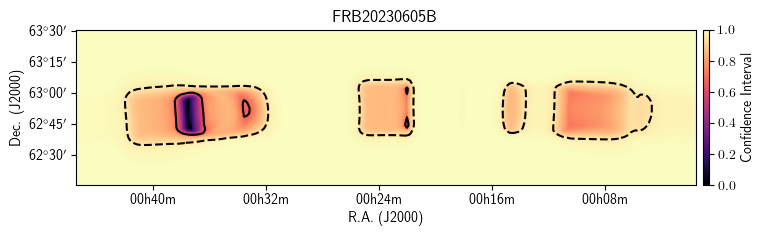} \\ 
\includegraphics[width=0.8\textwidth]{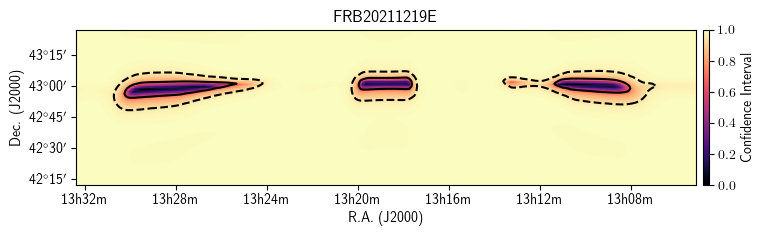} \\ 
\includegraphics[width=0.8\textwidth]{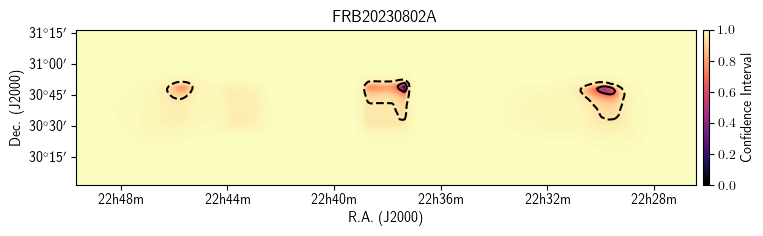} \\ 
\figcaption{Examples of event localization plots for four different beam detection patterns. From top to bottom: single beam, two-beams adjacent in E-W, two-beams adjacent in N-S, and four beams in a square. \label{fig:header_localization}}
\end{figure*}

The reported sky localization for each event is determined using the header
metadata measured in real-time by the L1 pipeline and stored in the L4 database. We follow the localization method detailed in \citet{abb+19c} and used previously in \citet{aab+21}. Predictions from the beam model are fit to ratios between per-beam \snr values for a grid of model sky locations and model intrinsic spectra. The mapping between $\Delta \chi^2$ and confidence interval is constructed using a set of pulsar events identified by the real-time system. For example, the $\Delta \chi^2$ for the 68\% confidence level corresponds to the contour level that encloses the true positions for 68\% of pulsar events.

The resulting localization error regions are presented as plots in Figure~\ref{fig:header_localization}. In the E-W direction, the grid of model locations is chosen to contain the main lobe of the primary beam. This span includes the first-order side lobes of the formed beams (commonly referred to as grating lobes), which leads to the disjointed uncertainty regions seen in
Figure~\ref{fig:header_localization}. CHIME/FRB occasionally detects events in higher-order grating lobes. The resulting header localizations for those events are therefore incorrect, as the model sky locations do not extend outward far enough in the E-W direction.  Such far sidelobe events can be identified through their ``spiky'' spectra \citep[see][]{lsn+23}.  We flag events visually identified as occurring in the far sidelobes using the \texttt{sidelobe\_flag} column in the catalog tabular data (Table~\ref{table:catalog_excerpt}) and do not present their header localizations, exposure estimates, flux density and fluence estimates, or sensitivity thresholds.

The localizations in Table~\ref{table:catalog_excerpt} are presented as the central location and extent of the 68\% confidence interval closest to the beam with the strongest detection. This single uncertainty region does not encompass the full localization error region. The full confidence level maps, i.e. the data plotted in Figure~\ref{fig:header_localization}, which include the near side lobes, can be found at \data.

\subsection{Event Morphology}
\label{sec:fitburst}

\begin{figure*}[p]
	\centering
\includegraphics[width=0.90\textwidth]{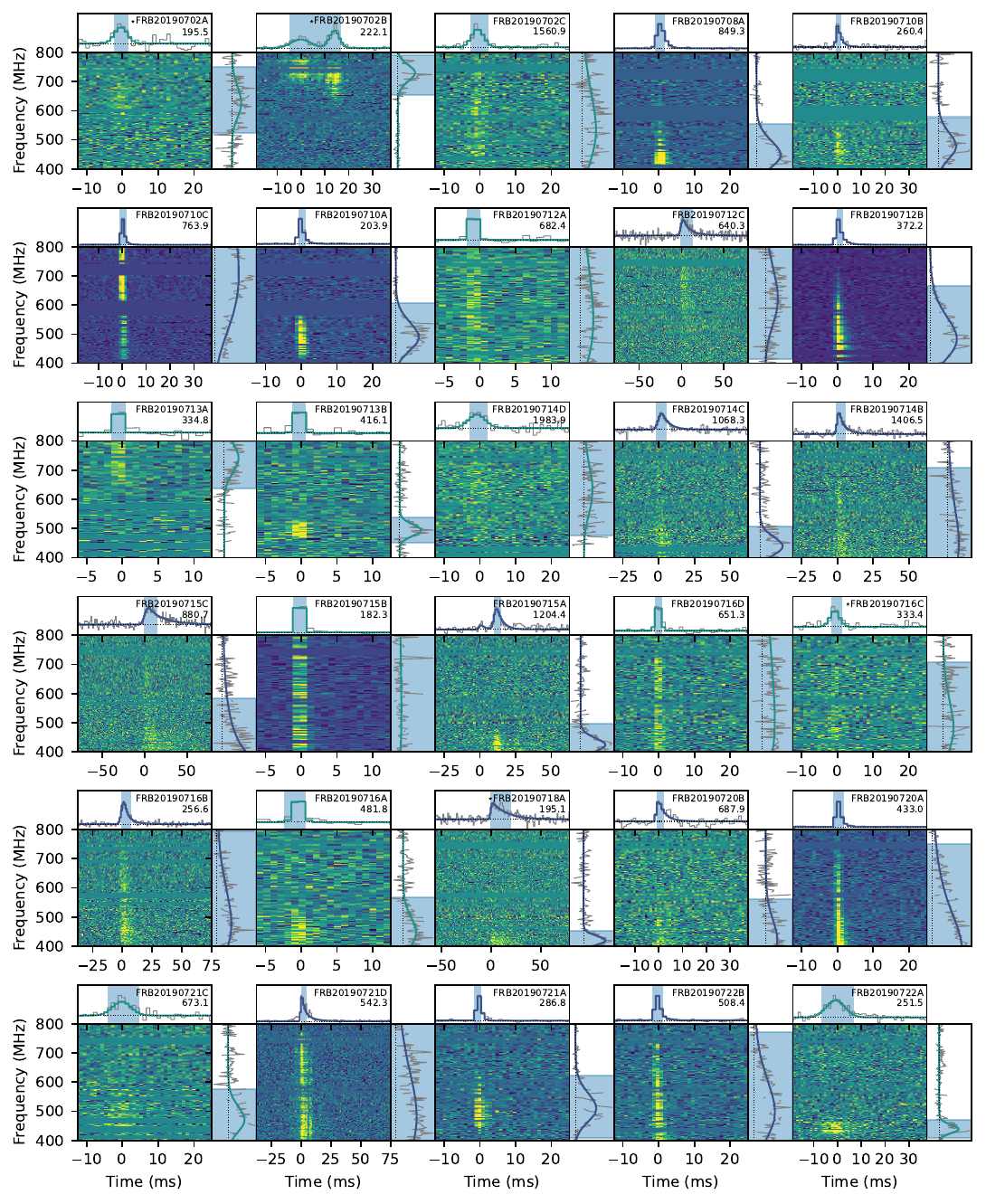}
\figcaption{Dynamic spectra (de-dispersed), frequency-averaged time series, and time-averaged spectra for the first 30 FRBs in \cattwo after the end of \catone (July 1, 2019), ordered by arrival time.  The TNS name and best-fit dispersion measure (DM, in pc cm$^{-3}$) are displayed in the upper right corner of each panel.  A star preceding the TNS name indicates that the burst is from a repeating source.  Model fits are overlaid on the time series and spectra, in green when scattering is insignificant and in blue when scattering is significant.  The blue shaded regions in the time series and spectra indicate the boxcar burst durations and full-width at tenth-maximum (FWTM) emission bandwidths, respectively. The frequency-averaged time series and time-averaged spectrum are computed using only frequencies and times within the blue shaded region.  The complete figure set containing all events in \cattwo with \fitburst{} results is available in the online journal (149 panels) and at \data.
\label{fig:waterfall_data}}
\end{figure*}

We employ the \fitburst{} analysis framework to generate models of all FRB dynamic spectra detected by CHIME/FRB as part of this second catalog \citep{fpb+24}. The resulting total-intensity dynamic spectra (at 0.983~ms temporal resolution and 24.4~kHz spectral resolution) and their best-fit \fitburst{} models are available for download from \data.  The \fitburst{} codebase assumes that each dispersed pulse is intrinsically Gaussian in temporal shape, with an intrinsic temporal width $\sigma$ and amplitude $A = 10^\alpha$ that is centered on the arrival time $t_0$. The spectral energy distribution (SED) of each pulse is characterized with a ``running" power-law distribution
\begin{equation}
    F_{k} = \left(\frac{\nu_k}{\nu_r}\right)^{\gamma + \beta\ln(\nu_k/\nu_r)} \ ,
    \label{eq:fitburst_sed}
\end{equation}
where $\nu_k$ is the electromagnetic frequency for channel $k$, $\nu_r > 0$ MHz is an arbitrary reference frequency, $\gamma$ is the spectral index, and $\beta$ is the spectral running.  This SED is an empirical model designed to flexibly describe both broadband power-law-like spectra and narrow-band Gaussian-like spectra within a single framework.  While it effectively captures observed spectral shapes, this SED does not have a well-established physical interpretation in the context of FRBs.

Pulse dispersion is modeled by adjusting the per-channel timeseries ($t_{kn}$) using the time delay associated with cold-plasma dispersion, i.e, requiring that
\begin{equation}
    t_{kn} = t_{rn} + k_{\rm DM}{\rm DM}(\nu_k^\epsilon - \nu_r^\epsilon) \ ,
\end{equation}
where $t_{rn}$ is the timeseries for data in reference channel $v_r$ and $k_{\rm DM}$ is a collection of physical constants. 

In the \fitburst{} framework, each pulsed feature (or ``sub-burst") is characterized with its own set of \{$\alpha_i$, $\beta_i$, $\gamma_i$, $\sigma_i$, $t_{0,i}$\} values, where $i = 0, 1, ..., N-1$, and $N$ is the total number of sub-bursts. However, all sub-bursts are assumed to share the same propagation-induced properties, i.e., they have the same DM and, if relevant, scattering timescale as measured at $\nu_r$ ($\tau_r$) of a one-sided exponential pulse broadening function predicted by the ``thin-screen" turbulence model \citep[e.g.,][]{mck14}. The value of $N$ is manually determined based on visual inspection of the data, along with a rudimentary peak-finding algorithm built into the \fitburst{} framework. The \fitburst{} framework also accounts for dispersion smearing within each channel by first computing a high-resolution model of the dispersed dynamic spectrum, and then downsampling to the resolution of the data. We neglect this smearing-correction procedure in the rare circumstance where $\sigma$ exhibits numerical instability while simultaneously fitting for $\tau_r \gg 0$ ms at $\nu_r$, which typically occurs when $\sigma \approx 0$ ms; these exceptions are noted with the {\tt intrachan\_flag} entry in Table~\ref{table:catalog_excerpt}.

We assume that the frequency dependence of both scatter-broadening and dispersion is a power law in form, parameterized by separate indexes of scatter-broadening ($\delta$) and dispersion ($\epsilon$).  We fix $\delta=-4$ and $\epsilon=-2$ for all \fitburst{} models in \cattwo because the diversity of FRB morphology and brightness makes robust optimization of these parameters difficult.  These values are chosen based on theoretical expectations from Kolmogorov turbulence ($\delta = -4$) and cold plasma dispersion ($\epsilon = -2$).

In keeping with \catone, we generate two \fitburst{} models for each recorded dynamic spectrum in \cattwo: one where $\tau_r=0$ is held fixed (i.e., scatter-broadening is insignificant); and a second where $\tau_r$ is subject to optimization. The preferred model for each FRB is then determined using an F-test with the best-fit statistics for each model, as described by \cite{fpb+24}. For initial guesses, we use the L1 estimates of DM and $t_0$ while assuming that $\sigma = 10$ ms and \{$\alpha$, $\beta$, $\gamma$\} are all initially zero, which are then optimized using a least-squares algorithm. For all multi-component FRBs, we use the L1 DM for all components but use the aforementioned peak-finding algorithm to estimate initial guesses for $t_{0,i}$, while setting $\sigma_i = 0.5$ ms and \{$\alpha_i$, $\beta_i$, $\gamma_i$\} all to zero. For models where $\tau_r$ is optimized, we use the best-fit, $\tau_r=0$ model as the initial guess and set the initial value of $\tau_r$ equal to $\sigma$.

Total-intensity dynamic spectra and best-fit \fitburst{} models for all \cattwo events with successful fits are shown in Figure~\ref{fig:waterfall_data}. The corresponding residual dynamic spectra are provided in Appendix~\ref{app:waterfall_residual}. For a subset of 90 events in \cattwo, the \fitburst{} procedure failed to produce a satisfactory fit, most often due to heavy RFI or rapid variations in the background sky intensity. For these events, we do not provide panels showing the data and residuals in the figure sets. However, we do release the corresponding de-dispersed total-intensity dynamic spectra using a larger time window centered on the arrival time inferred from the real-time detection pipeline, so that users can attempt alternative modeling or re-fitting if desired.

The \fitburst{} framework has undergone significant updates since its use in \catone, driven by the need to scale the analysis to a much larger sample of events in \cattwo. These updates include reparameterizations of key variables to improve convergence, revised handling of multi-component bursts, refinements to the numerical evaluation of the pulse broadening function, and a revised convention for referencing the time and frequency axes. Many of these improvements are described in \cite{fpb+24}, while additional, more recent updates were introduced as part of the present work. A complete summary of changes relative to the version used in \catone is provided in Appendix~\ref{app:fitburst}.

\subsection{Event Signal Strength}
\label{sec:flux}

To characterize signal strength for each event, we provide the \snr of the initial real-time pipeline detection, along with a flux density and fluence determined in off-line analyses. We also provide an estimate of \snr made with \fitburst{}, which is calculated for each dynamic spectrum in a manner described by \cite{fpb+24}. The description of the automated flux calibration pipeline, including an explanation of current limitations, is detailed in \cite{apb+23}. We summarize the procedure here.

As in \cite{aab+21}, the flux densities and fluences reported in this catalog are biased low and should be interpreted as lower limits, with an uncertainty on the limiting value. This bias arises because uncertainty in our burst localization (see \S\ref{sec:localization}) prevents us from reliably correcting for the complex and rapidly varying attenuation of the CHIME primary beam and formed beams.  As a result, we assume that each burst was detected along the meridian of the primary beam at the peak of a formed beam, corresponding to the location of highest sensitivity in its declination arc on the sky. The resulting bias for a subset of \catone events is illustrated in Figure~4 of \citet{aaa+24}, which compares fluences estimated from real-time intensity data to those obtained from baseband data, where a more precise localization is possible.

The conversion from beamformer units to Janskys (BF/Jy) as a function of frequency is determined using transit observations of steady sources with known spectral properties. We pair each burst with the calibration spectrum of the nearest steady-source transit, closest first in declination, then in time. We assume N--S beam symmetry, so that sources on both sides of zenith can be used for each burst. We apply the BF/Jy spectrum from the nearest calibrator transit to the total-intensity data for that burst to roughly correct for N--S primary beam variations, and convert the dynamic spectrum to physical units.

From this calibrated dynamic spectrum, we form a band-averaged time series by averaging over frequency channels across the usable band (after RFI excision).  We then integrate this band-averaged time series over the burst extent to obtain the fluence. The peak flux density is reported as the maximum value of the band-averaged time series (at 0.983~ms resolution) within the burst extent.  The uncertainties presented in this catalog account for the errors due to differences in the primary beam between the calibrator and the FRB location along the meridian, the error due to temporal variation in calculated BF/Jy spectra, and the RMS of the off-pulse region in the band-averaged time series. These uncertainties do not encapsulate the bias owing to our assumption that each burst is detected along the meridian of the primary beam, which causes our flux density and fluence measurements to be lower limits.

Results from \fitburst{} are used as inputs to the flux calibration pipeline. As a consequence, we only report flux densities and fluence measurements for events that were successfully processed by \fitburst{}.  Of the total \nfrbtot events, flux density and fluence measurements are not reported for 449 events due to a variety of issues, including missing \fitburst{} results, low-quality calibration, RFI contamination, and other factors. The specific reason for each event missing flux density and fluence measurements is provided in the \texttt{fluence\_notes} column in the  catalog tabular data (see Table~\ref{table:catalog_excerpt}).  Additional details on the causes of missing flux density and fluence measurements, along with an exploration of potential selection biases in the excluded sample, are provided in Appendix~\ref{app:bias}.  For 55 events missing flux density and fluence estimates in \cattwo, values were previously published in \catone. For these events, we adopt the Catalog 1 parameter values and indicate this by setting the \texttt{catalog1\_param\_flag} to \True.

In this catalog, we report flux densities and fluences for 4090 events. It is important to note that the burst extent used in the fluence calculation is not uniformly defined across the entire sample. We first identify a ``pulse emission region'' in our analysis pipelines, centered on the best-fit arrival time and extending to the best-fit Gaussian width. If scattering is fit, the post-arrival end is extended by the scattering timescale. For the majority of the events (3845), the burst extent is defined as the pulse emission region extended by double the Gaussian width on either end. This ensures that the entire signal is captured by the pipeline. The error introduced by additional noise is less than the quoted uncertainties.

To prevent computational errors in a subset of 245 events (see Table~\ref{table:missing} and Appendix~\ref{app:bias}), we instead define the burst extent to be the pulse emission region itself, without any additional extension (a more restricted signal window). This approach may exclude some signal near the leading and trailing edges of the burst; however, this loss of signal, and subsequent underestimation of fluence, is adequately represented within our measurement uncertainties. The two definitions are distinguished by the {\burstwindowextended} flag, {\True} indicating that the pulse emission region was extended, and {\False} indicating that the restricted (non-extended) window was used. Potential biases resulting from missing fluence measurements and non-uniform definition of burst extent are discussed in Appendix~\ref{app:bias}.

\section{Validation}
\label{sec:validation}

In this section, we describe a series of validation tests performed on the \cattwo sample to assess the impact of instrumental effects and methodological choices.

\subsection{Comparison to \catone}
\label{sec:compare_cat1}

\begin{figure*}
    \centering
    \includegraphics[width=0.32\linewidth]{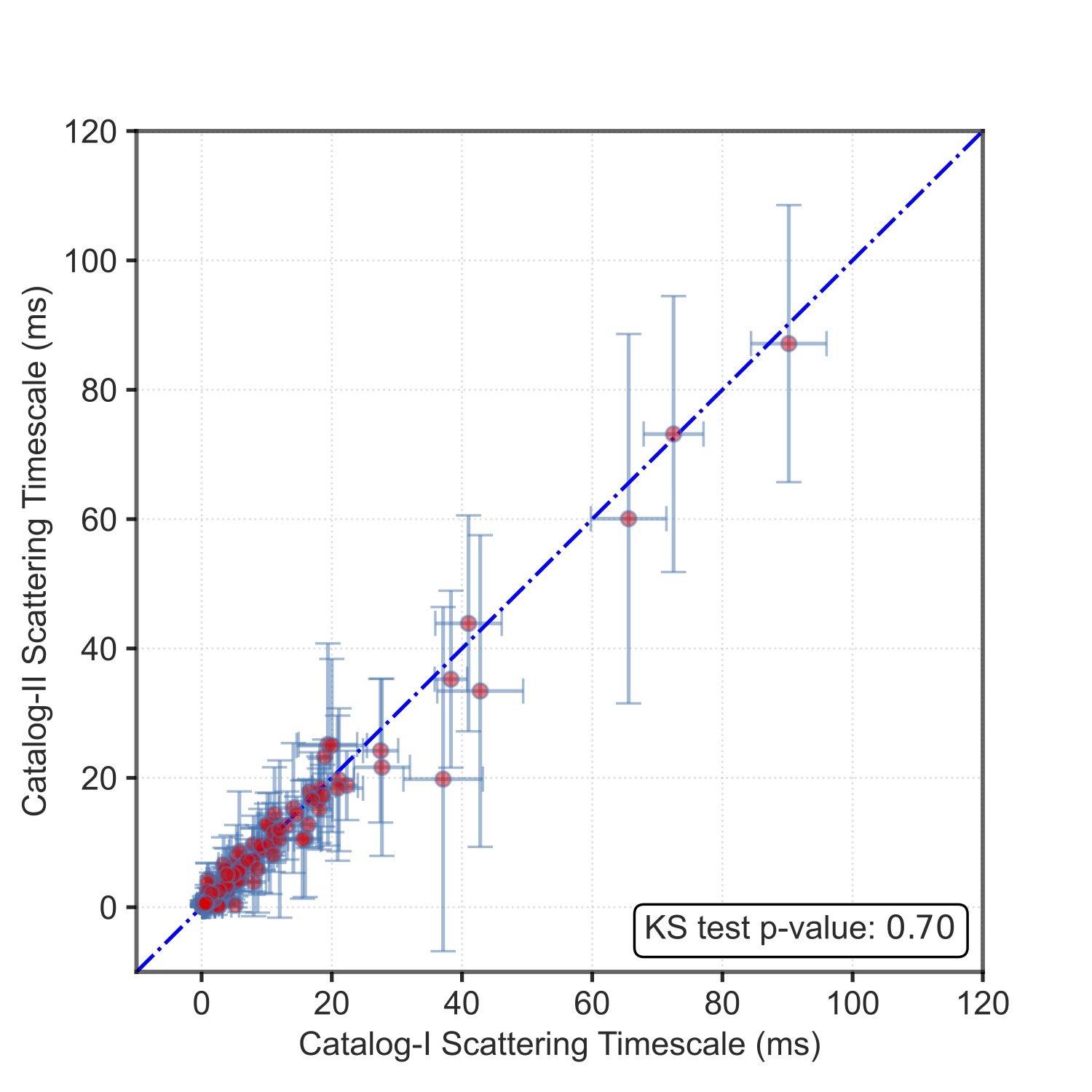}
    \includegraphics[width=0.32\linewidth]{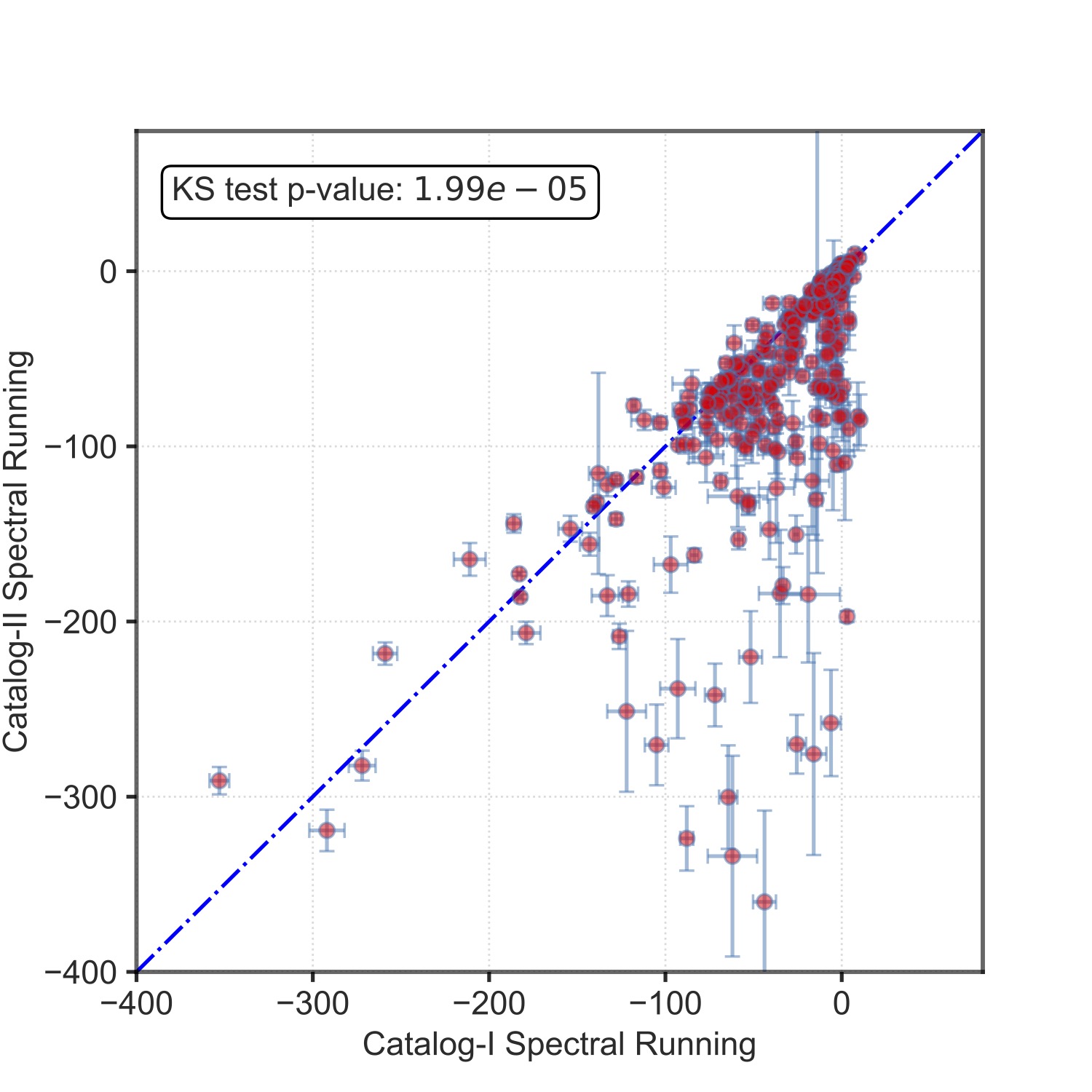}
    \includegraphics[width=0.32\linewidth]{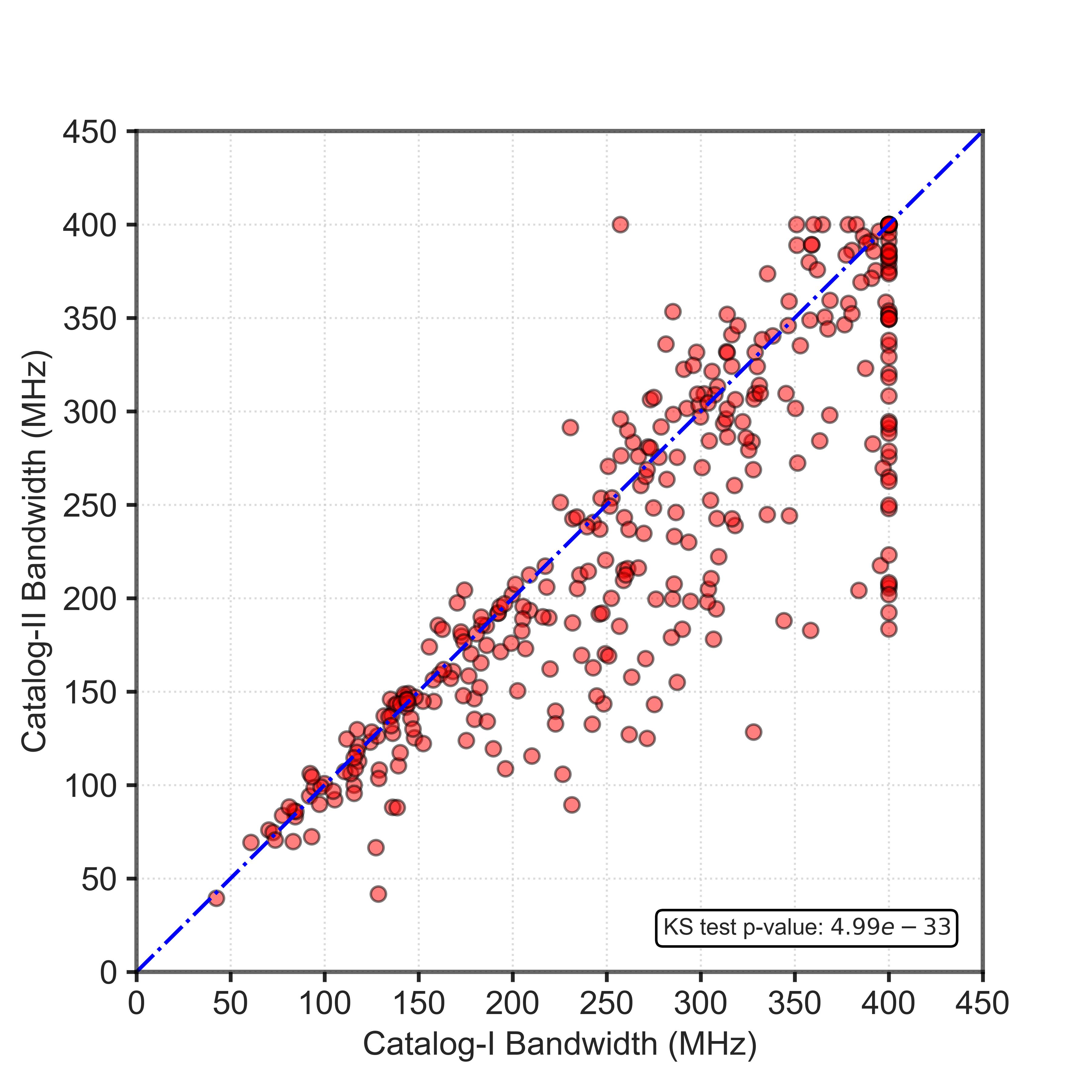}
    \caption{(Left.) A scatter plot of revised \fitburst{} estimates of scattering timescale (all referenced to 400.195 MHz) made for FRBs first presented in \catone; a Kolmogorov-Smirnov (KS) test between these two distributions yields a $p$-value of 0.7, indicating that the \catone and \cattwo estimates are statistically consistent despite changes to the \fitburst{} model described in \S\ref{sec:fitburst}. (Middle.) A similar scatter plot and KS test for \fitburst{} estimates of the spectral running between \catone and \cattwo. The statistically significant differences reflect changes in the \fitburst{} model that lead to improved fits against narrow-band (i.e., repeater-like) features in dynamic spectra, which tend to yield large magnitudes of spectral running. (Right.) A similar scatter plot and KS test for the full-width-at-tenth-maximum (FWTM) emission bandwidth, derived from best-fit \fitburst{} models using an algorithm described by \cite{p++21}.}
    \label{fig:fitburst}
\end{figure*}

The first validation test compares event lists from \catone and \cattwo for their overlapping time span (\start to \finishone). Of the 536 events from \catone, 533 meet the criteria outlined in \S\ref{sec:obs} and have been reprocessed in \cattwo. For the remaining three events —- FRB 20190415C, 20190422B, and 20190517D -- total-intensity callback data were lost.  To maintain consistency with previously published results, all three events are included in \cattwo. For any pipeline that relies on intensity data, we adopt the original \catone{} results, include a pipeline-specific note in the catalog tabular data, and set the corresponding \texttt{catalog1\_param\_flag} to \True.

Additionally, we identify eight events that occurred within the \catone time span, meet the \cattwo criteria, but were not included in \catone.  Four of these events -- FRB 20180912A, 20181022A, 20181208C, and 20190326B -- were excluded from \catone due to either low \snr or classification as faint or RFI during further inspection of the dynamic spectra by the catalog team.  For \cattwo, these events have been included to maintain consistent, well-defined human verification criteria and allow users to apply their own \snr thresholds.  One event, FRB 20181201D, was excluded because the corresponding total-intensity data could not be located at the time of \catone construction, and another, FRB 20181022F, was mistakenly flagged as a duplicate of a different burst; both events are now included in \cattwo.  Lastly, two events -- FRB 20190607C and 20190618B -- were mistakenly excluded from \catone for unknown reasons and have now been correctly included in \cattwo.

CHIME/FRB often detects bursts in multiple formed beams, each with an associated \snr, which quantifies detection strength. In some cases, differences in time of arrival or DM prevent the real-time pipeline from correctly grouping these detections, resulting in multiple recorded instances of the same event (i.e., ``duplicate events''; see \S\ref{sec:obs}). In \catone, parameter estimation for two events—FRB 20190308A and FRB 20181226C—was based on the formed beam with a lower \snr, whereas in \cattwo, we instead use the beam with the highest \snr in all cases.

We also regenerate morphological models for all FRBs in \cattwo that were first presented in \catone. This choice is made to compare the efficacy of our burst fitting software {\tt fitburst} following substantial changes to its modeling framework that were implemented after the publication of \catone, as described by \citet{fpb+24} and in \S\ref{sec:fitburst}. Direct comparisons in re-estimated {\tt fitburst} parameters for \catone FRBs show the following differences:

\begin{itemize}
    \item Of the original 536 FRBs published in \catone, six are missing \fitburst{} results due to lack of available total-intensity data (three cases) or heavily masked data (three cases).
    \item Of the remaining 530 FRBs, 18 have a different number of sub-bursts in \cattwo.
    \item Of the 512 FRBs with the same number of sub-bursts, 149 had their models changed in \cattwo between an unscattered Gaussian temporal shape and a pulse broadening function (PBF) as described in \S\ref{sec:fitburst}.
\end{itemize}

\noindent The above model changes occurred as a result of manual intervention and model redefinition. For example, all 18 models with different number of sub-bursts were manually changed as one or more sub-burst parameters in the \catone model were numerically unstable when subject to re-fitting using the reparameterized version of \fitburst.\footnote{The highly constrained nature of \catone  \ \fitburst{} models for multi-component FRBs -- which benefited from explicit parameter bounds that are no longer in use within \fitburst{} due to modeling improvements described by \cite{fpb+24} -- likely led to their dynamic spectra being overfitted.}  Furthermore, nearly 70\% of the 149 FRBs with changes in the PBF model yield F-test $p$-values that are larger than the 0.1\% used as a criterion for selecting PBF models over unscattered models; in other words, 70\% of these changes are consistent with statistical fluctuations in classification for data with marginal evidence for scattering (see left panel of Fig. \ref{fig:fitburst}).

After accounting for these differences, 363 \fitburst{} models remain effectively unchanged between \catone and \cattwo and can therefore be directly compared. Within this subset, 282 measurements (i.e., 78\%) of intrinsic temporal width are statistically consistent to within a factor of five\footnote{While arbitrary, this value is chosen to account for the difference in fit performance between \fitburst{} versions used in \catone{} and \cattwo that are not reflected in the reported statistical uncertainties.} times the largest of the two statistical uncertainties estimated in the two catalogs. The spectral parameters \{$\beta$, $\gamma$\} show larger differences between catalogs as shown in the right-hand panel of Figure~\ref{fig:fitburst}.  However, these differences are expected due to covariance with $\alpha$ which was redefined for optimal fitting after \catone, as described by \cite{fpb+24} and further discussed in Appendix \ref{app:fitburst}. Moreover, \cattwo estimates of \fitburst{} parameters correspond to goodness-of-fit values that are lower in magnitude than those first estimated for \catone; these lower values indicate that the changed parameterization described in \S\ref{sec:fitburst} produces better representations of the data.

This accounting shows that, despite changes in our analysis methods, the vast majority of comparable, morphological fits of \catone FRBs remain statistically consistent when reanalyzed for \cattwo. This fact is reflected in the consistency of fits for ``global" morphological parameters, such as $\tau_r$, as shown in the left-hand panel of Figure \ref{fig:fitburst}. 

\begin{figure*}
    \centering
    \includegraphics[width=0.45\linewidth]{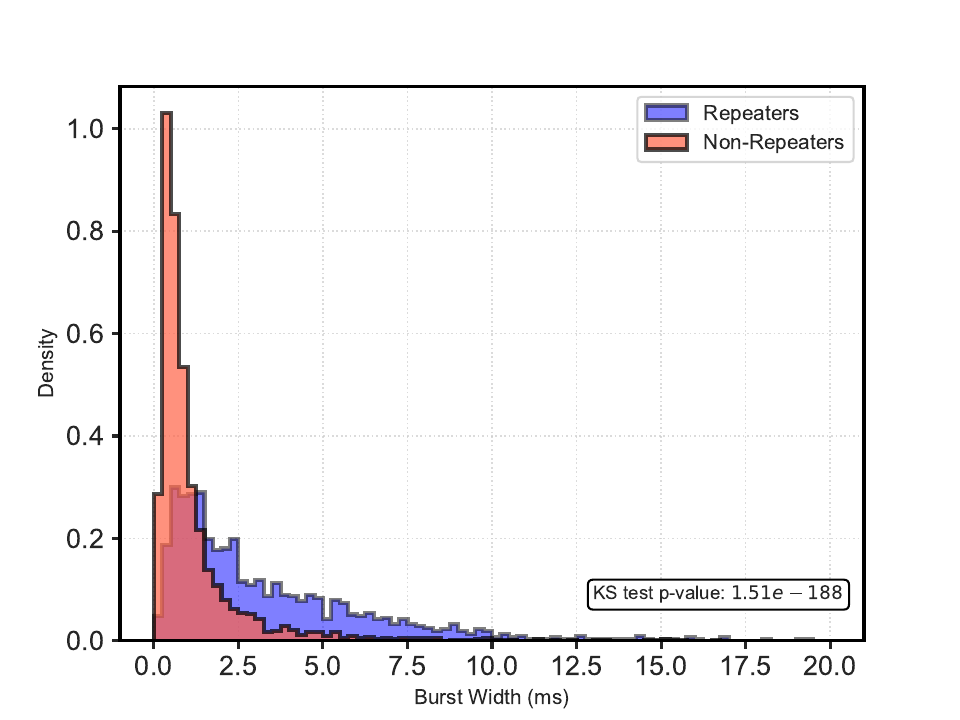}
    \includegraphics[width=0.45\linewidth]{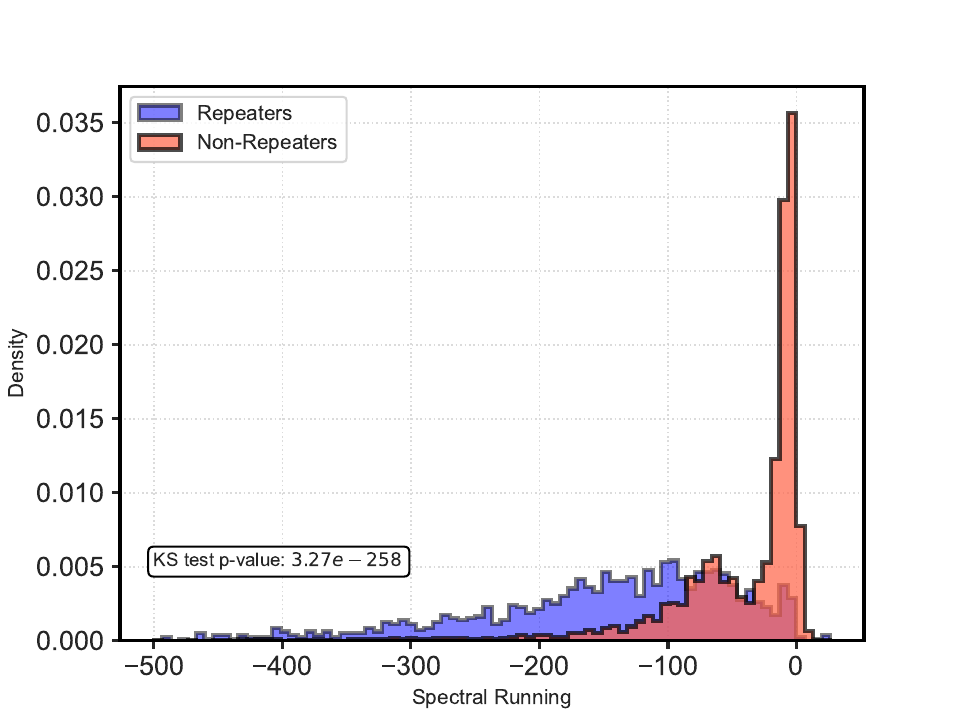}
    \caption{Histograms of best-fit values of burst width (left) and spectral running (right) as estimated by \fitburst{} for \cattwo, separated by classification of known repeating and apparently non-repeating FRB sources. Note the difference in burst width distributions between repeaters and apparent non-repeaters, as previously noted in \catone and by \citet{pgk+21}. Similarly, the spectral differences between the two classes persist, as in \catone and as detailed by \citet{pgk+21}. However, \cattwo data now reveal a clear bi-modality in the spectral running distribution for apparent non-repeaters.  As discussed in \S\ref{sec:compare_cat1}, this bi-modality is primarily due to an instrumental effect in which broadband bursts appear narrowband due to truncation by the frequency-dependent beam response.}
    \label{fig:fitburst_repeater}
\end{figure*}

Furthermore, the \cattwo estimates of burst properties continue to show clear differences between the spectra and morphologies of repeating and apparently non-repeating FRBs, consistent with those studied by \cite{pgk+21}. Comparisons of burst width and spectral running are shown in Figure~\ref{fig:fitburst_repeater}, left and right, respectively.  Bursts from repeating sources tend to have larger burst widths and steeper (more negative) spectral running.

Notably, we observe a new bi-modality in the spectral running distribution of apparent non-repeaters. While one possible interpretation is that sources in the left peak are undetected repeaters, an instrumental effect is likely responsible for the vast majority of these events. In what follows, the non-repeater sample is divided into two subsamples based on spectral running, approximately corresponding to the two peaks: the ``steep-running'' sample with $\beta < -35$ and the ``shallow-running'' sample with $\beta \geq -35$. The steep-running sample consists of bursts that are primarily narrowband, with a median (maximum) full-width-at-tenth-maximum (FWTM) bandwidth of 150 MHz (313 MHz), compared to 391 MHz ($\geq 400$ MHz) for the shallow-running sample. We find that 37.5\% of the non-repeaters are in the steep-running sample, a fraction that is roughly consistent with the $30.2 \pm 4.1\%$ of FRBs that are detected in the first grating lobe of the formed beam in an analysis of baseband data for 129 FRBs from \catone, where improved localization is possible \citep{aaa+24}.  The grating lobes arise due to spatial aliasing in the east-west direction, and their position shifts as a function of frequency. As a result, broadband bursts detected in a grating lobe have portions of their spectrum attenuated or completely suppressed as the burst’s position moves out of the grating lobe response at those frequencies, making these bursts appear artificially narrowband.

Further supporting this hypothesis is the distribution of intrinsic burst widths $\sigma$. The burst width distribution for the steep-running sample of non-repeaters is much more similar to that of the shallow-running sample (KS test statistic = 0.072, $p$-value $= 5.8 \times 10^{-4}$) than to the repeater sample (KS test statistic = 0.44, $p$-value $< 10^{-16}$). While the two non-repeater samples exhibit a statistically significant difference in their burst width distributions -- with the steep-running sample showing a slightly larger fraction of bursts with widths between 1 and 10~msec -- this may be due to the narrowband nature of these bursts making it more difficult to distinguish intrinsic pulse width from scatter broadening.  When restricting the analysis to events where a scatter-broadened profile is preferred, we find that the width distributions of the two non-repeater samples are consistent (KS test statistic = 0.051, $p$-value $= 0.26$).  Thus, while a small fraction of the sources currently classified as non-repeaters may be undiscovered repeaters, the overall bi-modality in spectral running is primarily a consequence of an instrumental effect in which broadband bursts are truncated due to frequency-dependent grating lobes. Future analysis of baseband data from \cattwo will provide improved localizations and enable a more definitive separation of instrumental effects from intrinsically narrowband emission.

\subsection{Examination of Source Rate}
\label{sec:source_rate}

This section analyzes the source rate as a function of various parameters to identify potential biases in the \cattwo selection function. The parameters include the year, month, day of week, and hour of day in which the source was detected; the solar hour angle and accumulated rainfall at the time of detection; the configuration of the real-time pipeline at the time of detection; and the \ra, zenith angle, Galactic longitude, and Galactic latitude of the source.  Since the intrinsic FRB rate is not expected to depend on these parameters, any observed non-constant source rate may indicate unaccounted variations in exposure, instrument sensitivity, or the analysis pipeline.

\paragraph{Event Selection Criteria} The analysis below excludes events detected prior to \expstart, corresponding to a commissioning period without reliable exposure estimates. Events detected during periods of non-nominal system sensitivity, those discovered by citizen scientists, and those detected in the far sidelobes of the primary beam (indicated by the \texttt{excluded\_flag}, \texttt{citizen\_science\_flag}, and \texttt{sidelobe\_flag}, respectively) are also excluded.  Events whose \snr at detection (\texttt{bonsai\_snr}) was less than 10 are also excluded to match the most stringent criteria used in the early data (see \S\ref{sec:obs}) and thus ensure a constant \snr threshold throughout.  Additionally, events missing header localization or real-time pipeline metadata are omitted. Finally, only the first detected event from each source is included, excluding repeat detections. This selection yields a total of $N = 2651$ sources of FRBs observed over a total exposure of $1387.1 \ \mbox{days}$, resulting in an observed rate of $\hat{R} = 1.91 \pm 0.037$ sources per day (Poisson uncertainty).

\begin{figure*}[p]
	\centering
\includegraphics{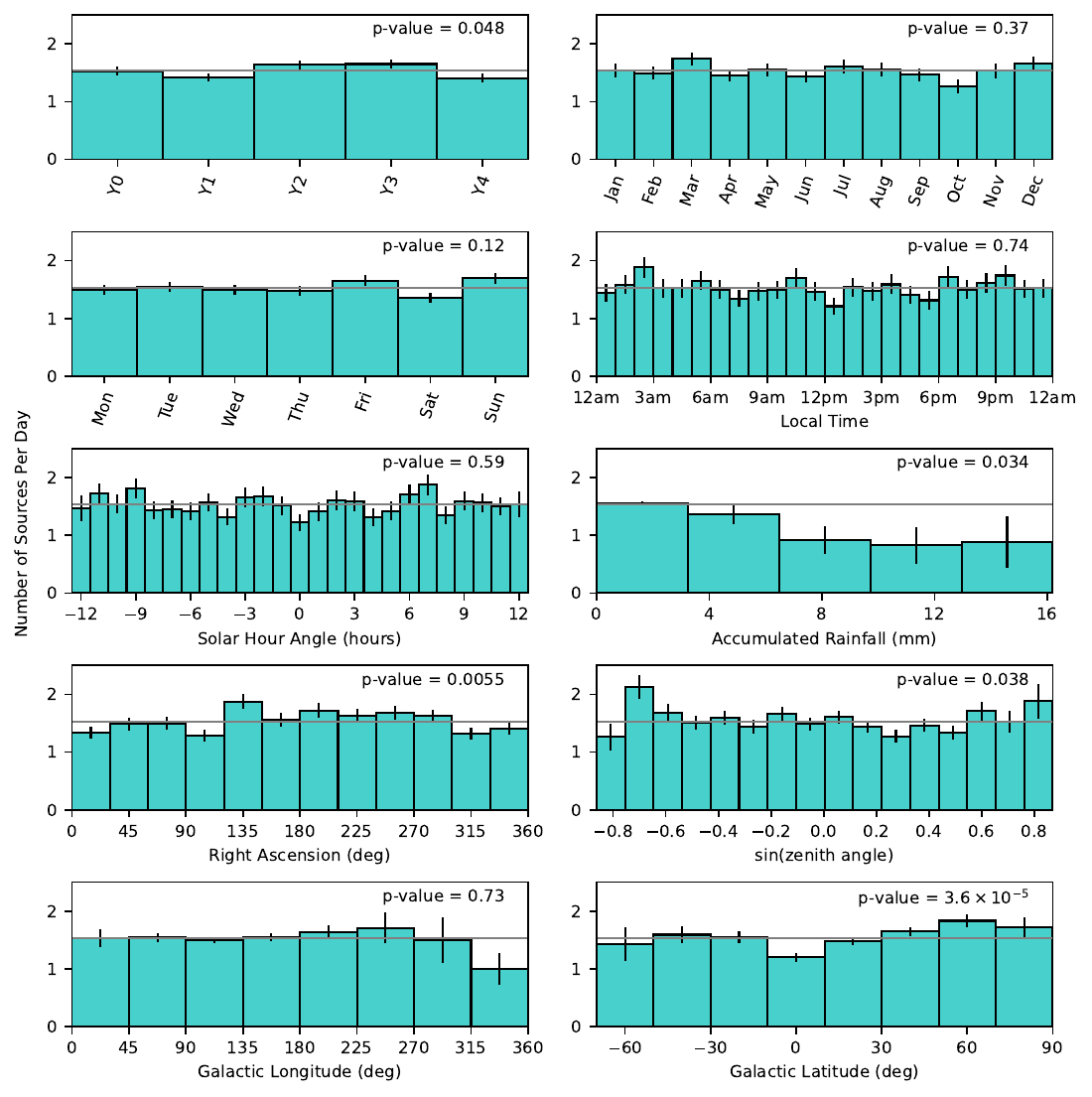}
\figcaption{Observed source rate as a function of various parameters: year of operation, month, day of the week, local time, solar hour angle, accumulated rainfall over the past 24 hours, \ra, sine of the zenith angle, Galactic longitude, and Galactic latitude, from left to right and top to bottom.  Only the first detected burst from each source is included, such that the rate reflects the number of unique sources per day.  Source rates have been corrected for variations in exposure and, approximately, for sensitivity using a variable \snr threshold based on an empirical proxy for system noise.  For $\sinza$, Galactic longitude, and Galactic latitude, the expected variation in source rate due to the primary beam response was estimated via Monte Carlo simulation and used to correct the observed rates.  The quoted $p$-values indicate the probability that the source rate remains constant across each parameter.  The gray horizontal line indicates the mean source rate over the full time span ($\hat{R} = $ \ratevarsnr) for reference.  Note that this value is lower than the raw number of sources detected per day by CHIME/FRB due to the selection criteria applied to ensure uniformity. \label{fig:consistency_check}}
\end{figure*}

\paragraph{Statistical Methodology} To test the null hypothesis that the source rate is constant across the parameter of interest, sources are grouped into $n$ bins of equal width based on that parameter. For each bin $i$, the total number of sources ($N_i$) and total exposure time ($t_i$) are calculated, with the exposure time first multiplied by the fraction of formed beams on-sky relative to the average number over the \cattwo period, which is 985 (out of 1024). Under both the null hypothesis ($H_0$) and the alternative hypothesis ($H_1$), the number of sources per bin ($N_i$) is assumed to follow a Poisson distribution.  Under $H_0$, the expected number of sources in bin $i$ is $R t_i$, where $R$ is a constant source rate estimated via maximum likelihood as $\hat{R} = \frac{\sum_i N_i}{\sum_i t_i}$. Under $H_1$, the source rate varies across bins, with the expected number of sources in bin $i$ given by $R_i t_i$, where the bin-dependent rate is estimated as $\hat{R}_i = \frac{N_i}{t_i}$.  The test statistic is the difference in the log-likelihoods between $H_1$ and $H_0$, given by:
\begin{align}
\Lambda & \equiv \log{\left[\mathcal{L}(H_{1})\right]} - \log{\left[\mathcal{L}(H_{0})\right]} \nonumber \\
& = 2 \sum_{i} \left[ N_i \log\left(\frac{N_i}{\hat{R} t_i}\right) - (N_i - \hat{R} t_i) \right].
\end{align}
The observed $\Lambda$ is compared to its distribution under $H_0$, estimated via 1 million Monte Carlo simulations. The $p$-value is the fraction of simulated $\Lambda$ values exceeding the observed $\Lambda$. Results are summarized in the column labelled ``$\snr \geq 10$'' in Table~\ref{table:consistency}.  A $p$-value below 0.05 is interpreted as statistically significant evidence against $H_0$, and below 0.01 as strong evidence.  No corrections for multiple hypothesis testing are applied to the reported $p$-values or significance thresholds.

\input{table_consistency_check}

\paragraph{Observed Variability in Source Rate} There is no evidence for variability in the source rate as a function of calendar month, day of week, local time, or solar hour angle. On the other hand, there is evidence for a dependence of the source rate on the amount of rainfall accumulated over the preceding day, which has previously been shown to degrade sky sensitivity in a modest fraction ($\sim$10\%) of antenna due to water pooling around focal line electronics \citep{abb+22}. The source rate is consistent before and after 19 October 2020, when a significant modification to the coarse-graining scheme of the real-time de-dispersion algorithm was deployed. In contrast, there is evidence for a decrease in the source rate after 16 May 2022, coinciding with the deployment of an upgraded RFI-sifting algorithm at the L2/L3 stage. More generally, there is evidence for variability in the source rate as a function of the year and month of operation. Additionally, variable source rates are strongly preferred as a function of \ra, \sinza, Galactic longitude, and Galactic latitude.

\paragraph{Correcting for Variations in Sensitivity} The consistency checks presented thus far have accounted for variations in exposure but not for variations in system sensitivity. To address this, a proxy for system noise is constructed using two auxiliary data products; this proxy is then used to apply a fixed cut in signal strength rather than \snr. The first component of the proxy consists of daily noise measurements inferred from pulsars, denoted as $\sigma_{\rm pulsar}(t)$, which are described in \S\ref{sec:exposure} and presented in Figure~\ref{fig:sens}. These measurements are smoothed using a rolling median filter with a 31-day window to mitigate short-timescale fluctuations while preserving long-term trends in the system noise. The second component, $\sigma_{\rm auto}(\mbox{R.A.})$, is derived from the feed-averaged autocorrelation measured at 10-second cadence by the CHIME cosmology backend \citep{abb+22}.  These measurements are rebinned onto a common grid in transiting \ra, averaged over all sidereal days covered by \cattwo, and averaged over frequency to capture the approximately $20\%$ variation in system noise over a sidereal day as the Galactic plane moves through the field of view.

The combined noise metric is defined as $\sigma(t) = \sigma_{\rm pulsar}(t) \times \sigma_{\rm auto}(\mbox{R.A.}(t))$, and it is used to apply a dynamic \snr threshold. Specifically, a source $s$ is included in the analysis if its detected \texttt{bonsai\_snr} exceeds $10 \times \max_{s}{\{\sigma(t_{s})\}} / \sigma(t_{s})$, where $t_{s}$ is the time of detection. This requirement ensures that sources are sufficiently bright to meet the $\snr > 10$ threshold even at the worst system sensitivity. The resulting \snr threshold varies between 10 and 13 across the sample, leading to the exclusion of an additional 526 sources compared to the fixed $\snr > 10$ threshold.  With this selection, the observed rate is $\hat{R} = 1.53 \pm 0.033$ sources per day (Poisson uncertainty).

The variable \snr cut does not account for the north-south response of the primary beam, which is relevant for three tests: $\sinza$, Galatic longitude, and Galactic latitude.  The primary beam exhibits significant variation across the dataset, with sensitivity differing by up to a factor of eight between the most and least sensitive formed beam locations.  A primary-beam-based \snr cut would require setting a threshold high enough to ensure detections even in the least sensitive locations.  However, this would impose an overly stringent cut over much of the sky, leading to the exclusion of a large fraction of detected FRBs.  Instead, the expected source rate is estimated using a Monte Carlo simulation assuming the selection-function-corrected FRB population model presented in \citet{aab+21} as well as models for the CHIME primary beam and system temperature \citep{abb+22}. The \sinza, Galactic latitude, and Galactic longitude tests then assesses deviations from this expected source rate.

\paragraph{Corrected Source Rate} Figure~\ref{fig:consistency_check} presents the binned source rate $\hat{R}_{i}$  after correcting for variations in system sensitivity.  The resulting $p$-values are reported in the column labelled ``$\snr \geq 10 / \sigma$'' in Table~\ref{table:consistency}.  The magnitudes of the previously mentioned discrepancies are reduced in all cases.  However, evidence persists for variability in the event rate as a function of accumulated rainfall and $\sinza$, and strong evidence remains for variability as a function of $\ra$ and Galactic latitude.  Additionally, evidence for variability persists as a function of the year of operation, the month of operation, and before and after the coarse-graining scheme modification.

Several instrumental and observational factors likely contribute to these residual discrepancies. The strongest observed deviation from a constant source rate occurs as a function of Galactic latitude, manifesting as a $\sim20\%$ reduction in the bin centered on the Galactic plane relative to the average rate, with a symmetric increase in detection rate toward higher Galactic latitudes, reaching a $10\%$ excess in the Galactic caps.  While our analysis corrects for the field-of-view-averaged variation in sky temperature, there remains an additional latitude-dependent increase in sky temperature toward the Galactic plane, which is not explicitly accounted for in our tests.  Additionally, excess dispersion and scattering imparted by propagation through the Galaxy further reduce detection efficiency at low Galactic latitudes.  Accounting for these propagation effects when constraining the source rate is beyond the scope of this paper and will be explored in future work.  The residual variations in source rate with \ra remain anti-correlated with the presence of the Galactic plane in the field of view, likely reflecting the same underlying Galactic-latitude-dependent effects at a reduced amplitude.

Similarly, a full treatment of the dependence on $\sin(\text{za})$ should account for the fraction of the band masked due to RFI, which has a complicated dependence on zenith angle and can introduce variations in sensitivity of up to $10\%$ \citep[see][]{jcc+21}. Furthermore, the simplified statistical treatment presented here does not incorporate uncertainties in the beam model or in the assumed FRB population used to construct the expected rate against which we compare.

\begin{figure*}[t]
\centering
\includegraphics{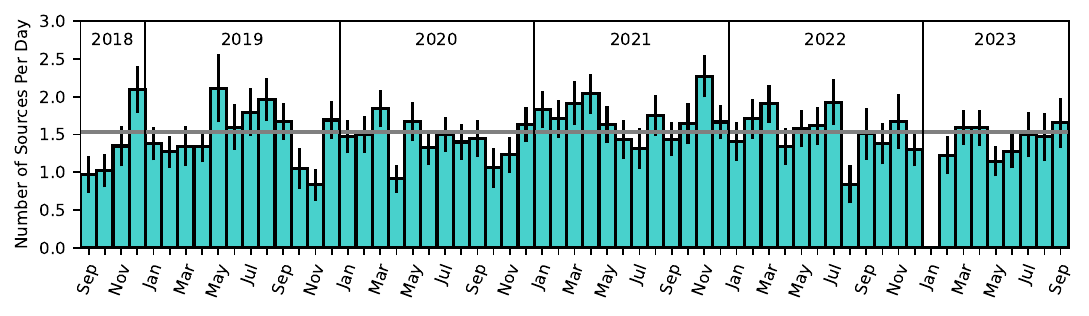}
\caption{Monthly variation in the observed source rate over the post-commissioning portion of the \cattwo period, spanning \expstart to \finish. Only the first detected burst from each source is included, such that the rate reflects the number of unique sources per day.  Source rates are corrected for variations in exposure and, approximately, for sensitivity using a variable \snr threshold based on an empirical proxy for system noise.  January 2023 has been excluded from the analysis due to significant snow and ice accumulation on the cylinders, which degraded system sensitivity.  The gray horizontal line indicates the mean source rate over the full time span ($\hat{R} = $ \ratevarsnr) for reference. Note that this value is lower than the raw CHIME/FRB detection rate due to the selection criteria applied to ensure uniformity. \label{fig:rate_versus_time}}
\end{figure*}

\paragraph{Impact of Temporal Outliers on Rate Consistency} Even after correcting for variations in sensitivity, there remains evidence for variability in the source rate over time, particularly across different years of operation and following modifications to the coarse-graining scheme. To investigate further, Figure~\ref{fig:rate_versus_time} shows the source rate as a function of month of operation over the full analysis period. When binned in this way, deviations from a constant-rate Poisson process are apparent: the corresponding $p$-value, reported in the final row of Table~\ref{table:consistency}, is 0.018. Excluding periods of non-negligible rainfall does not improve the consistency of the source rate across month of operation.

Using a Bayesian mixture model, we identified three months---November 2021, April 2020, and November 2019---as having elevated posterior probabilities ($20\text{--}30\%$) of being drawn from a distribution with excess variance relative to a constant-rate Poisson process. The strongest outlier is November 2021; removing this single month, which accounts for 2\% of the total exposure and 74 events, is sufficient to raise the $p$-value for the consistency test across month of operation above 0.05.  The results of all consistency tests with November 2021 excluded are presented in the final column of Table~\ref{table:consistency}.  Crucially, this single exclusion resolves the marginal inconsistencies related to the year of operation and the coarse-graining scheme. This suggests that the observed time-dependent variations in source rate are not due to systematic evolution in survey performance or changes in the underlying search software, but rather are driven by a small number of discrete time periods with anomalous behavior. We note that the dependence on accumulated rainfall remains marginal, with the $p$-value fluctuating near the 0.05 significance threshold regardless of whether one, two, or three potential outlier months are excluded. This persistence is consistent with a genuine physical reduction in sensitivity during wet conditions rather than an artifact of specific anomalous time periods.

Because the underlying cause of the anomalies in these specific months is unknown, we do not recommend excluding them from the catalog. Instead, to account for the observed non-Poissonian variation, we generally recommend estimating uncertainties in rate-derived parameters using bootstrap resampling. As a representative example, applying this method to the data binned by month of operation yields an uncertainty estimate on the mean rate that is approximately $20\%$ larger than the standard Poisson statistical error.

\paragraph{All-Sky FRB Rate} The source rate estimates presented above reflect the number of unique FRB sources detected per day by CHIME/FRB, after correcting for variations in exposure and approximately correcting for variations in sensitivity.  However, these measurements are specific to the instrument's sky coverage and detection thresholds.  We do not attempt here to infer the all-sky FRB rate, defined as the number of bursts per day across the full sky that exceed a given fluence threshold.  Estimating this requires detailed modeling of the CHIME/FRB detection efficiency as a function of sky position, fluence, scattering time, and other burst properties. Unlike in \catone \citep{aab+21}, we defer this analysis until the completion of a new, more comprehensive injection campaign. The resulting selection function will enable a robust estimate of the all-sky FRB rate in a forthcoming study.

\section{Discussion}
\label{sec:discussion}

\begin{figure*}[p]
\centering
\includegraphics[width=1.0\textwidth]{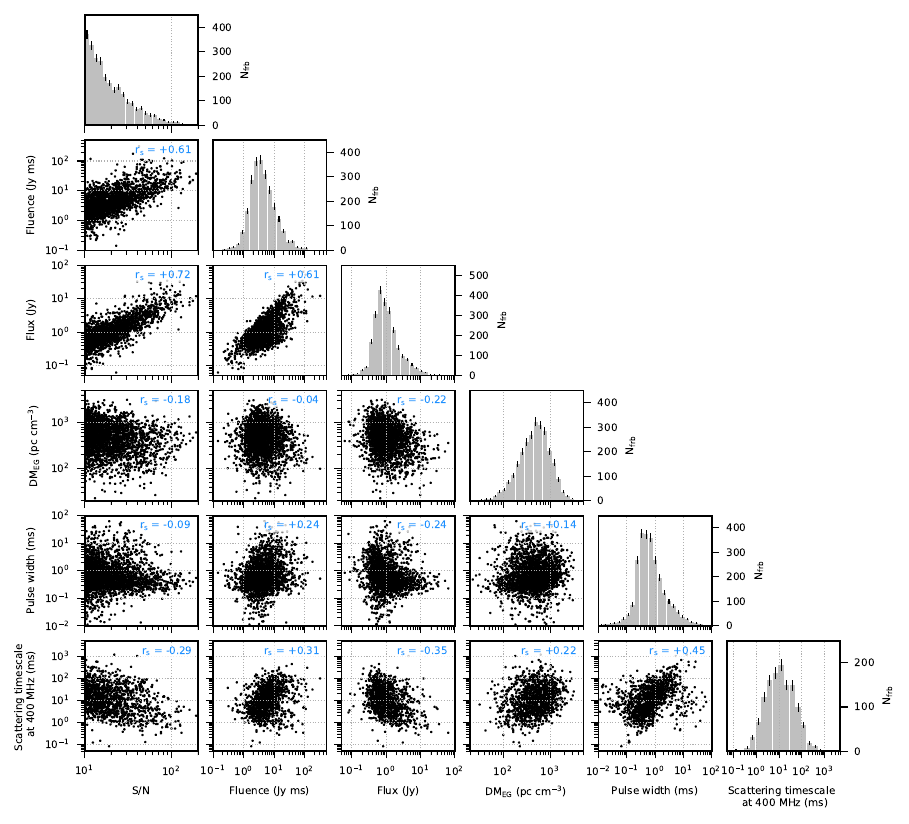}
\caption{Distribution of signal-to-noise ratio (\snr), fluence, peak flux density, extragalactic dispersion measure (\dmeg), pulse width, and scattering timescale at 400 MHz for sources that have not been observed to repeat.  Diagonal panels show 1D histograms of each parameter, where the y-axis represents the number of bursts per bin, with Poisson error bars.  Off-diagonal panels display 2D scatter plots of parameter pairs.  All parameter axes are log-scaled.  The blue text in the upper right corner of each scatter plot shows the Spearman rank correlation coefficient between the two parameters.  All correlations are statistically significant ($p$-value $\leq 10^{-5}$), with the exception of fluence-\dmeg ($p$-value $= 0.032$).  Extragalactic DM is obtained by subtracting the Galactic DM -- estimated using the NE2001 model \citep{ne2001} and a 30 pc cm$^{-3}$ contribution from the Galactic halo \citep{cbg+23} -- from the measured DM.  Scattering timescales are omitted for FRBs where the Gaussian profile was preferred over a scatter-broadened profile. \label{fig:parameter_distributions}}
\end{figure*}

In this section, we take a first look at the observed distributions of parameters that characterize the morphology and brightness of FRBs in \cattwo.  The analysis follows an event selection process similar to that described in \S\ref{sec:source_rate}.  Specifically, we exclude events detected prior to \expstart, as well as those observed during periods of non-nominal system sensitivity, those discovered by citizen scientists, and those detected in the far sidelobes of the primary beam.  These exclusions are implemented using the \texttt{excluded\_flag}, \texttt{citizen\_science\_flag}, and \texttt{sidelobe\_flag}, respectively.  Additionally, we require that the signal-to-noise ratio at detection (\texttt{bonsai\_snr}) be at least 10.

Unlike in \S\ref{sec:source_rate}, this analysis is restricted to sources that have not been observed to repeat, as there are well known differences in morphology between the repeating and non-repeating populations \citep{pgk+21,csp+25}.  A detailed comparison between these populations will be presented in a future work (Cook et al., \inprep).  After applying these criteria, the sample consists of 2588 bursts.  Among these, 38 lack morphological parameters (pulse width and scattering time) and 214 lack flux density and fluence estimates.  Bursts missing these properties are excluded from the corresponding distributions of those properties.

Figure~\ref{fig:parameter_distributions} presents the observed distributions of several key parameters, including \snr, fluence, peak flux density, extragalactic dispersion measure ($\dm_{\rm EG}$), pulse width ($\sigma$), and scattering time referenced to $400 \ \mbox{MHz}$ ($\tau_{400}$). To estimate the extragalactic DM, we subtract the predicted Galactic contribution from our measured DM, using the NE2001 model \citep{ne2001} and an additional 30 pc cm$^{-3}$ to account for the Galactic halo \citep{cbg+23}.  For pulse width, 132 events were best described by a multi-component burst morphology; in these cases, we use $(t_{0,{\rm last}} + \frac{1}{2} \sigma_{\rm last}) - (t_{0,{\rm first}} - \frac{1}{2} \sigma_{\rm first})$ to capture the full extent of the burst, where $t_0$ is the arrival time of a sub-burst, and ``first'' and ``last'' refer to the earliest and latest sub-burst in time, respectively.  Scattering times are not shown for the 1292 bursts where a Gaussian profile was preferred over the scatter-broadened profile.

Importantly, none of the distributions shown in Figure~\ref{fig:parameter_distributions} have been corrected for the significant instrumental biases of the CHIME/FRB detection system \citep[see][]{mts+23}.  An injection analysis is currently underway to better quantify these biases and will be presented in a forthcoming study (McGregor et al., \inprep).  In the meantime, we highlight a few preliminary observations based on the uncorrected data.

We quantify pairwise correlations between parameters using the Spearman rank correlation coefficient, $r_s$, which measures the strength and direction of a monotonic relationship between two variables by computing the Pearson correlation coefficient of their rank-transformed values. Because of the large sample size in our analysis, we are sensitive to even weak trends. Most parameter pairs in Figure~\ref{fig:parameter_distributions} have a $p$-value $\leq 10^{-6}$ under the null hypothesis of no correlation, indicating that the measured correlations are statistically significant; the notable exception is the fluence--\dmeg pair, whose lack of correlation ($p$-value $= 0.038$) we discuss below.

We begin by noting a strong correlation between \snr and both flux density ($r_s = 0.72$) and fluence ($r_s = 0.61$), even though our flux density and fluence estimates are lower limits. This correlation is expected since both quantities contribute to the detectability of a burst. We also observe modest but statistically significant anti-correlations between \dmeg and both \snr ($r_s = -0.18$) and flux density ($r_s = -0.22$), as one might expect if \dmeg is a proxy for distance and more distant sources are, on average, fainter. The modest amplitudes of these correlations likely reflect the large intrinsic scatter in \dmeg at a given redshift. We further note that \dmeg is slightly positively correlated with width ($r_s = 0.14$), which could arise from imperfect correction of intra-channel dispersion smearing, which preferentially broadens high-DM bursts. Because fluence is proportional to width~$\times$~flux density, the mild positive \dmeg--width trend partially cancels the negative \dmeg--flux density trend. The resulting correlation between fluence and \dmeg ($r_s = -0.04$) is not statistically significant.  Similar behaviour was seen in the \catone sample with baseband-derived fluences, where improved measurements also failed to reveal a significant trend between fluence and \dmeg \citep{scm+25,csp+25}, though the smaller sample size in that case may have limited sensitivity to subtle effects.

The apparent width-scattering correlation ($r_s = 0.45$) is unsurprising, likely at least in part a result of the measurability of each being linked to the magnitude of the other, plus that both are subject to sizable detection biases \citep[see][]{mts+23}.  Similarly the fluence-scattering correlation ($r_s = 0.31$)  must be strongly influenced by CHIME/FRB's difficulty in detecting highly scattered bursts; those with low fluences are generally missed.  The hint of a DM-scattering correlation ($r_s = 0.22$) likewise requires careful bias correction before any astrophysical inference can be made.  Our upcoming injection analysis will examine these trends critically and quantitatively.

\section{Conclusions and Future Work}
\label{sec:conclusions}

The Second CHIME/FRB Catalog offers the global astronomical community the opportunity to assist in the unravelling of the mystery of FRBs, as well as in their utilization as novel probes of the matter distribution in the Universe.

Proper analysis and interpretation of CHIME/FRB catalog results requires careful attention to our instrument's detection biases.  For the First CHIME/FRB Catalog \citep{aab+21}, we quantified our detection biases by injecting a population of fake FRBs and observing which our system detected as a function of different burst properties \citep{mts+23}.  One clear outcome of that analysis was the strong selection bias against heavily scattered events, a bias that remained largely unquantified in our first injection analysis. That shortcoming, together with significant evolution in our RFI environment since the time of \catone, has motivated a second, much larger scale injection analysis.  This is presently underway and will be reported on elsewhere (McGregor et al., \inprep).
Additionally, a large fraction of our \cattwo events have saved baseband data \citep[see][]{abb+18}.  These voluminous data are presently under analysis and will be published elsewhere, as
we did for \catone \citep{mmm+21,aaa+24}.  This analysis will update the catalog we present here with improved FRB sky localizations, higher time- and frequency-resolution waterfall plots, and importantly, actual fluence and flux density measurements, rather than the upper limits presented in this present work.   Also underway as an independent effort is a catalog of dozens of newly discovered repeater sources, based in part on the contents of the Catalog we present here, as we did for \catone \citep{abb+23}.  Careful statistical analysis to identify only those sources that have negligible probability of being due to chance is required to robustly identify true repeaters, given the large number of FRBs we have discovered \citep{cle+24}.

In addition, our team is presently undertaking a variety of studies making use of \cattwo, including: a cross-correlation of FRB dispersion measures with galaxy catalogs \citep{wma+25}; an analysis of FRB morphology using machine-learning methods \citep{kfb+25}; an analysis of the impact of Milky Way structures on our FRB sky distribution \citep{smf+25}; and a study of the impact of M31 and its circumgalactic medium on FRB dispersion measures and scattering timescales (Kahinga et al., \inprep, Leung et al., \inprep).  Other studies include population synthesis analyses to investigate FRB properties such as their energy function, redshift distribution, and the fraction of repeating sources.  We are also examining particularly interesting sources within the catalog, including those with low dispersion measures, strong scattering, or positions coincident with known structures such as galaxy clusters, as well as potential gravitationally lensed FRBs. We are certain that the community will come up with many complementary and creative additional uses for our work, which we strongly encourage.  We hope \cattwo serves as a major milestone in the continued study of FRBs.

Finally, we note that our efforts to understand the FRB mystery are part of a broader global endeavor. Instruments such as ASKAP \citep{mbb+10}, FAST \citep{nl13}, DSA-110 \citep{lsr+24}, realfast \citep{lbb+18}, and MeerKAT \citep{sta16} have already made important contributions to FRB discovery and characterization. In particular, a large number of new FRB host galaxies have been identified from the sample of FRBs discovered and localized by ASKAP \citep[e.g.,][]{bsp+20, hpf+20, gfk+23, gfs+24, gfd+25}, DSA-110 \citep[e.g.,][]{lsr+24, src+24}, and the CHIME/FRB–KKO Outrigger program \citep{aaa+25a}. MeerKAT has also discovered new FRBs \citep{rbc+22, jbc+23}, including the first non-CHIME discovery of a new source in a highly active repeating state \citep{tpr+25}. Data from these instruments have enabled high time resolution and polarimetric studies of FRBs, both as a sample and on remarkable individual sources, providing insight into plausible FRB emission mechanisms and origins \citep[e.g.,][]{dds+20, zlz+23, bjd+24, jxn+24, DSA_pol_2024, ddu+25, bjm+25, usg+26, sdb+25}. Population studies were conducted on bursts observed by these instruments to investigate FRB energy distributions \citep{aje+25, hjg+25}. Furthermore, FRBs discovered by these instruments have been used to probe the diffuse ionized gas in the Universe \citep[e.g.,][]{crs+25, fro2024, kal+24, alk+25, hck+25, hsk+25, lsm+25}. These existing instruments will soon be joined by upcoming facilities like CHORD \citep{vlg+19}, DSA-2000 \citep{hrw+19}, BURSTT \citep{lll+22}, HIRAX \citep{caa+22}, and the Square Kilometre Array \citep{dhsl+2009}. In addition, the CHIME/FRB experiment is currently commissioning a new operational phase where companion, ``outrigger" telescopes are used in tandem with CHIME to enable real-time, low-frequency VLBI of FRBs \citep{aaa+25}. The Second CHIME/FRB Catalog represents one of many forthcoming milestones in the study of fast radio bursts. The capabilities of the aforementioned experiments are highly complementary to those of CHIME/FRB, and together they will advance our collective knowledge of the FRB phenomenon.

\section*{ACKNOWLEDGEMENTS}

We thank the many students and short-term team members who have contributed to the Second CHIME/FRB Catalog, whether through data processing, software development, or monitoring telescope operations.  We are also grateful to the more than 5000 citizen scientists who participated in the \href{https://www.zooniverse.org/projects/mikewalmsley/bursts-from-space}{Bursts from Space} project on the Zooniverse platform, helping to identify FRBs that are included in this catalog.

We acknowledge that CHIME is located on the traditional, ancestral, and unceded territory of the Syilx/Okanagan people. We are grateful for the warm reception and skillful help we have received from the staff of the Dominion Radio Astrophysical Observatory, which is operated by the National Research Council of Canada.

CHIME operations are funded by a grant from the NSERC Alliance Program and by support from McGill University, University of British Columbia, and University of Toronto. CHIME was funded by a grant from the Canada Foundation for Innovation (CFI) 2012 Leading Edge Fund (Project 31170) and by contributions from the provinces of British Columbia, Québec and Ontario. The CHIME/FRB Project was funded by a grant from the CFI 2015 Innovation Fund (Project 33213) and by contributions from the provinces of British Columbia and Québec, and by the Dunlap Institute for Astronomy and Astrophysics at the University of Toronto. Additional support was provided by the Canadian Institute for Advanced Research (CIFAR), the Trottier Space Institute at McGill University, and the University of British Columbia.

The CHIME/FRB baseband recording system is funded in part by a CFI John R. Evans Leaders Fund award to IHS.  The Dunlap Institute is funded by an endowment established by the David Dunlap family and the University of Toronto. Research at Perimeter Institute is supported by the Government of Canada through Industry Canada and by the Province of Ontario through the Ministry of Research \& Innovation. The National Radio Astronomy Observatory is a facility of the National Science Foundation operated under cooperative agreement by Associated Universities, Inc.

\input{acknowledgements}

\facilities{CHIME(FRB)}

\pagebreak

\appendix

\section{Description of Catalog Data Fields}
\label{app:datafields}

Table~\ref{table:datafields} provides descriptions of each field included in CHIME/FRB \cattwo tabular data.

\input{table_data_fields}

\section{Residuals to {\tt Fitburst} Fits in Catalog 2}
\label{app:waterfall_residual}

\begin{figure*}[p]
	\centering
\includegraphics[width=0.90\textwidth]{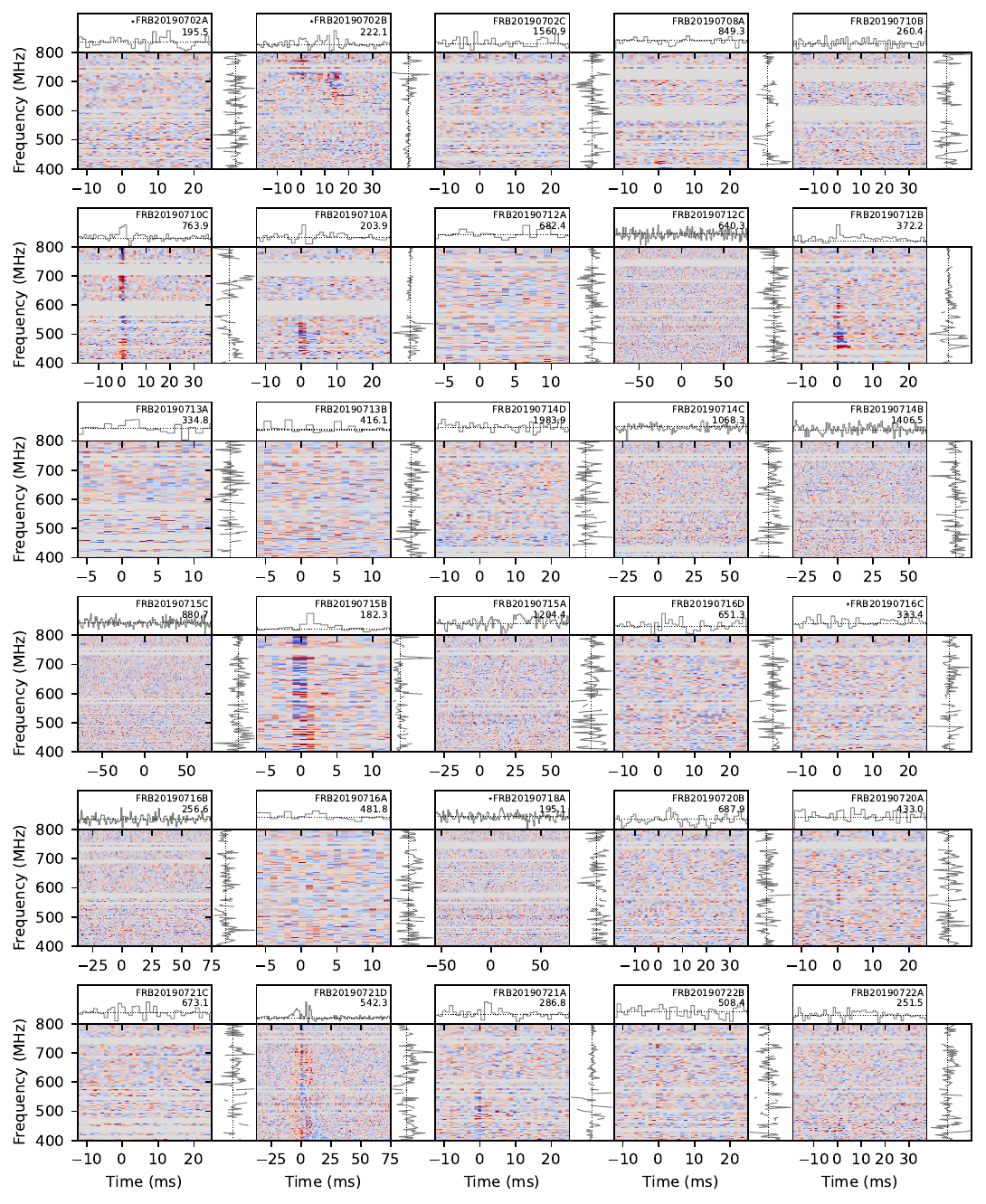}
\figcaption{Residual dynamic spectra (de-dispersed), frequency-averaged time series, and time-averaged spectra for the first 30 FRBs in \cattwo after the end of \catone (July 1, 2019), ordered by arrival time.  The best-fit model for each event has been subtracted.  The TNS name and best-fit dispersion measure (DM, in pc cm$^{-3}$) are displayed in the upper right corner of each panel.  A star preceding the TNS name indicates that the burst is from a repeating source.  The complete figure set containing all events in \cattwo with \fitburst{} results is available in the online journal (149 panels) and at \data.
\label{fig:waterfall_residual}}
\end{figure*}

Figure~\ref{fig:waterfall_residual} shows the residuals from the burst-morphology fits shown in Figure~\ref{fig:waterfall_data} and described in \S\ref{sec:fitburst}.

\section{Changes to {\tt FITBURST} Between CHIME/FRB Catalogs}
\label{app:fitburst}

\begin{figure}[p]
    \centering
    \includegraphics[width=0.75\linewidth]{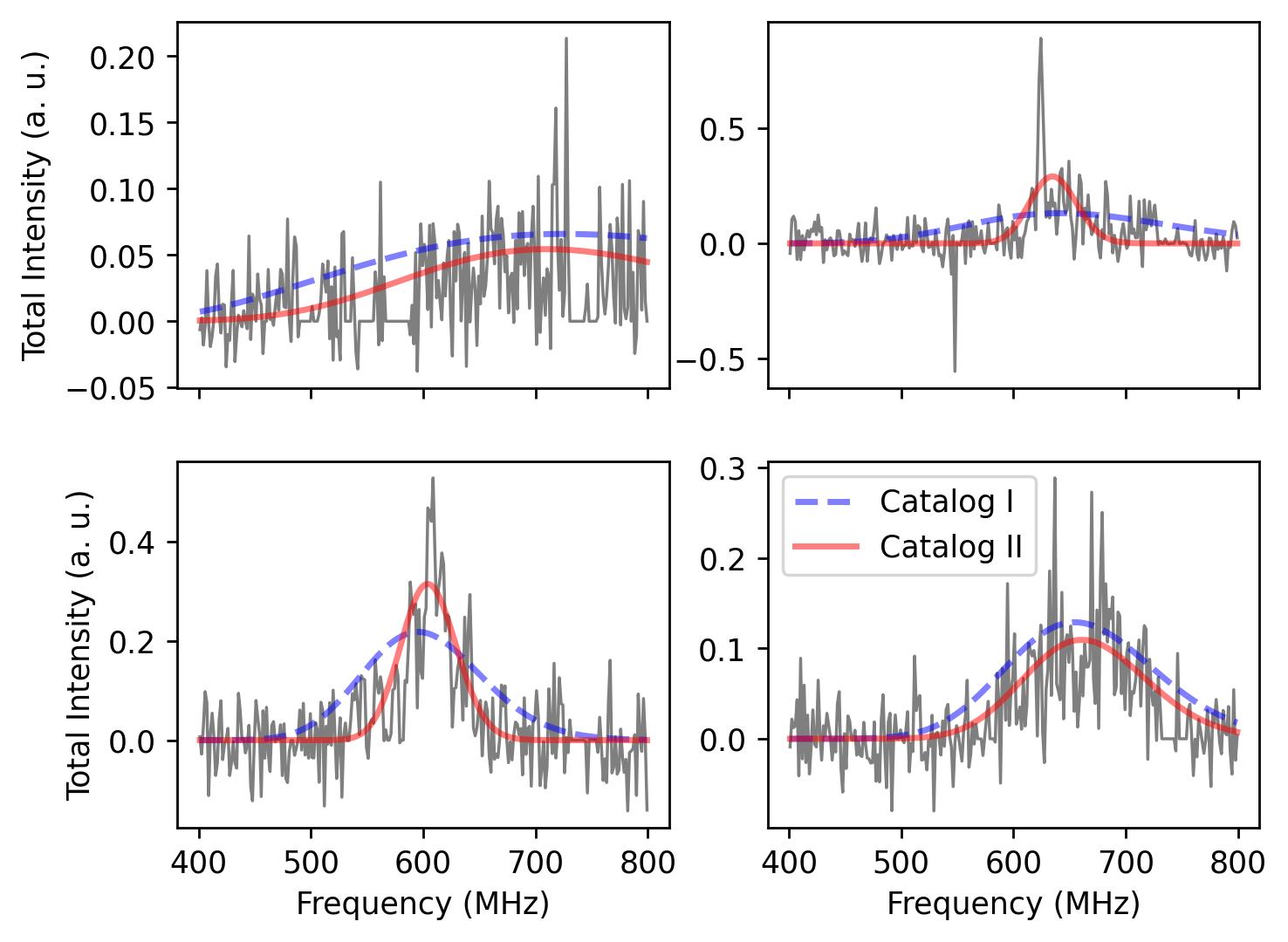}
    \caption{Spectral energy distributions for four FRBs detected by the CHIME/FRB backend as part of \catone. The grey lines are obtained by averaging the dedispersed dynamic spectrum over an amount of time equal to the best-fit value of $\sigma_i$ centered on $t_{0,i}$, and adding over all $N$ burst components. The red and blue lines represent the best-fit estimates of $\sum_i^N A_iF_{k,i}$, using Equation \ref{eq:fitburst_sed} and the best-fit estimates of \{$\alpha_i$, $\beta_i$, $\gamma_i$\} as determined with the versions of \fitburst{} developed for \catone and \cattwo. In all cases, the \cattwo{} \fitburst{} models yielded lower goodness-of-fit statistics than those first determined for \catone.}
    \label{fig:fitburst_comparisons_example}
\end{figure}

We have made a number of changes to the {\tt fitburst} codebase for \cattwo relative to the version used for \catone. These changes are implemented to (a) improve fitting quality against an increasingly diverse range of FRB morphology, and (b) ensure robust and comparable statistics derived from the optimized models. For example, as described by \cite{fpb+24}, we redefine the amplitude to be $10^\alpha$ instead of $A$. This change leads to unambiguous improvements in the generation of models for repeater-like bursts, due to the distinct values of \{$\alpha$, $\beta$, $\gamma$\} that better describe narrow-band bursts. This reparameterization has nonetheless led to differences in the best-fit values of \{$\alpha$, $\beta$, $\gamma$\} as described in \S\ref{sec:validation} and shown in Figure \ref{fig:fitburst_comparisons_example}.  Furthermore, the version of \fitburst{} used to generate \catone models relied on the manual use of parameter ``bounds" when modeling multi-component FRBs. These bounds were useful in expediting model convergence for multi-component FRBs, but were nonetheless {\it ad hoc} and not applied uniformly to all \catone data. Starting with \cattwo, we no longer use parameter bounds to ensure both uniformity and efficacy in our modeling efforts.  As a result of these upgrades, we find that the \cattwo pipeline achieves more robust convergence and consistently yields better goodness-of-fit statistics than the \catone version (see Figure \ref{fig:fitburst_comparisons_example}).

We have also found that the pulse broadening function -- referred to as $T_{kn}$ by \cite{fpb+24} -- exhibited numerical instability in the regime where the scattering timescale is much smaller than the intrinsic width ($\tau_k \ll \sigma$), where $\tau_k = \tau_r(\nu_k/\nu_r)^\delta$. We avoided this floating-point overflow in past CHIME/FRB works, including \catone, by setting a criterion where a model of the dynamic spectrum switched between a scatter-broadened and intrinsically-Gaussian shape at the value of $\nu_k$ where $(\tau_k / \sigma) < 0.15$, with the criterion value arbitrarily chosen. For \cattwo, we removed the need for this arbitrary threshold and improved numerical stability by using the scaled complementary error function ($\erfcx$) instead of the error function ($\erf$) when calculating $T_{kn}$:

\begin{align}
    T_{kn} = & \bigg(\frac{\nu_k}{\nu_r}\bigg)^{-\delta}\exp\bigg[-\frac{1}{2}\bigg(\frac{t_{kn} - t_0}{\sigma}\bigg)^2\bigg] \nonumber \\
    & \times \erfcx\bigg\{\frac{\sigma^2/\tau_k - (t_{kn}-t_0)}{\sqrt{2}\sigma}\bigg\}.
\end{align}

\noindent This form of $T_{kn}$ is analytically equivalent to the original definition presented by \cite{fpb+24}, but avoids numerical overflow in the exponential function noted above in the case where $\tau_k$ becomes small relative to $\sigma$.

In addition to the algorithmic changes described above, we also made two technical updates that improve the accuracy of model evaluation. First, we revised the convention used to define the dynamic spectrum axes. Whereas the version used in \catone referenced time to the start of each integration and frequency to the upper edge of each channel, we now reference them to the center of the time integration and the center of the frequency channel, respectively. Second, we modified how timestamps are handled during data extraction and model evaluation. In \catone, the timestamp at each frequency was reconstructed from a time axis defined at the reference frequency and the incoherent dispersion relation. However, because the data are discretely sampled, the extracted window at each frequency is centered on a nearby time sample rather than the exact nominal arrival time, introducing a small but unavoidable rounding offset. This offset was not accounted for when reconstructing the model timestamps, leading to small discrepancies between the data and the model. In \cattwo, we now extract and store the full 2D array of timestamps and evaluate the model at the exact times at which the corresponding data samples were recorded, eliminating this inconsistency.

\section{Discussion on Biases in Fluence Calculation}
\label{app:bias}

\begin{figure}[ht]
    \centering
\includegraphics[scale=0.35]{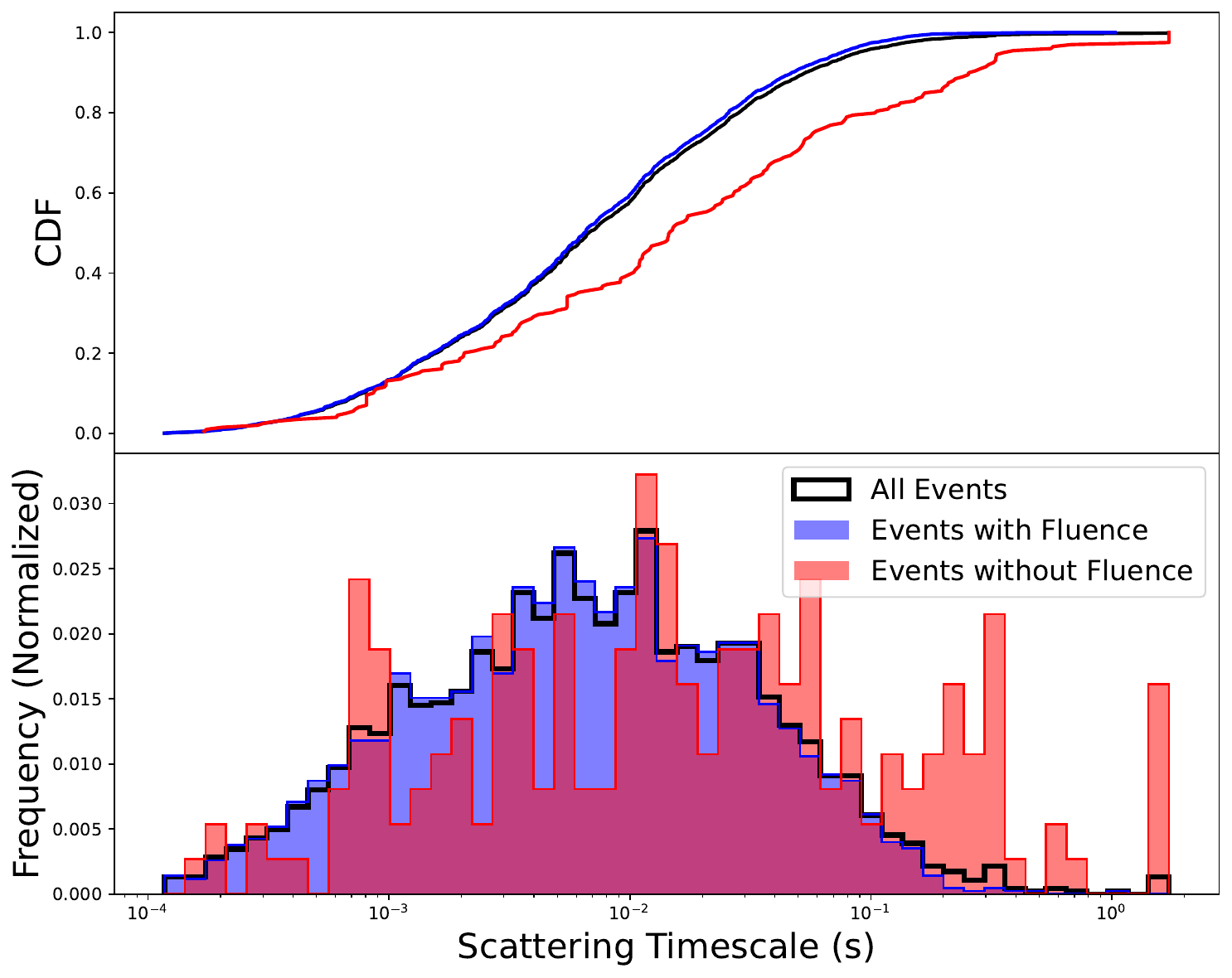}
\includegraphics[scale=0.35]{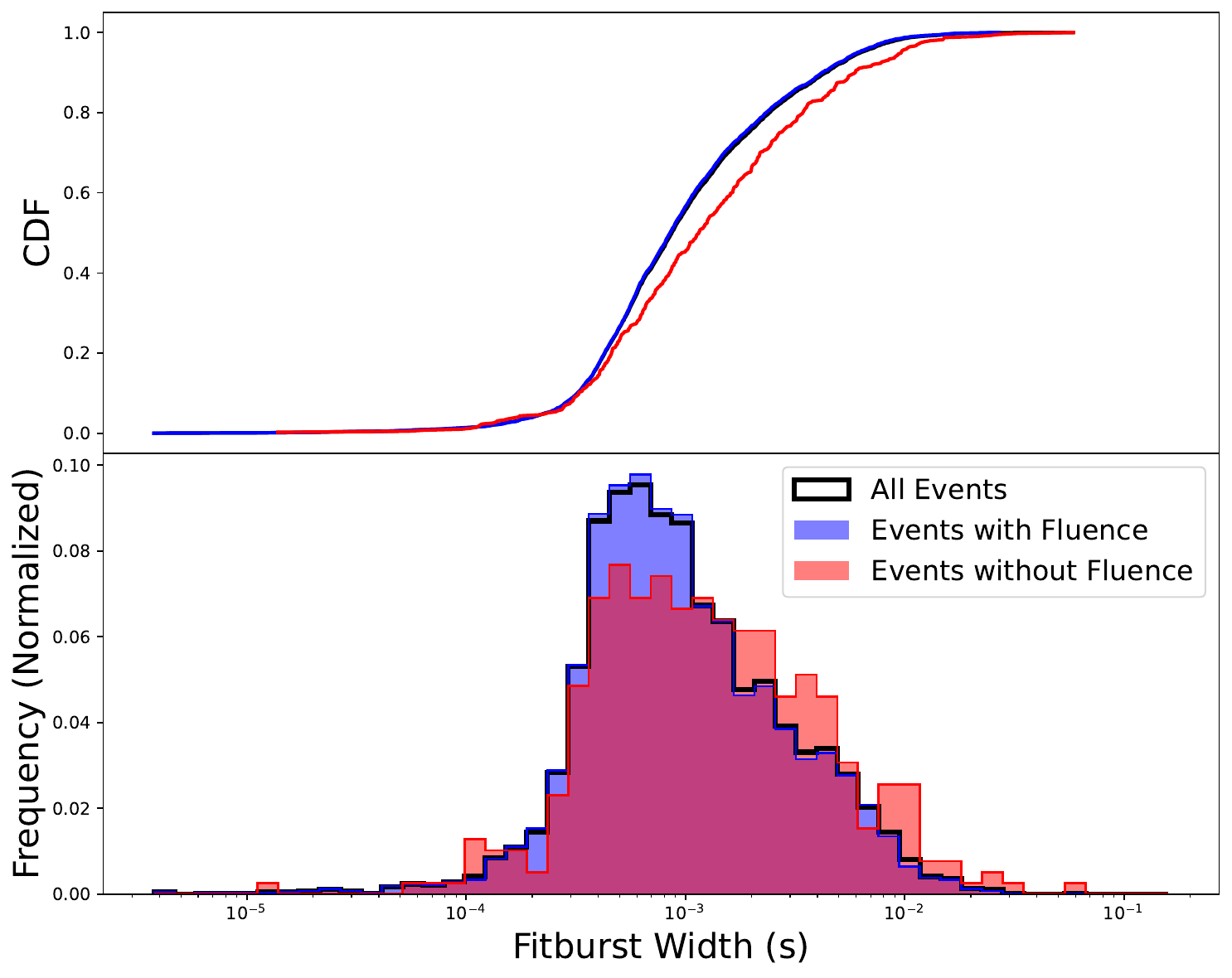}
\figcaption{Cumulative distribution function (top) and histogram (bottom) of scattering timescales (left) and \fitburst{} width (right) of events in this catalog. The histograms are normalized by the total counts in each sample. The width distributions include all 4455 events with valid \fitburst{} results (4085 with fluence, 370 without). The scattering distributions are restricted to the 1811 events where a scattering model was statistically preferred via an F-test (1645 with fluence, 166 without).  We find a bias in the failure to measure fluences for events with high scattering  (KS stat: 0.22, p-value: $3.68 \times 10^{-8}$) and large \fitburst{} widths (KS stat: 0.12, p-value: $2.93 \times 10^{-5}$). \label{fig:fluence_calculation_bias}}
\end{figure}

\begin{figure*}[ht]
	\centering
\includegraphics[scale=0.4]{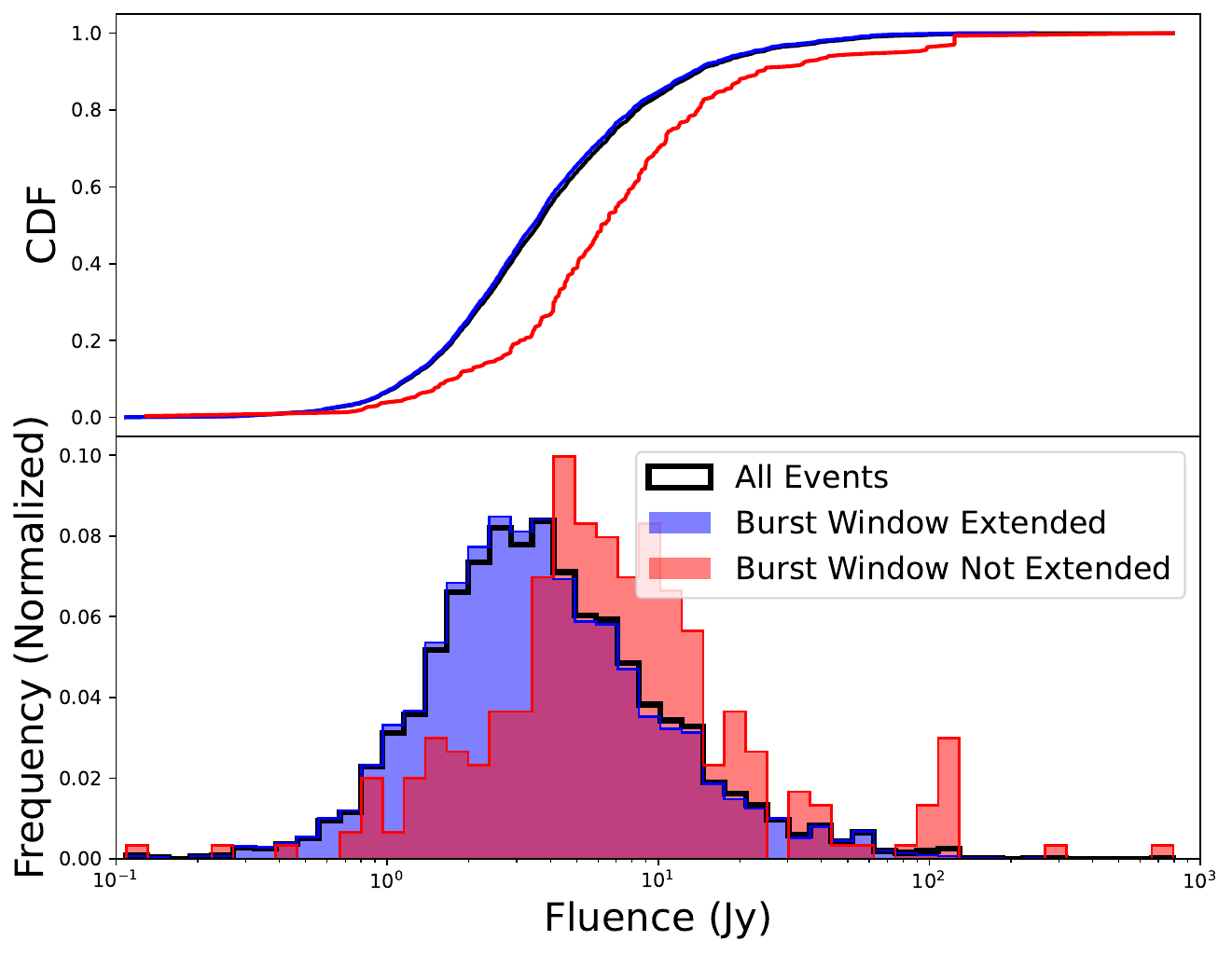}
\includegraphics[scale=0.35]{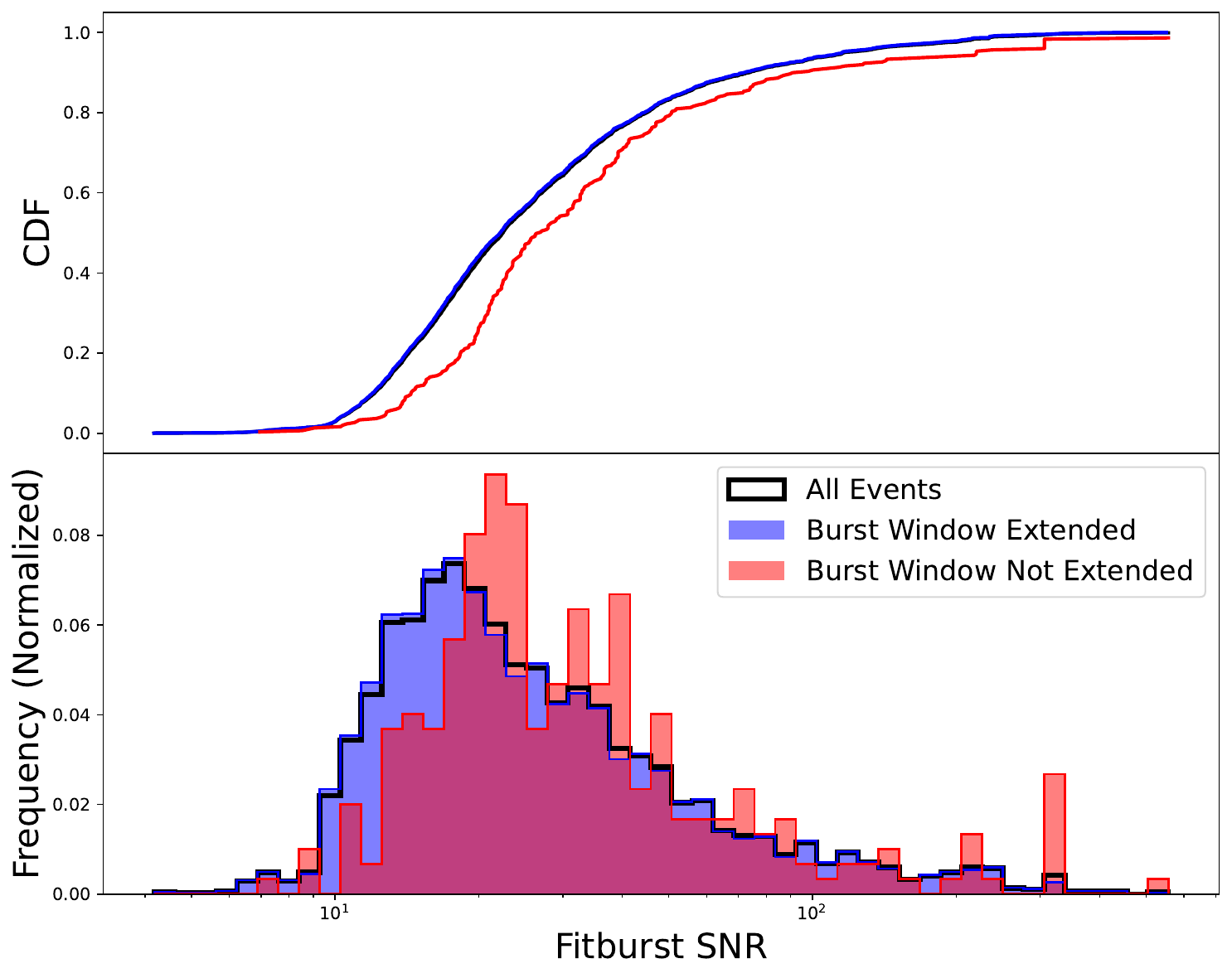}
\includegraphics[scale=0.35]{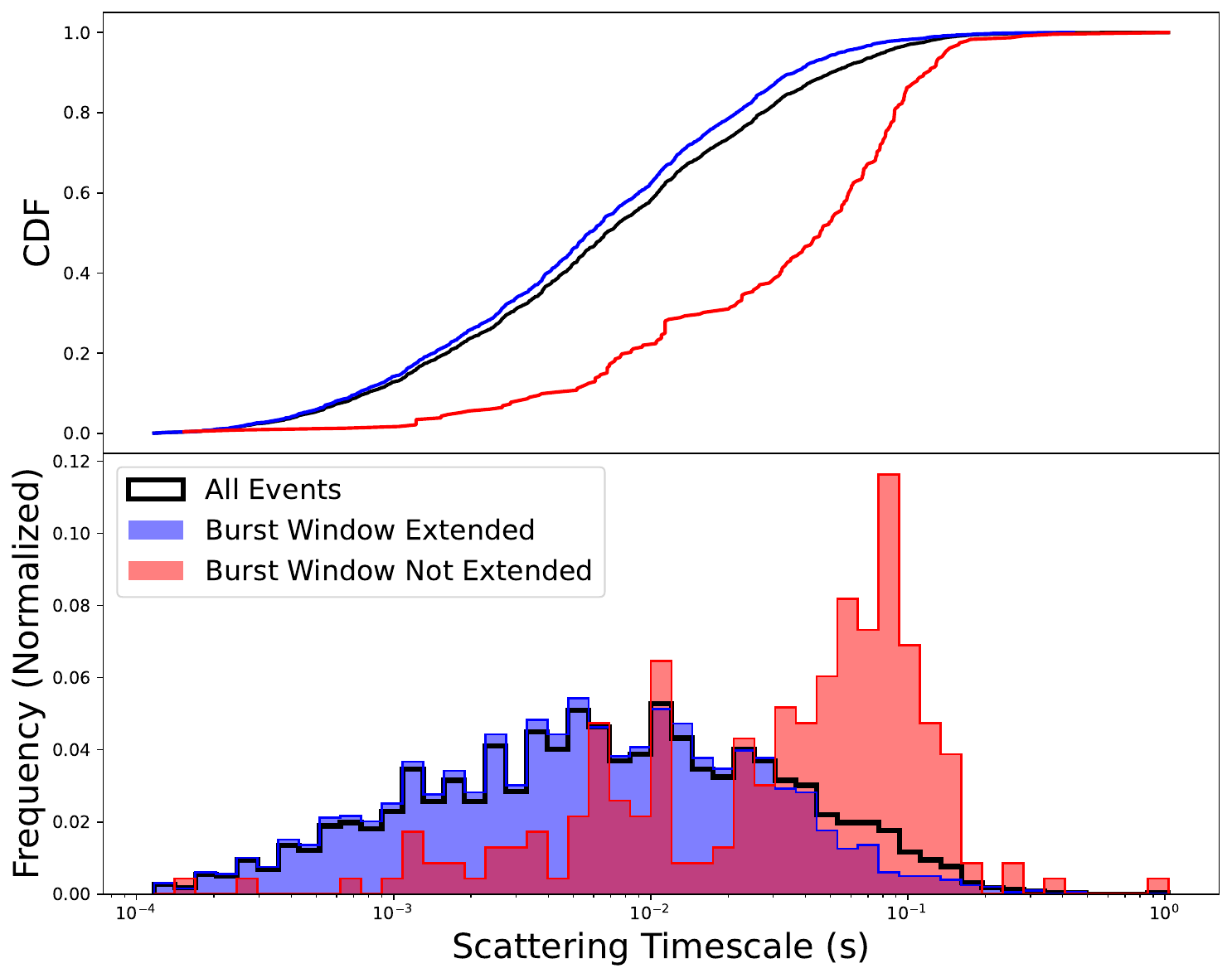}
\includegraphics[scale=0.35]{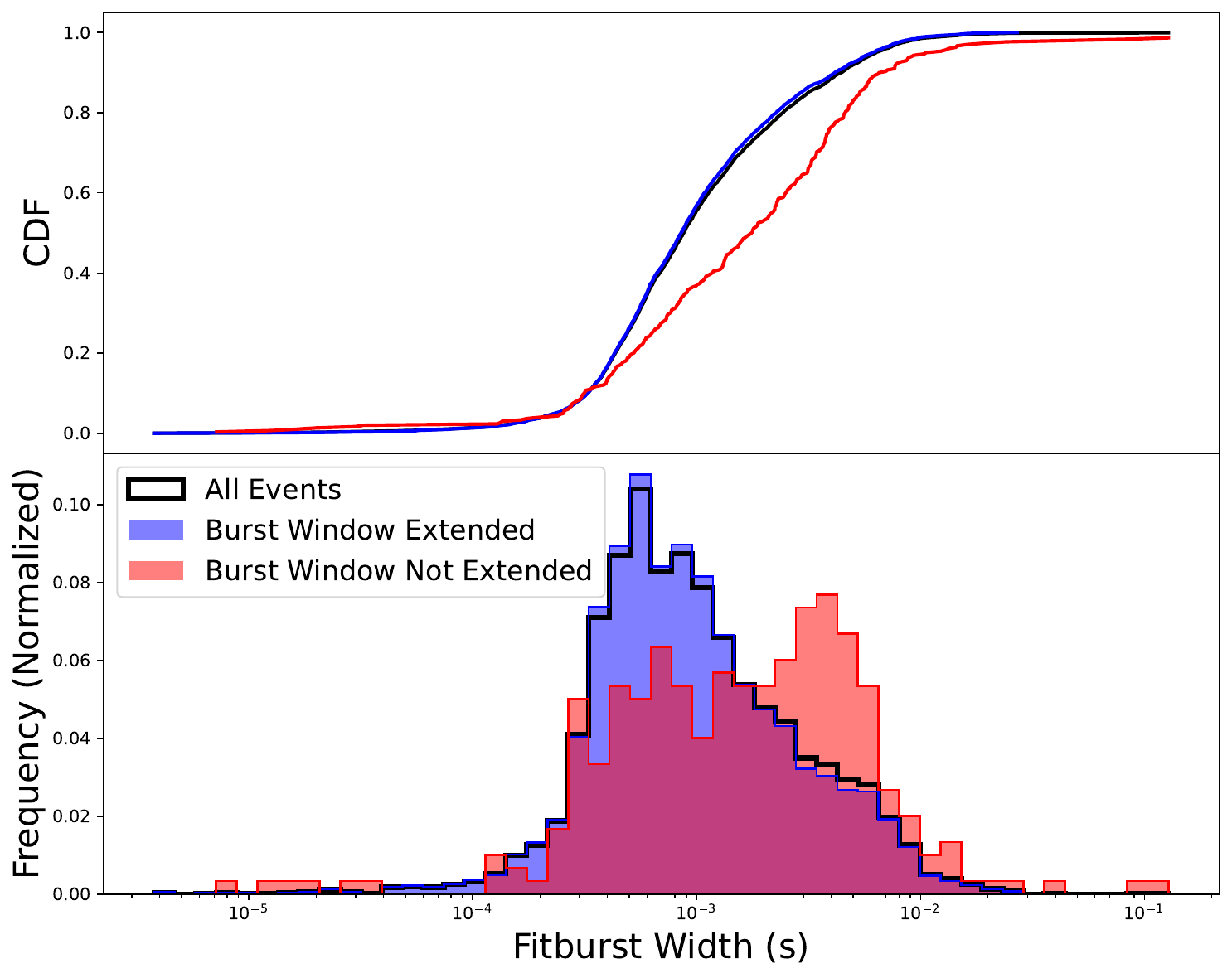}
\figcaption{Cumulative distribution functions and histograms of fluence (top left),  \fitburst{} \snr (top right), scattering timescale (bottom left), and \fitburst{} width (bottom right) for events in this catalog. The histograms are normalized by the total counts in each sample.  For fluence, \snr, and width, the sample includes all events with valid measurements, totaling 4085 events (3843 with extended fluence windows, 242 without). For scattering, the sample is restricted to the 1645 events where a scattering model was statistically preferred via F-test (1453 with extended windows, 192 without). We find a bias in the fluence window extension process, with extended events tending to have higher fluences (KS stat: 0.31, p-value: $7.77 \times 10^{-16}$), higher \snr (KS stat: 0.20, p-value: $2.10 \times 10^{-10}$), higher scattering (KS stat: 0.49, p-value: $1.22 \times 10^{-15}$), and larger \fitburst{} widths (KS stat: 0.24, p-value: $9.21 \times 10^{-15}$). \label{fig:fluence_burst_window_bias}}
\end{figure*}

The primary factors limiting fluence measurements are computational failures and large fractional errors arising from calibration issues or RFI contamination. A lack of bright, stable, and compact calibration sources -- particularly at high declinations -- is also responsible for a significant fraction of failed fluence calculations.
  
Figure~\ref{fig:fluence_calculation_bias} compares the distribution of scattering times and burst widths between events with and without flux density or fluence measurements. We find a significant bias in our failure modes towards wide and highly scattered bursts. No such biases are observed in DM, \snr, spectral running, or spectral index parameters. We also find no correlation between our ability to measure a fluence with number of sub-bursts and repeater identification of the event. 

In this catalog, we use two definitions of burst window for calculating fluences - the pulse emission region and the pulse emission region extended by double the Gaussian width on either end. Figure~\ref{fig:fluence_burst_window_bias} compares the distributions of fluence, \snr, burst width and scattering timescale between events with different burst window definitions. We find that a restricted burst window was needed for high \snr, highly scattered, and wide bursts that, as a consequence, have larger fluences compared to the other set. No such biases are observed in DM, spectral running, or spectral index parameters.

\bibliographystyle{aasjournalv7}

\bibliography{frbrefs}

\end{document}

%% file: preamble.tex
\usepackage{amssymb}
\usepackage{amsmath} 

\usepackage{epstopdf}
\usepackage{color}
\usepackage{amsmath}
\usepackage{graphicx}
\usepackage{graphics,graphicx,natbib}
\usepackage{multirow}
\usepackage{xspace}
\usepackage[dvipsnames]{xcolor}
\usepackage{makecell}

\newcommand{\nfrbtot}{4539\xspace}
\newcommand{\nfrbrep}{981\xspace}
\newcommand{\nfrbnorep}{3558\xspace}
\newcommand{\nrepeater}{83\xspace}
\newcommand{\nsource}{3641\xspace}

\newcommand{\ratevarsnr}{1.53 sources per day\xspace}

\newcommand{\start}{25 July 2018\xspace}
\newcommand{\finish}{15 September 2023\xspace}
\newcommand{\finishone}{1 July 2019\xspace}
\newcommand{\expstart}{4 September 2018\xspace}

\newcommand{\catone}{Catalog~1\xspace}
\newcommand{\cattwo}{Catalog~2\xspace}

\newcommand{\data}{\dataset[doi:10.11570/25.0066]{\doi{10.11570/25.0066}}\xspace}
\newcommand{\portal}{\dataset[CHIME/FRB Catalog 2 portal]{https://www.chime-frb.ca/catalog2}\xspace}

\newcommand{\dm}{\ensuremath{\text{DM}}\xspace}
\newcommand{\dmeg}{\ensuremath{\text{DM}_{\text{EG}}}\xspace}
\newcommand{\snr}{\ensuremath{\text{S/N}}\xspace}

\newcommand{\ra}{\ensuremath{\text{R.A.}}\xspace}
\newcommand{\dec}{\ensuremath{\text{decl.}}\xspace}

\newcommand{\sinza}{\ensuremath{\sin(\text{za})}\xspace}

\newcommand{\inprep}{in preparation\xspace}

\newcommand{\fitburst}[0]{{\tt fitburst}}
\newcommand{\burstwindowextended}[0]{{\tt fluence\_win\_extended}}
\newcommand{\True}[0]{1\xspace}
\newcommand{\False}[0]{0\xspace}
\newcommand{\missing}{{\tt null}\xspace}

\DeclareMathOperator\erf{erf}
\DeclareMathOperator\erfcx{erfcx}

\def\lapp{\ifmmode\stackrel{<}{_{\sim}}\else$\stackrel{<}{_{\sim}}$\fi}
\def\gapp{\ifmmode\stackrel{>}{_{\sim}}\else$\stackrel{>}{_{\sim}}$\fi}

\newcommand{\rev}[1]{\textbf{#1}}

%% file: authors.tex
\author[0000-0001-5002-0868, gname=Thomas, sname=Abbott]{Thomas Abbott}
    \email{thomas.abbott@mail.mcgill.ca}
    \affiliation{Department of Physics, McGill University, 3600 rue University, Montr\'eal, QC H3A 2T8, Canada}
    \affiliation{Trottier Space Institute, McGill University, 3550 rue University, Montr\'eal, QC H3A 2A7, Canada}
\author[0000-0001-5908-3152, gname=Bridget C., sname=Andersen]{Bridget C. Andersen}
    \email{bridget.andersen@mail.mcgill.ca}
    \affiliation{Department of Physics, McGill University, 3600 rue University, Montr\'eal, QC H3A 2T8, Canada}
    \affiliation{Trottier Space Institute, McGill University, 3550 rue University, Montr\'eal, QC H3A 2A7, Canada}
\author[0000-0002-3980-815X, gname=Shion, sname=Andrew]{Shion Andrew}
    \email{shiona@mit.edu}
    \affiliation{MIT Kavli Institute for Astrophysics and Space Research, Massachusetts Institute of Technology, 77 Massachusetts Ave, Cambridge, MA 02139, USA}
    \affiliation{Department of Physics, Massachusetts Institute of Technology, 77 Massachusetts Ave, Cambridge, MA 02139, USA}
\author[0000-0003-3772-2798, gname=Kevin, sname=Bandura]{Kevin Bandura}
    \email{kevin.bandura@mail.wvu.edu}
    \affiliation{Lane Department of Computer Science and Electrical Engineering, 1220 Evansdale Drive, PO Box 6109, Morgantown, WV 26506, USA}
    \affiliation{Center for Gravitational Waves and Cosmology, West Virginia University, Chestnut Ridge Research Building, Morgantown, WV 26505, USA}
\author[0000-0002-3615-3514, gname=Mohit, sname=Bhardwaj]{Mohit Bhardwaj}
    \email{mohitb@andrew.cmu.edu}
    \affiliation{McWilliams Center for Cosmology \& Astrophysics, Department of Physics, Carnegie Mellon University, Pittsburgh, PA 15213, USA}
\author[0000-0002-5342-163X, gname=Yash, sname=Bhusare]{Yash Bhusare}
    \email{ybhusare@ncra.tifr.res.in}
    \affiliation{National Centre for Radio Astrophysics, Post Bag 3, Ganeshkhind, Pune, 411007, India}
\author[0000-0002-1800-8233, gname=Charanjot, sname=Brar]{Charanjot Brar}
    \email{charanjot.brar@nrc-cnrc.gc.ca}
    \affiliation{National Research Council of Canada, Herzberg Astronomy and Astrophysics, 5071 West Saanich Road, Victoria, BC V9E 2E7, Canada}
\author[0000-0003-2047-5276, gname=Tomas, sname=Cassanelli]{Tomas Cassanelli}
    \email{tcassanelli@ing.uchile.cl}
    \affiliation{Department of Electrical Engineering, Universidad de Chile, Av. Tupper 2007, Santiago 8370451, Chile}
\author[0000-0002-2878-1502, gname=Shami, sname=Chatterjee]{Shami Chatterjee}
    \email{shami@astro.cornell.edu}
    \affiliation{Cornell Center for Astrophysics and Planetary Science, Cornell University, Ithaca, NY 14853, USA}
\author[0000-0001-6509-8430, gname=Jean-Francois, sname=Cliche]{Jean-Francois Cliche}
    \email{jfcliche@jfcliche.com}
    \affiliation{Department of Physics, McGill University, 3600 rue University, Montr\'eal, QC H3A 2T8, Canada}
    \affiliation{Trottier Space Institute, McGill University, 3550 rue University, Montr\'eal, QC H3A 2A7, Canada}
\author[0000-0001-6422-8125, gname=Amanda M., sname=Cook]{Amanda M. Cook}
    \email{amanda.cook@mcgill.ca}
    \affiliation{Department of Physics, McGill University, 3600 rue University, Montr\'eal, QC H3A 2T8, Canada}
    \affiliation{Trottier Space Institute, McGill University, 3550 rue University, Montr\'eal, QC H3A 2A7, Canada}
    \affiliation{Dunlap Institute for Astronomy and Astrophysics, 50 St. George Street, University of Toronto, ON M5S 3H4, Canada}
    \affiliation{David A.\ Dunlap Department of Astronomy and Astrophysics, 50 St. George Street, University of Toronto, ON M5S 3H4, Canada}
\author[0000-0002-8376-1563, gname=Alice, sname=Curtin]{Alice Curtin}
    \email{alice.curtin@mail.mcgill.ca}
    \affiliation{Department of Physics, McGill University, 3600 rue University, Montr\'eal, QC H3A 2T8, Canada}
    \affiliation{Trottier Space Institute, McGill University, 3550 rue University, Montr\'eal, QC H3A 2A7, Canada}
\author[0000-0001-7166-6422, gname=Matt , sname=Dobbs]{Matt  Dobbs}
    \email{Matt.Dobbs@mcgill.ca}
    \affiliation{Department of Physics, McGill University, 3600 rue University, Montr\'eal, QC H3A 2T8, Canada}
    \affiliation{Trottier Space Institute, McGill University, 3550 rue University, Montr\'eal, QC H3A 2A7, Canada}
\author[0000-0003-4098-5222, gname=Fengqiu Adam, sname=Dong]{Fengqiu Adam Dong}
    \email{fengqiu.dong@gmail.com}
    \affiliation{National Radio Astronomy Observatory, 520 Edgemont Rd, Charlottesville, VA 22903, USA}
\author[0000-0003-3734-8177, gname=Gwendolyn , sname=Eadie]{Gwendolyn  Eadie}
    \email{gwen.eadie@utoronto.ca}
    \affiliation{David A.\ Dunlap Department of Astronomy and Astrophysics, 50 St. George Street, University of Toronto, ON M5S 3H4, Canada}
    \affiliation{Department of Statistical Sciences, University of Toronto, Toronto, ON M5S 3G3, Canada}
    \affiliation{Data Sciences Institute, University of Toronto, 17th Floor, Ontario Power Building, 700 University Ave, Toronto, ON M5G 1Z5, Canada}
\author[0000-0003-0307-9984, gname=Tarraneh, sname=Eftekhari]{Tarraneh Eftekhari}
    \email{teftekhari@northwestern.edu}
    \affiliation{Center for Interdisciplinary Exploration and Research in Astronomy, Northwestern University, 1800 Sherman Avenue, Evanston, IL 60201, USA }
\author[0000-0001-8384-5049, gname=Emmanuel, sname=Fonseca]{Emmanuel Fonseca}
    \email{emmanuel.fonseca@mail.wvu.edu}
    \affiliation{Department of Physics and Astronomy, West Virginia University, PO Box 6315, Morgantown, WV 26506, USA }
    \affiliation{Center for Gravitational Waves and Cosmology, West Virginia University, Chestnut Ridge Research Building, Morgantown, WV 26505, USA}
\author[0000-0002-3382-9558, gname=B. M., sname=Gaensler]{B. M. Gaensler}
    \email{gaensler@ucsc.edu}
    \affiliation{Department of Astronomy and Astrophysics, University of California, Santa Cruz, 1156 High Street, Santa Cruz, CA 95060, USA}
    \affiliation{Dunlap Institute for Astronomy and Astrophysics, 50 St. George Street, University of Toronto, ON M5S 3H4, Canada}
    \affiliation{David A.\ Dunlap Department of Astronomy and Astrophysics, 50 St. George Street, University of Toronto, ON M5S 3H4, Canada}
\author[0000-0003-1884-348X, gname=Deborah, sname=Good]{Deborah Good}
    \email{deborah.good@umontana.edu}
    \affiliation{Department of Physics and Astronomy, University of Montana, 32 Campus Drive, Missoula, MT 59812, USA}
\author[0000-0002-1760-0868, gname=Mark, sname=Halpern]{Mark Halpern}
    \email{halpern@physics.ubc.ca}
    \affiliation{Department of Physics and Astronomy, University of British Columbia, 6224 Agricultural Road, Vancouver, BC V6T 1Z1 Canada}
\author[0000-0003-2317-1446, gname=Jason W. T., sname=Hessels]{Jason W. T. Hessels}
    \email{jason.hessels@mcgill.ca}
    \affiliation{Department of Physics, McGill University, 3600 rue University, Montr\'eal, QC H3A 2T8, Canada}
    \affiliation{Trottier Space Institute, McGill University, 3550 rue University, Montr\'eal, QC H3A 2A7, Canada}
    \affiliation{Anton Pannekoek Institute for Astronomy, University of Amsterdam, Science Park 904, 1098 XH Amsterdam, The Netherlands}
    \affiliation{ASTRON, Netherlands Institute for Radio Astronomy, Oude Hoogeveensedijk 4, 7991 PD Dwingeloo, The Netherlands}
\author[0000-0003-2405-2967, gname=Adaeze, sname=Ibik]{Adaeze Ibik}
    \email{adaeze.ibik@mail.utoronto.ca}
    \affiliation{Dunlap Institute for Astronomy and Astrophysics, 50 St. George Street, University of Toronto, ON M5S 3H4, Canada}
    \affiliation{David A.\ Dunlap Department of Astronomy and Astrophysics, 50 St. George Street, University of Toronto, ON M5S 3H4, Canada}
\author[0009-0009-0938-1595, gname=Naman, sname=Jain]{Naman Jain}
    \email{naman.jain@mail.mcgill.ca}
    \affiliation{Department of Physics, McGill University, 3600 rue University, Montr\'eal, QC H3A 2T8, Canada}
    \affiliation{Trottier Space Institute, McGill University, 3550 rue University, Montr\'eal, QC H3A 2A7, Canada}
\author[0000-0003-3457-4670, gname=Ronniy C., sname=Joseph]{Ronniy C. Joseph}
    \email{ronniy.joseph@mcgill.ca}
    \affiliation{Trottier Space Institute, McGill University, 3550 rue University, Montr\'eal, QC H3A 2A7, Canada}
    \affiliation{Department of Physics, McGill University, 3600 rue University, Montr\'eal, QC H3A 2T8, Canada}
\author[0000-0003-2739-5869, gname=Zarif, sname=Kader]{Zarif Kader}
    \email{zarif.kader@maill.mcgill.ca}
    \affiliation{Department of Physics, McGill University, 3600 rue University, Montr\'eal, QC H3A 2T8, Canada}
    \affiliation{Trottier Space Institute, McGill University, 3550 rue University, Montr\'eal, QC H3A 2A7, Canada}
\author[0000-0001-9345-0307, gname=Victoria M., sname=Kaspi]{Victoria M. Kaspi}
    \email{victoria.kaspi@mcgill.ca}
    \affiliation{Department of Physics, McGill University, 3600 rue University, Montr\'eal, QC H3A 2T8, Canada}
    \affiliation{Trottier Space Institute, McGill University, 3550 rue University, Montr\'eal, QC H3A 2A7, Canada}
\author[0009-0004-4176-0062, gname=Afrokk, sname=Khan]{Afrokk Khan}
    \email{afrasiyab.khan@mcgill.ca}
    \affiliation{Department of Physics, McGill University, 3600 rue University, Montr\'eal, QC H3A 2T8, Canada}
    \affiliation{Trottier Space Institute, McGill University, 3550 rue University, Montr\'eal, QC H3A 2A7, Canada}
\author[0009-0008-6166-1095, gname=Bikash, sname=Kharel]{Bikash Kharel}
    \email{bk0055@mix.wvu.edu}
    \affiliation{Department of Physics and Astronomy, West Virginia University, PO Box 6315, Morgantown, WV 26506, USA }
    \affiliation{Center for Gravitational Waves and Cosmology, West Virginia University, Chestnut Ridge Research Building, Morgantown, WV 26505, USA}
\author[0009-0002-0330-9188, gname=Ajay, sname=Kumar]{Ajay Kumar}
    \email{akumar@ncra.tifr.res.in}
    \affiliation{National Centre for Radio Astrophysics, Post Bag 3, Ganeshkhind, Pune, 411007, India}
\author[0000-0003-1455-2546, gname=T.L., sname=Landecker]{T.L. Landecker}
    \email{tom.landecker.drao@gmail.com}
    \affiliation{Dominion Radio Astrophysical Observatory, Herzberg Research Centre for Astronomy and Astrophysics, National Research Council Canada, PO Box 248, Penticton, BC V2A 6J9, Canada}
\author[0000-0002-1172-0754, gname=Dustin, sname=Lang]{Dustin Lang}
    \email{dlang@perimeterinstitute.ca}
    \affiliation{Perimeter Institute of Theoretical Physics, 31 Caroline Street North, Waterloo, ON N2L 2Y5, Canada}
\author[0000-0003-2116-3573, gname=Adam E., sname=Lanman]{Adam E. Lanman}
    \email{alanman@mit.edu}
    \affiliation{MIT Kavli Institute for Astrophysics and Space Research, Massachusetts Institute of Technology, 77 Massachusetts Ave, Cambridge, MA 02139, USA}
    \affiliation{Department of Physics, Massachusetts Institute of Technology, 77 Massachusetts Ave, Cambridge, MA 02139, USA}
\author[0000-0001-5523-6051, gname=Magnus, sname=L'Argent]{Magnus L'Argent}
    \email{magnus.largent@mail.mcgill.ca}
    \affiliation{Department of Physics, McGill University, 3600 rue University, Montr\'eal, QC H3A 2T8, Canada}
    \affiliation{Trottier Space Institute, McGill University, 3550 rue University, Montr\'eal, QC H3A 2A7, Canada}
\author[0000-0002-5857-4264, gname=Mattias, sname=Lazda]{Mattias Lazda}
    \email{mattias.lazda@mail.utoronto.ca}
    \affiliation{Dunlap Institute for Astronomy and Astrophysics, 50 St. George Street, University of Toronto, ON M5S 3H4, Canada}
    \affiliation{David A.\ Dunlap Department of Astronomy and Astrophysics, 50 St. George Street, University of Toronto, ON M5S 3H4, Canada}
\author[0000-0002-4209-7408, gname=Calvin, sname=Leung]{Calvin Leung}
    \email{calvin_leung@berkeley.edu}
    \affiliation{Department of Astronomy, University of California, Berkeley, CA 94720, United States}
    \affiliation{Miller Institute for Basic Research, Stanley Hall, Room 206B, Berkeley, CA 94720, USA}
\author[0000-0001-7931-0607, gname=Dong Zi, sname=Li]{Dong Zi Li}
    \email{dongzili@princeton.edu}
    \affiliation{Department of Astrophysical Sciences, Princeton University, Princeton, NJ 08544, USA}
\author[0000-0001-5578-359X, gname=Chris J., sname=Lintott]{Chris J. Lintott}
    \email{chris.lintott@physics.ox.ac.uk}
    \affiliation{Department of Physics, University of Oxford, Denys Wilkinson Building, Keble Road, Oxford, OX1 3RH, UK}
\author[0000-0002-7164-9507, gname=Robert, sname=Main]{Robert Main}
    \email{robert.main@mcgill.ca}
    \affiliation{Department of Physics, McGill University, 3600 rue University, Montr\'eal, QC H3A 2T8, Canada}
    \affiliation{Trottier Space Institute, McGill University, 3550 rue University, Montr\'eal, QC H3A 2A7, Canada}
\author[0000-0002-4279-6946, gname=Kiyoshi W., sname=Masui]{Kiyoshi W. Masui}
    \email{kmasui@mit.edu}
    \affiliation{MIT Kavli Institute for Astrophysics and Space Research, Massachusetts Institute of Technology, 77 Massachusetts Ave, Cambridge, MA 02139, USA}
    \affiliation{Department of Physics, Massachusetts Institute of Technology, 77 Massachusetts Ave, Cambridge, MA 02139, USA}
\author[0000-0001-5536-4635, gname=Sujay, sname=Mate]{Sujay Mate}
    \email{sujay.mate@rrimail.rri.res.in}
    \affiliation{Raman Research Institute, C. V. Raman Avenue, Sadashivanagar, Bangalore, Karnataka - 560080, India}
    \affiliation{Department of Astronomy and Astrophysics, Tata Institute of Fundamental Research, Mumbai, 400005, India}
\author[0000-0003-2111-3437, gname=Kyle, sname=McGregor]{Kyle McGregor}
    \email{kyle.mcgregor@mail.mcgill.ca}
    \affiliation{Department of Physics, McGill University, 3600 rue University, Montr\'eal, QC H3A 2T8, Canada}
    \affiliation{Trottier Space Institute, McGill University, 3550 rue University, Montr\'eal, QC H3A 2A7, Canada}
\author[0000-0001-7348-6900, gname=Ryan, sname=Mckinven]{Ryan Mckinven}
    \email{ryan.mckinven@mcgill.ca}
    \affiliation{Department of Physics, McGill University, 3600 rue University, Montr\'eal, QC H3A 2T8, Canada}
    \affiliation{Trottier Space Institute, McGill University, 3550 rue University, Montr\'eal, QC H3A 2A7, Canada}
\author[0000-0002-0772-9326, gname=Juan, sname=Mena-Parra]{Juan Mena-Parra}
    \email{juan.menaparra@utoronto.ca}
    \affiliation{Dunlap Institute for Astronomy and Astrophysics, 50 St. George Street, University of Toronto, ON M5S 3H4, Canada}
    \affiliation{David A.\ Dunlap Department of Astronomy and Astrophysics, 50 St. George Street, University of Toronto, ON M5S 3H4, Canada}
\author[0000-0001-8845-1225, gname=Bradley W., sname=Meyers]{Bradley W. Meyers}
    \email{bradley.meyers@curtin.edu.au}
    \affiliation{International Centre for Radio Astronomy Research (ICRAR), Curtin University, Bentley WA 6102, Australia}
    \affiliation{Australian SKA Regional Centre (AusSRC), Curtin University, Bentley WA 6102, Australia}
\author[0000-0002-2551-7554, gname=Daniele, sname=Michilli]{Daniele Michilli}
    \email{danielemichilli@gmail.com}
    \affiliation{Laboratoire d'Astrophysique de Marseille, Aix-Marseille Univ., CNRS, CNES, Marseille, France}
\author[0000-0002-3616-5160, gname=Cherry, sname=Ng]{Cherry Ng}
    \email{cherry.ng-guiheneuf@cnrs-orleans.fr}
    \affiliation{Laboratoire de Physique et Chimie de l'Environnement et de l'Espace - Université d'Orléans/CNRS, 45071, Orléans Cedex 02, France}
\author[0000-0002-0940-6563, gname=Mason, sname=Ng]{Mason Ng}
    \email{mason.ng@mcgill.ca}
    \affiliation{Department of Physics, McGill University, 3600 rue University, Montr\'eal, QC H3A 2T8, Canada}
    \affiliation{Trottier Space Institute, McGill University, 3550 rue University, Montr\'eal, QC H3A 2A7, Canada}
\author[0000-0003-0510-0740, gname=Kenzie, sname=Nimmo]{Kenzie Nimmo}
    \email{knimmo@mit.edu}
    \affiliation{MIT Kavli Institute for Astrophysics and Space Research, Massachusetts Institute of Technology, 77 Massachusetts Ave, Cambridge, MA 02139, USA}
\author[0000-0002-5254-243X, gname=Gavin, sname=Noble]{Gavin Noble}
    \email{gavin.noble@mail.utoronto.ca}
    \affiliation{David A.\ Dunlap Department of Astronomy and Astrophysics, 50 St. George Street, University of Toronto, ON M5S 3H4, Canada}
    \affiliation{Dunlap Institute for Astronomy and Astrophysics, 50 St. George Street, University of Toronto, ON M5S 3H4, Canada}
\author[0000-0002-8897-1973, gname=Ayush, sname=Pandhi]{Ayush Pandhi}
    \email{ayush.pandhi@mail.utoronto.ca}
    \affiliation{David A.\ Dunlap Department of Astronomy and Astrophysics, 50 St. George Street, University of Toronto, ON M5S 3H4, Canada}
    \affiliation{Dunlap Institute for Astronomy and Astrophysics, 50 St. George Street, University of Toronto, ON M5S 3H4, Canada}
\author[0009-0008-7264-1778, gname=Swarali S., sname=Patil]{Swarali S. Patil}
    \email{sp00049@mix.wvu.edu}
    \affiliation{Department of Physics and Astronomy, West Virginia University, PO Box 6315, Morgantown, WV 26506, USA }
    \affiliation{Center for Gravitational Waves and Cosmology, West Virginia University, Chestnut Ridge Research Building, Morgantown, WV 26505, USA}
\author[0000-0002-8912-0732, gname=Aaron B., sname=Pearlman]{Aaron B. Pearlman}
    \email{aaron.b.pearlman@physics.mcgill.ca}
    \affiliation{Department of Physics, McGill University, 3600 rue University, Montr\'eal, QC H3A 2T8, Canada}
    \affiliation{Trottier Space Institute, McGill University, 3550 rue University, Montr\'eal, QC H3A 2A7, Canada}
\author[0000-0003-2155-9578, gname=Ue-Li, sname=Pen]{Ue-Li Pen}
    \email{pen@cita.utoronto.ca}
    \affiliation{Institute of Astronomy and Astrophysics, Academia Sinica, Astronomy-Mathematics Building, No. 1, Sec. 4, Roosevelt Road, Taipei 10617, Taiwan}
    \affiliation{Canadian Institute for Theoretical Astrophysics, 60 St.~George Street, Toronto, ON M5S 3H8, Canada}
    \affiliation{Canadian Institute for Advanced Research, 180 Dundas St West, Toronto, ON M5G 1Z8, Canada}
    \affiliation{Dunlap Institute for Astronomy and Astrophysics, 50 St. George Street, University of Toronto, ON M5S 3H4, Canada}
    \affiliation{Perimeter Institute of Theoretical Physics, 31 Caroline Street North, Waterloo, ON N2L 2Y5, Canada}
\author[0000-0002-4795-697X, gname=Ziggy, sname=Pleunis]{Ziggy Pleunis}
    \email{z.pleunis@uva.nl}
    \affiliation{Anton Pannekoek Institute for Astronomy, University of Amsterdam, Science Park 904, 1098 XH Amsterdam, The Netherlands}
    \affiliation{ASTRON, Netherlands Institute for Radio Astronomy, Oude Hoogeveensedijk 4, 7991 PD Dwingeloo, The Netherlands}
\author[0000-0002-7738-6875, gname=J. Xavier, sname=Prochaska]{J. Xavier Prochaska}
    \email{jxp@ucsc.edu}
    \affiliation{Department of Astronomy and Astrophysics, University of California, Santa Cruz, 1156 High Street, Santa Cruz, CA 95060, USA}
    \affiliation{Kavli Institute for the Physics and Mathematics of the Universe (Kavli IPMU), 5-1-5 Kashiwanoha, Kashiwa, 277-8583, Japan}
    \affiliation{Division of Science, National Astronomical Observatory of Japan, 2-21-1 Osawa, Mitaka, Tokyo 181-8588, Japan}
\author[0000-0001-7694-6650, gname=Masoud, sname=Rafiei-Ravandi]{Masoud Rafiei-Ravandi}
    \email{masoudrafieiravandi@gmail.com}
    \affiliation{Department of Physics, McGill University, 3600 rue University, Montr\'eal, QC H3A 2T8, Canada}
\author[0000-0001-5799-9714, gname=Scott, sname=Ransom]{Scott Ransom}
    \email{sransom@nrao.edu}
    \affiliation{National Radio Astronomy Observatory, 520 Edgemont Rd, Charlottesville, VA 22903, USA}
\author[0000-0003-3463-7918, gname=Andre, sname=Renard]{Andre Renard}
    \email{andre@renard.io}
    \affiliation{Dunlap Institute for Astronomy and Astrophysics, 50 St. George Street, University of Toronto, ON M5S 3H4, Canada}
\author[0000-0002-4623-5329, gname=Mawson W., sname=Sammons]{Mawson W. Sammons}
    \email{mawson.sammons@mcgill.ca}
    \affiliation{Department of Physics, McGill University, 3600 rue University, Montr\'eal, QC H3A 2T8, Canada}
    \affiliation{Trottier Space Institute, McGill University, 3550 rue University, Montr\'eal, QC H3A 2A7, Canada}
\author[0000-0003-3154-3676, gname=Ketan R., sname=Sand]{Ketan R. Sand}
    \email{ketan.sand@mail.mcgill.ca}
    \affiliation{Department of Physics, McGill University, 3600 rue University, Montr\'eal, QC H3A 2T8, Canada}
    \affiliation{Trottier Space Institute, McGill University, 3550 rue University, Montr\'eal, QC H3A 2A7, Canada}
\author[0000-0002-7374-7119, gname=Paul, sname=Scholz]{Paul Scholz}
    \email{pscholz@yorku.ca}
    \affiliation{Department of Physics and Astronomy, York University, 4700 Keele Street, Toronto, ON MJ3 1P3, Canada}
    \affiliation{Dunlap Institute for Astronomy and Astrophysics, 50 St. George Street, University of Toronto, ON M5S 3H4, Canada}
\author[0000-0002-4823-1946, gname=Vishwangi, sname=Shah]{Vishwangi Shah}
    \email{vishwangi.shah@mail.mcgill.ca}
    \affiliation{Department of Physics, McGill University, 3600 rue University, Montr\'eal, QC H3A 2T8, Canada}
    \affiliation{Trottier Space Institute, McGill University, 3550 rue University, Montr\'eal, QC H3A 2A7, Canada}
\author[0000-0002-6823-2073, gname=Kaitlyn, sname=Shin]{Kaitlyn Shin}
    \email{kshin@mit.edu}
    \affiliation{MIT Kavli Institute for Astrophysics and Space Research, Massachusetts Institute of Technology, 77 Massachusetts Ave, Cambridge, MA 02139, USA}
    \affiliation{Department of Physics, Massachusetts Institute of Technology, 77 Massachusetts Ave, Cambridge, MA 02139, USA}
\author[0000-0003-2631-6217, gname=Seth R., sname=Siegel]{Seth R. Siegel}
    \email{ssiegel@perimeterinstitute.ca}
    \affiliation{Perimeter Institute of Theoretical Physics, 31 Caroline Street North, Waterloo, ON N2L 2Y5, Canada}
    \affiliation{Department of Physics, McGill University, 3600 rue University, Montr\'eal, QC H3A 2T8, Canada}
    \affiliation{Trottier Space Institute, McGill University, 3550 rue University, Montr\'eal, QC H3A 2A7, Canada}
\author[0009-0008-6718-172X, gname=Sloane, sname=Sirota]{Sloane Sirota}
    \email{sloane.sirota@mail.mcgill.ca}
    \affiliation{Department of Physics, McGill University, 3600 rue University, Montr\'eal, QC H3A 2T8, Canada}
    \affiliation{Trottier Space Institute, McGill University, 3550 rue University, Montr\'eal, QC H3A 2A7, Canada}
    \affiliation{Department of Physics and Astronomy, West Virginia University, PO Box 6315, Morgantown, WV 26506, USA }
    \affiliation{Center for Gravitational Waves and Cosmology, West Virginia University, Chestnut Ridge Research Building, Morgantown, WV 26505, USA}
\author[0000-0002-2088-3125, gname=Kendrick, sname=Smith]{Kendrick Smith}
    \email{kmsmith@perimeterinstitute.ca}
    \affiliation{Perimeter Institute of Theoretical Physics, 31 Caroline Street North, Waterloo, ON N2L 2Y5, Canada}
\author[0000-0001-9784-8670, gname=Ingrid, sname=Stairs]{Ingrid Stairs}
    \email{stairs@astro.ubc.ca}
    \affiliation{Department of Physics and Astronomy, University of British Columbia, 6224 Agricultural Road, Vancouver, BC V6T 1Z1 Canada}
\author[0000-0002-9761-4353, gname=David C., sname=Stenning]{David C. Stenning}
    \email{dstennin@sfu.ca}
    \affiliation{Department of Statistics and Actuarial Science, 8888 University Dr W, Burnaby, BC V5A 1S6, Canada}
\author[0000-0003-2548-2926, gname=Shriharsh P., sname=Tendulkar]{Shriharsh P. Tendulkar}
    \email{shriharsh@gmail.com}
    \affiliation{Department of Astronomy and Astrophysics, Tata Institute of Fundamental Research, Mumbai, 400005, India}
    \affiliation{National Centre for Radio Astrophysics, Post Bag 3, Ganeshkhind, Pune, 411007, India}
\author[0000-0003-4535-9378, gname=Keith, sname=Vanderlinde]{Keith Vanderlinde}
    \email{keith.vanderlinde@utoronto.ca}
    \affiliation{Dunlap Institute for Astronomy and Astrophysics, 50 St. George Street, University of Toronto, ON M5S 3H4, Canada}
    \affiliation{David A.\ Dunlap Department of Astronomy and Astrophysics, 50 St. George Street, University of Toronto, ON M5S 3H4, Canada}
\author[0000-0002-6408-4181, gname=Mike, sname=Walmsley]{Mike Walmsley}
    \email{m.walmsley@utoronto.ca}
    \affiliation{Dunlap Institute for Astronomy and Astrophysics, 50 St. George Street, University of Toronto, ON M5S 3H4, Canada}
\author[0000-0002-1491-3738, gname=Haochen, sname=Wang]{Haochen Wang}
    \email{hcwang96@mit.edu}
    \affiliation{MIT Kavli Institute for Astrophysics and Space Research, Massachusetts Institute of Technology, 77 Massachusetts Ave, Cambridge, MA 02139, USA}
    \affiliation{Department of Physics, Massachusetts Institute of Technology, 77 Massachusetts Ave, Cambridge, MA 02139, USA}
\author[0000-0001-7314-9496, gname=Dallas, sname=Wulf]{Dallas Wulf}
    \email{dallas.wulf@mcgill.ca}
    \affiliation{Department of Physics, McGill University, 3600 rue University, Montr\'eal, QC H3A 2T8, Canada}
    \affiliation{Trottier Space Institute, McGill University, 3550 rue University, Montr\'eal, QC H3A 2A7, Canada}

%% file: table_excerpt.tex
\begin{table*} 
\scriptsize 
\begin{center} 
\caption{Excerpt from \cattwo} 
\label{table:catalog_excerpt} 
\hspace{-1.5cm} 
\begin{tabular}{ccccccccc}
\Xhline{4\arrayrulewidth}
\texttt{tns\_name} & \texttt{previous\_name} & \texttt{repeater\_name} & \texttt{event\_id} & \texttt{sub\_num} & \texttt{ra} & \texttt{ra\_err} & \texttt{dec} & \texttt{dec\_err} \\
 &  &  &  &  & (degrees) & (degrees) & (degrees) & (degrees) \\
\hline 
FRB20190701D & ... & ... & 43246206 & 0 & 112.0 & 0.17 & 66.7 & 0.16 \\
FRB20190701E & ... & ... & 43253151 & 0 & 138.5 & 0.18 & 61.71 & 0.034 \\
FRB20190702A & ... & FRB20180908B & 43263682 & 0 & 188.0 & 0.24 & 74.2 & 0.28 \\
FRB20190702B & ... & FRB20190303A & 43265588 & 0 & 208.1 & 0.19 & 48.3 & 0.19 \\
FRB20190702B & ... & FRB20190303A & 43265588 & 1 & 208.1 & 0.19 & 48.3 & 0.19 \\
\end{tabular}
\begin{tabular}{cccccccccc}
\Xhline{4\arrayrulewidth}
\texttt{ra\_dec\_notes} & \texttt{gl} & \texttt{gb} & \texttt{exp\_up} & \texttt{exp\_up\_err} & \texttt{exp\_low} & \texttt{exp\_low\_err} & \texttt{exp\_notes} & \texttt{bonsai\_snr} & \texttt{bonsai\_dm} \\
 & (degrees) & (degrees) & (hour) & (hour) & (hour) & (hour) &  &  & (pc cm$^{-3}$) \\
\hline 
... & 149.28 & 28.34 & 220.0 & 92 & 0.0 & 0.0 & ... & 34.4 & 934.9 \\
... & 153.28 & 40.34 & 100.0 & 89 & 0.0 & 0.0 & ... & 15.2 & 888.0 \\
... & 124.74 & 42.84 & 300.0 & 150 & 200.0 & 120 & ... & 8.0 & 195.7 \\
... & 97.61 & 65.76 & 110.0 & 58 & 0.0 & 0.0 & ... & 18.5 & 223.2 \\
... & 97.61 & 65.76 & 110.0 & 58 & 0.0 & 0.0 & ... & 18.5 & 223.2 \\
\end{tabular}
\begin{tabular}{ccccccccc}
\Xhline{4\arrayrulewidth}
\texttt{low\_ft\_68} & \texttt{up\_ft\_68} & \texttt{low\_ft\_95} & \texttt{up\_ft\_95} & \texttt{snr\_fitb} & \texttt{dm\_fitb} & \texttt{dm\_fitb\_err} & \texttt{dm\_exc\_ne2001} & \texttt{dm\_exc\_ymw16} \\
(Jy ms) & (Jy ms) & (Jy ms) & (Jy ms) &  & (pc cm$^{-3}$) & (pc cm$^{-3}$) & (pc cm$^{-3}$) & (pc cm$^{-3}$) \\
\hline 
... & 3.0 & ... & 5.6 & 45.4 & 933.37 & 0.012 & 877.3 & 879.3 \\
... & 1.1 & ... & 2.1 & 17.0 & 890.49 & 0.015 & 848.0 & 856.9 \\
31.7 & 4.8 & 96.0 & 13.7 & 10.3 & 195.5 & 0.17 & 157.1 & 164.8 \\
0.0 & 3.5 & 0.0 & 9.3 & 53.8 & 222.1 & 0.14 & 192.7 & 200.3 \\
0.0 & 3.5 & 0.0 & 9.3 & 53.8 & 222.1 & 0.14 & 192.7 & 200.3 \\
\end{tabular}
\begin{tabular}{ccccccccc}
\Xhline{4\arrayrulewidth}
\texttt{bc\_width} & \texttt{scat\_time} & \texttt{scat\_time\_err} & \texttt{flux} & \texttt{flux\_err} & \texttt{fluence} & \texttt{fluence\_err} & \texttt{fluence\_notes} & \texttt{fluence\_win\_extended} \\
(s) & (s) & (s) & (Jy) & (Jy) & (Jy ms) & (Jy ms) &  &  \\
\hline 
0.00983 & 0.0112 & 0.00054 & 1.3 & 0.74 & 8.0 & 3.7 & ... & 1 \\
0.00295 & 0.00000 & ... & 0.7 & 0.24 & 2.0 & 0.58 & ... & 1 \\
0.00393 & 0.00000 & ... & 0.4 & 0.23 & 1.6 & 0.54 & ... & 1 \\
0.02163 & 0.00000 & ... & 0.9 & 0.31 & 8.0 & 1.6 & ... & 1 \\
0.02163 & 0.00000 & ... & 0.9 & 0.31 & 8.0 & 1.6 & ... & 1 \\
\end{tabular}
\begin{tabular}{cccccccc}
\Xhline{4\arrayrulewidth}
\texttt{mjd\_400} & \texttt{mjd\_400\_err} & \texttt{mjd\_inf} & \texttt{mjd\_inf\_err} & \texttt{width\_fitb} & \texttt{width\_fitb\_err} & \texttt{sp\_idx} & \texttt{sp\_idx\_err} \\
(MJD) & (MJD) & (MJD) & (MJD) & (s) & (s) &  &  \\
\hline 
58665.870924726 & 0.0000000016 & 58665.870644841 & 0.0000000039 & 0.0005 & 0.000066 & 5.3 & 0.67 \\
58665.938930197 & 0.0000000018 & 58665.938663169 & 0.0000000049 & 0.00094 & 0.000068 & 3.0 & 1.4 \\
58666.0821143 & 0.000000030 & 58666.08205567 & 0.000000059 & 0.0018 & 0.00023 & 60.0 & 19 \\
58666.13521923 & 0.000000030 & 58666.13515263 & 0.000000052 & 0.0041 & 0.00023 & 580.0 & 41 \\
58666.1352194 & 0.000000029 & 58666.13515279 & 0.000000052 & 0.00207 & 0.000088 & 350.0 & 32 \\
\end{tabular}
\begin{tabular}{ccccccccc}
\Xhline{4\arrayrulewidth}
\texttt{sp\_run} & \texttt{sp\_run\_err} & \texttt{high\_freq} & \texttt{low\_freq} & \texttt{peak\_freq} & \texttt{chi\_sq} & \texttt{dof} & \texttt{flag\_frac} & \texttt{notes\_fitb} \\
 &  & (MHz) & (MHz) & (MHz) &  &  &  &  \\
\hline 
-19.0 & 1.4 & 650.7 & 400.2 & 459.9 & 1946309.9 & 1948367 & 0.266 & ... \\
-9.0 & 2.7 & 777.4 & 400.2 & 471.2 & 1974167.7 & 1974450 & 0.256 & ... \\
-70.0 & 21 & 749.3 & 523.6 & 626.4 & 1950374.6 & 1950474 & 0.265 & ... \\
-480.0 & 33 & 788.4 & 686.0 & 735.4 & 2026805.9 & 2029687 & 0.235 & ... \\
-310.0 & 27 & 778.3 & 654.2 & 713.5 & 2026805.9 & 2029687 & 0.235 & ... \\
\end{tabular}
\begin{tabular}{cccccc}
\Xhline{4\arrayrulewidth}
\texttt{intrachan\_flag} & \texttt{excluded\_flag} & \texttt{sidelobe\_flag} & \texttt{citizen\_science\_flag} & \texttt{catalog1\_flag} & \texttt{catalog1\_param\_flag} \\
 &  &  &  &  &  \\
\hline 
0 & 0 & 0 & 0 & 1 & 0 \\
0 & 0 & 0 & 0 & 1 & 0 \\
0 & 0 & 0 & 0 & 0 & 0 \\
0 & 0 & 0 & 0 & 0 & 0 \\
0 & 0 & 0 & 0 & 0 & 0 \\
\hline
\end{tabular}
\vspace{6pt}
\tablecomments{The full machine-readable version of this table is available in the online journal and at the \portal. A description of each data field is provided in Table~\ref{table:datafields}.  \rev{Associated data products, including total-intensity dynamic spectra, localization contours, and full-sky exposure estimates are available at the Canadian Advanced Network for Astronomical Research (CANFAR) via \data.}}
\end{center}
\end{table*}

%% file: table_missing_flagged.tex
\begin{deluxetable}{l c}
\tablecolumns{2}
\tablewidth{\columnwidth}
\setlength{\tabcolsep}{20pt}
\tabletypesize{\normalsize}
\tablecaption{Summary of Missing Data Products and Flagged Events in \cattwo \label{table:missing}}
\tablehead{\colhead{ } & \colhead{$N_{\mathrm{FRB}}$}}
\startdata
\sidehead{\emph{Missing Data Products}}
\quad Localization & 30 \\
\quad Sky Exposure & 31 \\
\quad Morphology & 84 \\
\quad Signal Strength & 454 \\
\quad Sensitivity Threshold & 479 \\
\sidehead{\emph{Flagged}}
\quad \texttt{excluded\_flag} & 369 \\
\quad \texttt{sidelobe\_flag} & 30 \\
\quad \texttt{citizen\_science\_flag} & 57 \\
\quad \texttt{intrachan\_flag} & 56 \\
\quad \texttt{catalog1\_param\_flag} & 61 \\
\quad $\neg$ \texttt{fluence\_win\_extended} & 242 \\
\enddata
\tablecomments{Out of the \nfrbtot total events in \cattwo, this table presents the number that are missing various data products or are flagged as having occurred during a period of non-nominal operations or undergone non-standard processing.  The reasons for missing values and the interpretations of flags are provided in \S\ref{sec:obs} and \S\ref{sec:catalog}.}
\end{deluxetable}

%% file: table_consistency_check.tex
\begin{deluxetable}{l c c c}[h!]
\tablecolumns{4}
\tablewidth{0.95 \textwidth}
\tabletypesize{\small}
\tablecaption{Tests for Consistency of the Observed FRB Source Rate Across Different Parameters \label{table:consistency}}
\tablehead{ & \multicolumn{3}{c}{$p$-value} \\
 \cline{2-4}  &   &   & Exclude \\
  &   &   & Nov 2021 \\
\colhead{Parameter} &
\colhead{$\frac{\mbox{S}}{\mbox{N}} \geq 10$} &
\colhead{$\frac{\mbox{S}}{\mbox{N}} \geq \frac{10}{\sigma}$} &
\colhead{$\frac{\mbox{S}}{\mbox{N}} \geq \frac{10}{\sigma}$}}
\startdata
Year of Operation & 0.011 & 0.048 & 0.12 \\
Calendar Month & 0.091 & 0.37 & 0.14 \\
Day of Week & 0.52 & 0.12 & 0.25 \\
Local Time & 0.62 & 0.74 & 0.65 \\
Solar Hour Angle & 0.40 & 0.59 & 0.59 \\
Accumulated Rainfall & 0.018 & 0.034 & 0.062 \\
Coarse Graining\tablenotemark{a} & 0.11 & 0.043 & 0.1 \\
Upgraded RFI Sifter\tablenotemark{b} & 0.026 & 0.23 & 0.34 \\
Right Ascension & $\leq 10^{-6}$ & 0.0055 & 0.011 \\
sin(zenith angle) & $\leq 10^{-6}$ & 0.038 & 0.068 \\
Galactic Longitude & $\leq 10^{-6}$ & 0.73 & \ldots \\
Galactic Latitude & $\leq 10^{-6}$ & $3.6 \times 10^{-5}$ & \ldots \\
Month of Operation\tablenotemark{c} & 0.0094 & 0.018 & 0.067 \\
\enddata
\tablecomments{The last three columns show results using different selection criteria: a fixed $\snr \geq 10$ threshold, a variable $\snr \geq 10 / \sigma$ threshold that accounts for instrument noise variations, and the same $\snr \geq 10 / \sigma$ threshold applied after excluding a single outlier month, November 2021. Here, $\sigma$ is an empirical noise proxy derived from daily pulsar measurements and sidereal autocorrelation variations, incorporating long-term sensitivity trends and Galactic transit effects. In the last two columns, variations in the source rate as a function of $\sin$(zenith angle), Galactic latitude, and Galactic longitude -- due to the primary beam response -- have been estimated via Monte Carlo simulation and accounted for in the analysis.}
\tablenotetext{a}{Detections are grouped into two time bins: before and after 19 Oct 2020, when the coarse graining in the de-dispersion algorithm changed.}
\tablenotetext{b}{Detections are grouped into two time bins: before and after 16 May 2022, when the L2/L3 RFI sifting algorithm was upgraded.}
\tablenotetext{c}{As shown in Figure~\ref{fig:rate_versus_time}.}
\end{deluxetable}

%% file: acknowledgements.tex
B.C.A. is supported by a Fonds de Recherche du Quebec—Nature et Technologies (FRQNT) Doctoral Research Award. M.B. is a McWilliams fellow and an International Astronomical Union Gruber fellow. M.B. also receives support from the McWilliams seed grant. A.M.C. is a Banting Postdoctoral Researcher. A.P.C. is a Vanier Canada Graduate Scholar. M.D. is supported by a CRC Chair, NSERC Discovery Grant, CIFAR, and by the FRQNT Centre de Recherche en Astrophysique du Qu\'ebec (CRAQ). F.A.D. is supported by the Jansky Fellowship. G.M.E. acknowledges funding from a Collaborative Research Team grant from the Canadian Statistical Sciences Institute (CANSSI), which is supported by NSERC. G.M.E. also acknowledges the support of NSERC through Discovery Grant RGPIN-2020-04554. E.F. and S.S.P. are supported by the National Science Foundation (NSF) under grant number AST-2407399. D.C.G. is supported by NSF Astronomy and Astrophysics Grant (AAG) award 2406919. J.W.T.H. and the AstroFlash research group acknowledge support from a Canada Excellence Research Chair in Transient Astrophysics (CERC-2022-00009); the European Research Council (ERC) under the European Union’s Horizon 2020 research and innovation programme (`EuroFlash'; Grant agreement No. 101098079); and an NWO-Vici grant (`AstroFlash'; VI.C.192.045). V.M.K. holds the Lorne Trottier Chair in Astrophysics \& Cosmology, a Distinguished James McGill Professorship, and receives support from an NSERC Discovery grant (RGPIN 228738-13). C. L. is supported by the Miller Institute for Basic Research at UC Berkeley. K.W.M. holds the Adam J. Burgasser Chair in Astrophysics and is supported by an NSF grant (2018490). J.M.P. acknowledges the support of an NSERC Discovery Grant (RGPIN-2023-05373). D.M. acknowledges support from the French government under the France 2030 investment plan, as part of the Initiative d'Excellence d'Aix-Marseille Universit\'e -- A*MIDEX (AMX-23-CEI-088). M.N. is a Fonds de Recherche du Quebec – Nature et Technologies (FRQNT) postdoctoral fellow. K.N. is an MIT Kavli Fellow. A.P. is funded by the NSERC Canada Graduate Scholarships -- Doctoral program. A.B.P. is a Banting Fellow, a McGill Space Institute~(MSI) Fellow, and a Fonds de Recherche du Quebec -- Nature et Technologies~(FRQNT) postdoctoral fellow. U.P.  is supported by the Natural Sciences and Engineering Research Council of Canada (NSERC), [funding reference number RGPIN-2019-06770, ALLRP 586559-23],  Canadian Institute for Advanced Research (CIFAR), AMD AI Quantum Astro. Z.P. is supported by an NWO Veni fellowship (VI.Veni.222.295). The Fast and Fortunate for FRB Follow-up team acknowledges support from NSF grants AST-1911140, AST-1910471, and AST-2206490. The National Radio Astronomy Observatory is a facility of the National Science Foundation operated under cooperative agreement by Associated Universities, Inc. SMR is a CIFAR Fellow and is supported by the NSF Physics Frontiers Center award 2020265. M.W.S. acknowledges support from the Trottier Space Institute Fellowship program. P.S. acknowledges the support of an NSERC Discovery Grant (RGPIN-2024-06266). V.S. is supported by a Fonds de Recherche du Quebec—Nature et Technologies (FRQNT) Doctoral Research Award. K.S. is supported by the NSF Graduate Research Fellowship Program. FRB Research at UBC is supported by an NSERC Discovery Grant and by the Canadian Institute for Advanced Research. D.C.S. is supported by an NSERC Discovery Grant (RGPIN-2021-03985) and by a Canadian Statistical Sciences Institute (CANSSI) Collaborative Research Team Grant. The Dunlap Institute is funded through an endowment established by the David Dunlap family and the University of Toronto. B.M.G. acknowledges the support of the Natural Sciences and Engineering Research Council of Canada (NSERC) through grant RGPIN-2015-05948, and of the Canada Research Chairs program. This publication uses data generated via the Zooniverse.org platform, development of which is funded by generous support, including a Global Impact Award from Google, and by a grant from the Alfred P. Sloan Foundation. The Dunlap Institute is funded through an endowment established by the David Dunlap family and the University of Toronto. 

%% file: table_data_fields.tex
\startlongtable\begin{deluxetable*}{l l l l}
\tablecolumns{4}
\tabletypesize{\scriptsize}
\tablewidth{\textwidth}
\tablecaption{Description of CHIME/FRB Catalog~2 Data Fields \label{table:datafields}}
\tablehead{\colhead{Column Number} & \colhead{Units} & \colhead{Label} & \colhead{Explanations}}
\startdata
0 & ... & \texttt{tns\_name} & Transient Name Server name\\
1 & ... & \texttt{previous\_name} & Previous name, if applicable\\
2 & ... & \texttt{repeater\_name} & Associated repeater name, if applicable\\
3 & ... & \texttt{event\_id} & CHIME/FRB internal event identifier\\
4 & ... & \texttt{sub\_num} & Sub-burst number (0 for single-burst FRBs; otherwise, assigned in chronological order)\\
5 & degrees & \texttt{ra} & Right ascension in decimal degrees (J2000)\\
6 & degrees & \texttt{ra\_err} & Uncertainty in \texttt{ra} in degrees of arc\\
7 & degrees & \texttt{dec} & Declination in decimal degrees (J2000)\\
8 & degrees & \texttt{dec\_err} & Uncertainty in \texttt{dec}\\
9 & ... & \texttt{ra\_dec\_notes} & Notes on source localization\\
10 & degrees & \texttt{gl} & Galactic longitude\\
11 & degrees & \texttt{gb} & Galactic latitude\\
12 & hour & \texttt{exp\_up} & Exposure for upper transit of source\\
13 & hour & \texttt{exp\_up\_err} & Uncertainty in \texttt{exp\_up}\\
14 & hour & \texttt{exp\_low} & Exposure for lower transit of source\\
15 & hour & \texttt{exp\_low\_err} & Uncertainty in \texttt{exp\_low}\\
16 & ... & \texttt{exp\_notes} & Notes on exposure estimates\\
17 & ... & \texttt{bonsai\_snr} & Detection S/N\\
18 & pc cm$^{-3}$ & \texttt{bonsai\_dm} & Detection dispersion measure\\
19 & Jy ms & \texttt{low\_ft\_68} & Fluence above which 68\% of lower-transit bursts would have been detected, lower limit\\
20 & Jy ms & \texttt{up\_ft\_68} & Fluence above which 68\% of upper-transit bursts would have been detected, lower limit\\
21 & Jy ms & \texttt{low\_ft\_95} & Fluence above which 95\% of lower-transit bursts would have been detected, lower limit\\
22 & Jy ms & \texttt{up\_ft\_95} & Fluence above which 95\% of upper-transit bursts would have been detected, lower limit\\
23 & ... & \texttt{snr\_fitb} & S/N determined using fitting algorithm \texttt{fitburst}\\
24 & pc cm$^{-3}$ & \texttt{dm\_fitb} & Dispersion measure determined using fitting algorithm \texttt{fitburst}\\
25 & pc cm$^{-3}$ & \texttt{dm\_fitb\_err} & Uncertainty in \texttt{dm\_fitb}\\
26 & pc cm$^{-3}$ & \texttt{dm\_exc\_ne2001} & \multirow{2}{*}{\parbox{4in}{Dispersion measure excess between \texttt{dm\_fitb}, or \texttt{bonsai\_dm} if \texttt{dm\_fitb} unavailable, and NE2001 assuming best-fit sky position of source}}\\
 &  &  & \\
27 & pc cm$^{-3}$ & \texttt{dm\_exc\_ymw16} & \multirow{2}{*}{\parbox{4in}{Dispersion measure excess between \texttt{dm\_fitb}, or \texttt{bonsai\_dm} if \texttt{dm\_fitb} unavailable, and YMW16 assuming best-fit sky position of source}}\\
 &  &  & \\
28 & s & \texttt{bc\_width} & Boxcar width of pulse\\
29 & s & \texttt{scat\_time} & Scattering time at 400.1953125 MHz\\
30 & s & \texttt{scat\_time\_err} & Uncertainty in \texttt{scat\_time}\\
31 & Jy & \texttt{flux} & Maximum of band-averaged flux-density time series within burst window, lower limit\\
32 & Jy & \texttt{flux\_err} & Uncertainty in \texttt{flux}\\
33 & Jy ms & \texttt{fluence} & Integral of band-averaged flux-density time series over burst window, lower limit\\
34 & Jy ms & \texttt{fluence\_err} & Uncertainty in \texttt{fluence}\\
35 & ... & \texttt{fluence\_notes} & Notes on flux and fluence estimates\\
36 & ... & \texttt{fluence\_win\_extended} & \multirow{2}{*}{\parbox{4in}{Flag for events where burst window was extended during flux and fluence estimation (1 = window extended, 0 = window not extended)}}\\
 &  &  & \\
37 & MJD & \texttt{mjd\_400} & \multirow{2}{*}{\parbox{4in}{Time of arrival at topocentric CHIME location referenced to 400.1953125 MHz for specific sub-burst; MJD UTC}}\\
 &  &  & \\
38 & MJD & \texttt{mjd\_400\_err} & Uncertainty in \texttt{mjd\_400}\\
39 & MJD & \texttt{mjd\_inf} & \multirow{2}{*}{\parbox{4in}{Time of arrival at topocentric CHIME location referenced to infinite frequency for specific sub-burst; MJD UTC}}\\
 &  &  & \\
40 & MJD & \texttt{mjd\_inf\_err} & Uncertainty in \texttt{mjd\_inf}\\
41 & s & \texttt{width\_fitb} & Width of sub-burst using \texttt{fitburst}\\
42 & s & \texttt{width\_fitb\_err} & Uncertainty in \texttt{width\_fitb}\\
43 & ... & \texttt{sp\_idx} & Spectral index for sub-burst\\
44 & ... & \texttt{sp\_idx\_err} & Uncertainty in \texttt{sp\_idx}\\
45 & ... & \texttt{sp\_run} & Spectral running for sub-burst\\
46 & ... & \texttt{sp\_run\_err} & Uncertainty in \texttt{sp\_run}\\
47 & MHz & \texttt{high\_freq} & \multirow{2}{*}{\parbox{4in}{Highest frequency where best-fit spectral model for sub-burst exceeds 10\% of maximum, clipped at band edge}}\\
 &  &  & \\
48 & MHz & \texttt{low\_freq} & \multirow{2}{*}{\parbox{4in}{Lowest frequency where best-fit spectral model for sub-burst exceeds 10\% of maximum, clipped at band edge}}\\
 &  &  & \\
49 & MHz & \texttt{peak\_freq} & Frequency of maximum of best-fit spectral model for sub-burst\\
50 & ... & \texttt{chi\_sq} & chi-squared from \texttt{fitburst}\\
51 & ... & \texttt{dof} & Number of degrees of freedom in \texttt{fitburst}\\
52 & ... & \texttt{flag\_frac} & Fraction of spectral channels flagged in \texttt{fitburst}\\
53 & ... & \texttt{notes\_fitb} & Notes or comments on \texttt{fitburst} results\\
54 & ... & \texttt{intrachan\_flag} & \multirow{2}{*}{\parbox{4in}{Flag for events where intrachannel dispersion was not modeled during fit (1 = unmodeled, 0 = modeled)}}\\
 &  &  & \\
55 & ... & \texttt{excluded\_flag} & \multirow{2}{*}{\parbox{4in}{Flag for events that should be excluded from parameter inference due to non-nominal telescope operation (1 = excluded, 0 = included)}}\\
 &  &  & \\
56 & ... & \texttt{sidelobe\_flag} & \multirow{2}{*}{\parbox{4in}{Flag for events that were detected in sidelobes of CHIME primary beam (1 = sidelobe, 0 = mainlobe)}}\\
 &  &  & \\
57 & ... & \texttt{citizen\_science\_flag} & \multirow{2}{*}{\parbox{4in}{Flag for low S/N events first identified by citizen scientists on Zooniverse platform (1 = citizen science detection, 0 = regular detection)}}\\
 &  &  & \\
58 & ... & \texttt{catalog1\_flag} & Flag indicating whether event was included in Catalog 1 (1 = included, 0 = not included)\\
59 & ... & \texttt{catalog1\_param\_flag} & \multirow{2}{*}{\parbox{4in}{Flag indicating whether \texttt{fitburst} and/or flux/fluence pipeline results are taken from Catalog 1  (1 = Catalog 1, 0 = Catalog 2)}}\\
 &  &  & \\
\enddata
\end{deluxetable*}